\documentclass[preprint]{aastex}
\shorttitle{Photometry of SNe~Ia}
\shortauthors{Contreras et al.}

\begin{document}

\title{The Carnegie Supernova Project:\\ First Photometry Data Release of
Low-Redshift Type~Ia Supernovae\altaffilmark{1}}

\author{Carlos~Contreras\altaffilmark{2},
Mario~Hamuy\altaffilmark{3},
M.~M.~Phillips\altaffilmark{2},
Gast\'on~Folatelli\altaffilmark{2,3},
Nicholas~B.~Suntzeff\altaffilmark{4,5},
S.~E.~Persson\altaffilmark{6},
Maximilian Stritzinger\altaffilmark{2,7},
Luis~Boldt\altaffilmark{2},
Sergio~Gonz\'alez\altaffilmark{2},
Wojtek~Krzeminski\altaffilmark{2},
Nidia~Morrell\altaffilmark{2}, 
Miguel~Roth\altaffilmark{2},
Francisco~Salgado\altaffilmark{2,3},
Mar\'{i}a Jos\'{e} Maureira\altaffilmark{3},
Christopher~R.~Burns\altaffilmark{6},
W.~L.~Freedman\altaffilmark{6},
Barry~F.~Madore\altaffilmark{6,8},
David Murphy\altaffilmark{6},
Pamela Wyatt\altaffilmark{8},
Weidong Li\altaffilmark{9}, and
Alexei~V.~Filippenko\altaffilmark{9}
}

\altaffiltext{1}{This paper includes data gathered with the 
6.5-m Magellan Telescopes located at Las Campanas Observatory, Chile.}
\altaffiltext{2}{Las Campanas Observatory, Carnegie Observatories,
  Casilla 601, La Serena, Chile.}
\altaffiltext{3}{Universidad de Chile, Departamento de Astronom\'{\i}a,
  Casilla 36-D, Santiago, Chile.}
\altaffiltext{4}{Texas A\&M University, Physics Department, College
  Station, TX 77843-4242.}
\altaffiltext{5}{Mitchell Institute for Fundamental Physics and Astronomy.}
\altaffiltext{6}{Observatories of the Carnegie Institution of
  Washington, 813 Santa Barbara St., Pasadena, CA 91101.}
\altaffiltext{7}{Dark Cosmology Centre, Niels Bohr Institute, University
of Copenhagen, Juliane Maries Vej 30, 2100 Copenhagen \O, Denmark.}
\altaffiltext{8}{Infrared Processing and Analysis Center, Caltech/Jet
  Propulsion Laboratory, Pasadena, CA 91125.}
\altaffiltext{9}{Department of Astronomy, University of California,
  Berkeley, CA 94720-3411.}

\begin{abstract}
  \noindent 
The Carnegie Supernova Project is a five-year survey being carried out
at the Las Campanas Observatory to obtain 
high-quality light curves of $\sim$100 low-redshift Type~Ia supernovae in 
a well-defined photometric system.  Here we present the first release of 
photometric data that contains the optical light curves of 35 Type~Ia 
supernovae, and near-infrared light curves for a subset of 25 events. 
The data comprise 5559 optical ($ugriBV$) and 1043 near-infrared ($YJHK_s$)
data points in the natural system of the Swope telescope.
Twenty-eight supernovae have pre-maximum data, and for 15 of these, the 
observations begin at least 5 days before $B$ maximum.  This is one of the 
most accurate datasets of low-redshift Type~Ia supernovae published to date. 
When completed, the CSP dataset will constitute a fundamental reference for 
precise determinations of cosmological parameters, and serve as a rich 
resource for comparison with models of Type~Ia supernovae.
\end{abstract}

\keywords{galaxies: distances and redshifts -- supernovae: general}

\section{INTRODUCTION}
\label{sec:intro}

The observation that the expansion rate of the Universe is currently 
accelerating is arguably one of the most profound discoveries in modern 
astrophysics.  The first direct evidence of this phenomenon was provided a 
decade ago by the Hubble diagram of high-redshift Type~Ia supernovae 
\citep[SNe~Ia;][]{riess98,perlmutter99}. Since then, a wealth of 
data obtained from surveys of nearby and distant SNe~Ia has confirmed 
this conclusion, as have other methods such as the X-ray cluster distances
\citep[e.g.,][]{allen07} and the late-time integrated Sachs-Wolfe effect using 
cosmic microwave background radiation observations \citep[e.g.,][]{giannantonio08}.
These findings suggest that a new form of energy permeates the Universe, or 
that the theory of General Relativity breaks down on cosmological scales.

Today, the major challenge is to determine the nature of this mysterious 
energy (commonly referred to as ``dark energy'') by measuring its 
equation-of-state parameter, $w = P/(\rho c^2)$, 
and the corresponding time derivative  
$\dot w$.  SNe~Ia are playing an essential role in the endeavor to measure 
$w$. Both the SN Legacy Survey \citep{astier06} and the ESSENCE project 
\citep{wood-vasey07} have recently provided independent constraints on $w$ 
that favor a cosmological constant ($w = -1$).  The determination of $\dot w$, 
however, will require new and extensive samples of both low- and high-redshift 
SNe~Ia with systematic errors below 1--2\%.  This will require
the construction of a database of low-redshift light curves with excellent
 photometric precision and temporal coverage

The pioneering work of producing such a low-redshift sample of SNe~Ia 
was carried out by the Cal{\' a}n/Tololo Survey, which published 
$BVRI$ light curves of 
29 events \citep{hamuy96}. This is the local sample that was 
used by \citet{riess98} and \citet{perlmutter99} to detect the 
accelerating expansion of the Universe.  In 1999, the CfA Supernova Group 
released a set of 22 $BVRI$ light curves \citep{riess99}.  Subsequently, 
this group has published $UBVRI$ light curves of 44 events \citep{jha06},
and $UBVRIr'i'$ light curves of another 185 \citep{hicken09}.
Other significant samples of nearby SNe~Ia are being
produced by the Lick Observatory Supernova Search 
\citep[LOSS;][]{filippenko01,filippenko05,filippenko09}
and the Nearby Supernova Factory \citep{aldering02}.
In addition, the SDSS-II Supernova Survey \citep{frieman08} has
obtained $u'g'r'i'z'$ light curves of a sample of $\sim$500 
spectroscopically confirmed SNe~Ia in the redshift range of 
$0.05 < z < 0.35$. 

In September 2004, the Carnegie Supernova Project (hereafter CSP)
began a five-year program to obtain densely sampled optical {\em and} 
near-infrared light curves of $\sim$100 nearby SNe~Ia.  The overriding 
goal has been to obtain the highest possible precision in a stable, 
well-understood photometric system.  An additional objective is to use 
the broad wavelength leverage afforded by these observations to set 
stringent constraints on the properties of host-galaxy extinction 
\citep[see][]{folatelli09}.  In this paper, we present final photometry 
of the first 35 SNe~Ia observed in the years 2004--2006. This dataset consists 
of optical ($u'g'r'i'BV$) light curves of all 35 SNe, and near-infrared 
($YJHK_s$) light curves for a subset of 25.  In total, 5559 optical and 
1043 near-infrared measurements were obtained at a typical precision of 
0.01--0.03~mag, making this one of the most homogeneous and accurate 
sets of nearby SN~Ia light curves yet obtained.  A distinguishing aspect 
of the CSP is the quantity and quality of the near-infrared photometry, 
which is matched only by the recently published PAIRITEL results
\citep{wood08}.  

Previously, \citet[][hereafter H06]{hamuy06} gave a detailed description of 
the CSP observing methodology and data reduction procedures, and showed several 
examples of light curves obtained during the first campaign.  In a second 
paper, \citet{folatelli06} reported on the unique Type Ib SN~2005bf,
while the third CSP paper presented a detailed analysis of the peculiar 
Type~Ia SN 2005hk \citep{phillips07}.  A fourth paper on the distance to the 
Antennae Galaxies (NGC~4038/39) based on the Type~Ia SN~2007sr was presented 
by \citet{schweizer08}.  CSP photometry has also been included in case 
studies of the subluminous Type~Ia SN~2005bl \citep{taubenberger08} and 
the Type~Ia/IIn SN~2005gj \citep{prieto07}. Most recently, 
\citet{stritzinger09} published a comprehensive study on the 
Type~Ib SN~2007Y.

The organization of this paper is as follows. Section~\ref{sec:obs} describes
the observing and basic data reduction procedures, Section~\ref{sec:seq} 
reviews the establishment of the local photometric sequences, 
Section~\ref{sec:phot} gives details of the measurement and calibration of the
SN photometry, and Section~\ref{sec:results} presents the final light
curves.  In two Appendices, we present our most up-to-date information on
the sensitivity functions of our $u'g'r'i'BV$ bandpasses, and near-infrared 
photometry of Feige~16 obtained over the course of our observing campaigns 
which we use to check the validity of our $Y$-band calibration.

Accompanying this paper are two other papers. 
The first gives an analysis of the data presented here \citep{folatelli09}.
The second combines the data of this article with that of 35 high-redshift 
SNe~Ia, observed 
in the near-infrared with the Magellan Baade 6.5-m telescope, in order to 
construct the first rest-frame $i$-band Hubble diagram out to
$z \approx 0.7$ \citep{freedman09}.

\section{OBSERVATIONS AND BASIC DATA REDUCTIONS}
\label{sec:obs}

The CSP carries out one nine-month campaign each year from September through
May. All the data presented below were obtained during the first three
campaigns from 2004 through 2006. In H06, the observing procedures and data 
reduction techniques were described in detail; below we briefly summarize.  
Since the publication of H06, the optical photometry reductions 
have been modified so that 
the magnitudes of each SN~Ia are computed in the {\em natural} system of the 
Swope telescope CCD imager. This new approach is discussed in 
Section \ref{sec:nat}.

\subsection{Sample Selection}

A list of the general properties of the 35 SNe~Ia (17, 14, and 4 
in the first, second, and third campaigns, respectively) 
whose light curves are 
presented in this paper is given in Table \ref{tab:sne1}. This table 
includes (1) the  equatorial coordinates of each supernova, 
(2) the name, morphology, and heliocentric redshift of the host galaxy, 
(3) the discovery reference, and (4) the discovery group or individual.
As is seen, approximately one third of the SNe were discovered
by LOSS with the Katzman Automatic Imaging Telescope (KAIT), while the remainder
are drawn from a number of sources.  In general, the only criteria used
by the CSP to select which SN to follow were (1) discovery before or
near maximum brightness, and (2) a magnitude no fainter than 18--18.5 at 
the time of discovery. However, each of the discovery groups or individuals 
listed in Table \ref{tab:sne1} employs different search strategies 
(cadence, depth, target selection, etc.).  In a future paper, we plan to 
explore the possible biases introduced by such a heterogeneous 
sample.  Here, we briefly describe the completeness
of the LOSS discoveries and how this relates to Malmquist bias.

LOSS is a targeted search which annually monitors a sample of
$\sim14,000$ galaxies \citep[][and references therein]{filippenko09}.  
In 2004, the search was deliberately adjusted to include more galaxies south 
of the celestial equator, to enlarge the potential sample accessible to 
the CSP.  In order 
to study the completeness of the survey, we constructed an observed 
luminosity function for LOSS SNe~Ia from 56 discoveries within 
80~Mpc.    Based on this luminosity function, we conclude that 
LOSS has missed very few SNe~Ia in
its sample galaxies within $z \leq 0.04$.  Over most of the observed
luminosity range, the expected completeness is $> 80$\% 
at $z = 0.04$.  Since most of the LOSS galaxies are within 
100~Mpc ($z = 0.024$), the actual missing fraction of SNe~Ia is 
quite small.

Hence, a program that follows all of the
LOSS SN~Ia discoveries should end up with a fairly complete 
sample, with minimal Malmquist bias.
Of course, the CSP is only able to follow those LOSS SNe which
are accessible from the southern hemisphere.  For these galaxies,
which are located at relatively large zenith distances at Lick 
Observatory, the completeness is probably somewhat smaller due 
to the difficulty of the observations (seeing conditions, airmass, etc.).
Nevertheless, we expect that Malmquist bias for the 
subset of the LOSS SNe~Ia
at $z \leq 0.04$ followed by the CSP should be small.
Of the 12 LOSS SNe included in Table \ref{tab:sne1}, only one
(SN~2005ag) lies at $z > 0.04$.

\subsection{Observations}

Our follow-up program focuses on supernovae discovered by professional 
and amateur astronomers who report them  to the IAU Circulars (IAUC), the 
IAU Electronic Telegrams (CBET), or occasionally, directly to us via 
electronic mail.  When word arrives of a new candidate, we place it in our 
follow-up queue.  Photometric monitoring is obtained entirely at the Las 
Campanas Observatory (LCO).  The optical follow-up observations are conducted 
with the SITe3 CCD camera attached to the Swope 1-m telescope, which has been 
available to the CSP $\sim$250 nights per campaign. During the first campaign, 
the near-infrared follow-up was carried out solely with the Wide Field 
Infrared Camera (WIRC) attached to the du Pont 2.5-m telescope.  However, by 
the start of the second campaign, a new near-infrared camera called RetroCam, 
built by S.E.P. and D.M. for this program, became available on the Swope telescope. 
The addition of RetroCam significantly enhanced our ability to obtain detailed 
$YJH$-band light curves.  Nevertheless, we have continued to use WIRC to 
observe the fainter targets. In several cases, supplementary imaging has 
also been performed with the near-infrared imager, PANIC, attached to the 
Magellan Baade telescope.

In addition to follow-up observations, the du Pont telescope is used 
to obtain optical and near-infrared template images of the host galaxies. 
These templates are used to subtract away the host-galaxy contamination in 
each of our science images before instrumental magnitudes are computed 
(see $\S$~\ref{sec:phot1}).

Figure~\ref{fig:fcharts} presents a mosaic constructed with a
$V$-band image of each supernova. In Table \ref{tab:sne2}, the
following photometric and spectroscopic 
information is collected for each
supernova:  (1) an estimate of the decline-rate parameter 
$\Delta m_{15}(B)$ \citep{phillips93}, (2) the number of optical 
and near-infrared epochs each event was observed, (3) the number of 
photometric nights that the optical and near-infrared local sequences were 
calibrated, and (4) the spectral 
subtype of each event and the epoch of the spectrum used to make the identification.  

Spectral identification was performed using the SuperNova IDentification 
tool SNID \citep{blondin07}.
Twenty-six events are identified as normal, four as SN 1991bg-like, two 
as SN~1991T-like, and one as SN~2006gz-like.
The  identification of a dozen of the normal subtypes was obtained with
spectra sufficiently past maximum that a  SN~1991T-like identification 
cannot be excluded. However, out of these dozen events, 
only SN~2005lu  ($\Delta m_{15}(B) = 0.88 \pm 0.03$) has a $B$-band 
decline rate that is similar to that of other SN~1991T-like events, 
which typically exhibit $\Delta m_{15}(B)$ values of 0.9 or less.

Figure~\ref{fig:spectra} compares our day $-$3 spectrum of SN~2004gu to 
similar-epoch spectra of the peculiar SNe 1999ac \citep{phillips06} and 
2006gz \citep{hicken07,maeda09}.  As is evident, SN~2004gu resembled both of these 
events, but clearly a better overall match is with SN~2006gz.  The spectra of 
SNe~2004gu and 2006gz show an unusually strong \ion{Si}{3} line at 4560 \AA\, 
while the \ion{Ca}{2} H\&K feature appears to be more prevalent in SN~2004gu. 
Interestingly, SNe 2004gu and 2006gz have nearly identical decline rates of
$\Delta m_{15}(B) \approx 0.70$--0.80 mag.  Moreover, for a reddening law of 
$R_{V} = 1.2$, we compute very similar absolute peak magnitudes of 
$M_{B} = -19.63 \pm 0.09$ for SN~2004gu \citep{folatelli09} and 
$M_{B} = -19.57 \pm 0.16$ for SN~2006gz \citep{hicken07}.

\subsection{Data Reduction}

Basic data reductions are generally carried out the day following the
observation. This allows us to measure preliminary photometry to
obtain a prompt estimate of the age and brightness of the SN.
This information is particularly useful to help us decide whether
or not a new target merits extensive follow-up observations.

The optical images are reduced with the IRAF\footnote[9]{IRAF is distributed 
by the National Optical Astronomy Observatories, which are operated by the 
Association of Universities for Research in Astronomy, Inc., under 
cooperative agreement with the National Science Foundation.} task CCDPROC. 
The basic reduction process consists of (1) bias subtraction, 
(2) flat-fielding, (3) application of a linearity correction appropriate 
for the SITe3 CCD, and (4) an exposure-time correction that corrects for a 
shutter time delay.  Both the linearity and exposure-time corrections were 
monitored frequently during the first three campaigns, and the results are 
completely consistent with those given in Appendices A and B of H06.

For basic reduction of the near-infrared imaging, we have developed a specific 
data reduction pipeline (written in C that uses IRAF libraries) for each 
instrument at LCO.  These pipelines apply to each image corrections for 
detector linearity, electronic bias, pixel-to-pixel variations of the 
detector's sensitivity, and sky background.  During the September--May
period, the atmospheric conditions at LCO are typically quite stable.  Through 
the course of 
the CSP, we found that significant variations in seeing did not 
normally occur between the individual dithered images. Hence, as a 
general rule, once all of the dithered images were fully 
reduced, they were aligned and stacked without convolving the
image quality to make a master frame which was
then used to compute photometry.

\section{PHOTOMETRIC SEQUENCES}
\label{sec:seq}

The brightness of each supernova is obtained differentially with respect to a 
set of local sequence stars.  Absolute photometry of each local sequence is 
derived using observations of photometric standards generally 
obtained on a minimum of three photometric nights (see columns 5 and 6 of 
Table~\ref{tab:sne2}). In the optical, we use standards  that are in 
common with both the  \citet{landolt92} and \citet{smith02} catalogs.  
Aperture photometry of the standard stars is computed using
an aperture with a radius of 7$\arcsec$.

To place the instrumental magnitudes of the local sequence stars 
on the standard \citeauthor{landolt92} $BV$ and \citeauthor{smith02} 
(USNO40) $u'g'r'i'$ photometric systems, we use the transformation 
equations 1--6 given in H06.  Figures~\ref{fig:extinc} and 
\ref{fig:colorc} show plots of the extinction and color coefficients 
employed in these transformation equations, which were derived on 
photometric nights during the first four CSP campaigns. These figures 
illustrate the stability of our photometric system, as well as the excellent 
quality of the LCO site where, during September--May,
conditions are photometric $\sim65$\% of the time. 
When determining absolute photometry of the 
local sequences, we have used the average values of the color coefficients 
obtained during the first campaign (see Figure 4 of H06). These values 
differ only slightly from the four-year averages shown in 
Figure \ref{fig:colorc}.  
The final magnitudes reported for each 
sequence star were determined from an average of measurements that were 
calibrated to standard fields typically on a minimum of three photometric nights.

The color terms shown in Figure~\ref{fig:colorc} can be used to 
improve our knowledge of the response functions of the CSP optical
filters.  In Appendix~\ref{afilters}, we describe such an analysis.  In
summary, we find that matching the color terms for the $BV$ and $g'r'i'$
filters requires only minor modifications of the bandpasses given in 
H06.  However, the response function of the CSP $u'$ filter clearly
differs significantly from that provided in H06.

Absolute photometry of the near-infrared local sequences was calibrated
with observations of \citet{persson98} standards.  Standard $Y$-band 
magnitudes of the \citeauthor{persson98} standards were derived from  
$J$- and $K_s$-band measurements as described by equation C2 of H06.  
As detailed in Appendix~\ref{feige16}, observations of the A0V star 
Feige~16 obtained during the course of the CSP campaigns confirm that the 
$Y$-band calibration given in H06 is consistent with the \citet{elias82} 
system, where $\alpha$~Lyrae has 0.00 color.  As in H06, we neglect any 
color term in the near-infrared (see equations 8--11 in H06) since 
we have been using essentially the same filters employed by 
\citeauthor{persson98} to establish the standard system.  This assumption 
will be examined and tested in a future paper.

Final magnitudes of the local sequences in the {\em standard}
system are listed in Table \ref{tab:optir_stds}. The accompanying
uncertainties are weighted averages of the instrumental errors 
computed from multiple measurements of each sequence star 
obtained on different photometric nights.

As many of our near-infrared local standards are also included in the 2MASS 
catalog\footnote[10]{http://www.ipac.caltech.edu/2mass/ .}, we have carried out 
a comparison of photometry.  A total of 984 stars were identified with 
$J$- and $H$-band measurements in both systems, whereas for $K_s$ we found 
only 41 stars in common.  Generally speaking, the uncertainties in the 2MASS 
measurements are a factor of $\sim$2 larger than those in the CSP photometry.
Concentrating on the $J$ and $H$ bands, we find the following mean offsets 
between the zero points of the CSP and 2MASS magnitudes:

\begin{displaymath}
J_{\rm CSP} - J_{\rm 2MASS} = 0.010 \pm 0.003~{\rm mag,}
\end{displaymath}
\begin{displaymath}
H_{\rm CSP} - H_{\rm 2MASS} = 0.043 \pm 0.003~{\rm mag}.
\end{displaymath}

\noindent
To determine if there are any significant color terms, we looked 
for possible correlations of $J_{\rm CSP}-J_{\rm 2MASS}$ 
and $H_{\rm CSP}-H_{\rm 2MASS}$ vs.  ($J-H$)$_{\rm CSP}$.  
Over the range of color covered by the stars 
[$0.2 <$ ($J-H$)$_{\rm CSP} < 0.7$ mag], we obtained the following
fits:

\begin{displaymath}
J_{\rm CSP}-J_{\rm 2MASS} = (-0.045\pm0.008) \times (J-H)_{\rm CSP} + (0.035\pm0.067)
\end{displaymath}
\begin{displaymath}
H_{\rm CSP}-H_{\rm 2MASS} =  (0.005\pm0.006) \times (J-H)_{\rm CSP} + (0.038\pm0.080).
\end{displaymath}

\noindent
Thus, there is some indication of a small color term in the $J$ band, but not 
in $H$.

\section{SUPERNOVA PHOTOMETRY}
\label{sec:phot}

\subsection{Host-Galaxy Subtraction}
\label{sec:phot1}
Before final photometry of each SN is computed, a template image of the 
host galaxy is subtracted from each science frame. The templates are
obtained a year or more after the last follow-up image 
with the du Pont telescope using the same 
filters employed to take the original science images. 
By this time the brightness of the SN has decreased by at least 
$\sim$5 mag below the last follow-up measurement. 
Useful templates are $\sim$3 mag deeper than the Swope images, 
and are always obtained under seeing conditions that either exceed or 
match those of the best science frames. 

Although the template and science images are obtained with the same filters,
a  legitimate concern is whether there is a difference between the global response function of the Swope and du Pont instrument/telescope systems, which could affect the quality of our subtractions.
To ascertain if such a problem exists, we compared photometry of the 
local sequence stars derived from both Swope and du Pont images that were observed 
during photometric nights.  This exercise revealed no indication 
of any relative color terms.  However,  in the future, when
a more statistically significant 
sample of template images becomes available, we will revisit this issue. 

 To perform the galaxy subtraction, the template images are geometrically
transformed to each science image, convolved with a two-dimensional kernel 
function to match the point-spread functions (PSFs),  and then scaled in flux.
The modified template is then subtracted from the follow-up images
around the position of the SN with care taken not to subtract away any
local sequence stars. Figure~\ref{fig:subtractions} displays 
two cases of galaxy subtractions where the SN is located in a region
with substantial host-galaxy contamination. 
These  examples demonstrate the excellent subtractions that we are
able to obtain 
despite the presence of a significant background. 
Experiments done by adding a sequence of artificial stars  to the
template-subtracted images surrounding the location of the SN 
indicate that in essentially all cases
the maximum systematic errors incurred from the template subtractions
average $< 0.01$~mag.

Nevertheless, despite the excellent quality of the template subtractions,
it is still possible for the template-subtraction process to introduce a 
systematic error {\em at the specific location of the SN}.  
Such an error could, in principle, lead to
a correlated error in the light-curve photometry since the same
template is employed in the host-galaxy subtractions for all of the
SN images. We note that even in the ideal case of a perfect
subtraction, there will be a correlated error caused by the finite
signal-to-noise ratio in the template image.
We have estimated the magnitude of such correlations using the
photometry of artificial stars located at fixed positions around the
SN, for those cases with the highest background. As explained by
\citet{folatelli09}, we found negligible variations in the
light-curve fit parameters and their uncertainties when considering these
correlations.

\subsection{Photometry}

\subsubsection{Optical}
\label{sec:nat}

At this point our procedure supersedes what was described in H06.  Given 
the non-stellar nature of the spectral energy distribution (SED) of SNe~Ia, 
the conversion of instrumental magnitudes to the {\em standard} system by 
application of the color term derived from the photometric standard stars does 
not guarantee that the SN photometry will actually be on the {\em standard} 
system.  On the contrary, significant systematic errors can occur between two 
sets of observations of the same SN put on the same photometric system through 
application of color terms if the filter response functions are very different 
\citep[e.g., see Figure~2 in][]{krisciunas03}.  It is possible to place 
the SN photometry in the {\em standard} (or any other desired) photometric 
system only via application of an additional time-dependent term computed 
synthetically from template SN spectra, often referred to as an ``S-correction'' 
\citep{suntzeff00,stritzinger02}.  We emphasize that this physical effect is 
important to consider when CSP photometry is to be combined with other
optical datasets, {\em regardless of whether the photometry has been color-corrected}.

A sensible alternative which has been adopted by several other groups
(e.g., SN Legacy Survey, ESSENCE, and the CfA Supernova Group)
is to not apply the color-term and S-corrections, and explicitly
calculate the SN magnitudes in the {\em natural} photometric system
used for the observations. We thus provide our measurements in the
$ugriBV$\footnote[11]{From this point 
forward, the optical natural photometry is referred to as $ugriBV$ as opposed 
to $u'g'r'i'BV$.} natural system of the Swope 
telescope.  Photometry in the natural system is the ``purest'' form of the 
data and will facilitate in a more transparent way the combination of the CSP 
photometry with datasets from other groups. 
Moreover, photometry in this format allows one to easily incorporate 
improved SN spectral template sequences and/or passbands.

The specific procedure we have adopted to compute photometry in the natural 
system is as follows.

\noindent
(1) The magnitudes of the photometric standards given in the catalogs of 
\citet{landolt92} and \citet{smith02} are used to 
calculate new magnitudes of these stars in the natural photometric system of the 
Swope telescope using the following formulae:

\begin{equation}
u~=~u'_{\rm std}~-~ct_u \times (u'_{\rm std}~-~g'_{\rm std}),
\label{u_eq}
\end{equation}
\begin{equation}
g~=~g'_{\rm std}~-~ct_g  \times (g'_{\rm std}~-~r'_{\rm std}),
\label{g_eq}
\end{equation}
\begin{equation}
r~=~r'_{\rm std}~-~ct_r \times (r'_{\rm std}~-~i'_{\rm std}),
\label{r_eq}
\end{equation}
\begin{equation}
i~=~i'_{\rm std}~-~ct_i \times (r'_{\rm std}~-~i'_{\rm std}),
\label{i_eq}
\end{equation}
\begin{equation}
B~=~B_{\rm std}~-~ct_b \times (B_{\rm std}~-~V_{\rm std}),~{\rm and}
\label{B_eq}
\end{equation}
\begin{equation}
V~=~V_{\rm std}~-~ct_v \times (V_{\rm std}~-~i'_{\rm std}),
\label{V_eq}
\end{equation}

\noindent
where $ct_x$ are the color terms defined in equations 1-6 of H06
(and listed in Figure 4 of H06), and the magnitudes with the ``std''
 subscript are the catalog magnitudes. 
 In constructing our catalog of photometric standards, 
 great care was taken to select stars that cover a broad range in color.
 Consequently, the optical standards observed by the CSP cover a color range of  
 $-0.32 < B-V < 1.60$ mag. For comparison, the vast majority 
of local sequence stars fall
 well within this color range.
 
 \noindent
(2) The new catalog of revised photometry is then used to generate 
natural magnitudes for the local sequence of  each SN field from observations 
obtained on photometric nights.  The transformation equation in each band 
contains an extinction term and a zero point,

\begin{displaymath}
m_{nat}~=~m_{inst}~-~k_i~X_i~+~zp_i~ .
\end{displaymath}

\noindent
where $k_i$ is the extinction coefficient and $X_i$ is the effective
airmass. Note that, unlike equations 1-6 of H06, this equation 
does not include a color term. 

\noindent
(3) Instrumental magnitudes of the local sequence stars and the SN are 
measured on each galaxy-subtracted frame via PSF fitting.  Next these 
measurements of the local sequence are compared to the calibrated values 
in the natural system in order to derive a nightly zero point.  This 
zero point is then used to place the extinction-corrected ($k_{i}X$) 
instrumental magnitude of the SN  in the natural system.  

The treatment of uncertainties is improved with respect to the
procedures described in H06. The adopted uncertainty in each
magnitude measurement is the result of the sum in quadrature of two
components: (a) the uncertainty in the instrumental magnitude, as
estimated from the Poisson noise model of the flux of the SN and the
surrounding background, and (b) the uncertainty in the fit of the zero
point of the image. By considering the latter term we no longer need
to adopt an uncertainty minimum of 0.015 mag, as was done in H06.

\subsubsection{Near-Infrared}

In the near-infrared, we continue to use the procedures described in H06 
to measure the final SN magnitudes.  As mentioned in Section \ref{sec:seq}, 
we neglect any color terms in transforming the measurements made at the
Swope  and du Pont telescopes to the \citet{persson98} photometric system.  
Therefore, the local sequence is used only to compute the nightly
zero point. Regarding the magnitude uncertainties, we adopt the same
procedure as described in the previous section for the optical data:
we sum in quadrature the uncertainties in the instrumental
magnitude and in the zero-point fit.

\section{FINAL LIGHT CURVES}
\label{sec:results}

The final $ugriBV$ and $YJHK_s$ magnitudes of the 35 SNe~Ia presented
in this paper are listed in Tables~\ref{tab:opt_sne} and \ref{tab:ir_sne}, 
respectively.  These data are being made available through the
NASA/IPAC Extragalactic Database (NED), and can also be accessed on the 
CSP web site\footnote[12]{http://www.ociw.edu/csp/ .}.

Figure \ref{fig:flcurves} shows the optical and near-infrared light curves, 
along with fits carried out with SNOOPy (SuperNovae in Object-Oriented
Python; Burns et al. 2009).  The template-fitting algorithm implemented in 
SNOOPy is very simple.  We begin by selecting a sample of well-observed CSP 
supernovae that have clear maxima.  Using spline fits, we measure the time 
of $B$ maximum $t_{\rm max}$, peak magnitude $m_{\rm max}$ in each filter, and 
decline-rate parameter $\Delta m_{15}(B)$.  Each light curve is 
normalized to its peak magnitude and shifted in time such that the $B$ band 
peaks at epoch $t=0$.  Once this procedure has been done, each data point 
can be placed in a three-dimensional space based on its epoch, normalized 
magnitude, and $\Delta m_{15}(B)$.  The normalized magnitudes then define 
a sparsely sampled surface in this space.  Generating a template for a 
particular $\Delta m_{15}(B)$ and set of epochs therefore amounts to 
interpolating on this surface.  For this task, we have chosen a variation 
of the ``gloess'' algorithm \citep{madore01}, 
which is a Gaussian-windowed and error-weighted 
extension of data-smoothing methods outlined by \citet{cleveland79}, and 
first implemented by \citet{persson04} for fitting Cepheid light curves.
A subset of points situated 
around the interpolation point are selected and assigned weights using an 
adaptive elliptical Gaussian.  These points are then fit with a two-dimensional 
quadratic using weighted linear least squares.  The resulting quadratic is 
then used to interpolate the point of interest.

Given the non-linear nature of this 
interpolation algorithm, we use the Levenberg-Marquardt least squares 
algorithm to fit the light-curve templates to the data.  In doing so, we 
determine the values of $t_{\rm max}$, $\Delta m_{15}(B)$, and the peak 
magnitudes for each filter, as well as a complete covariance matrix for 
these parameters.  The light-curve fitting is performed in flux space.
Appendix~\ref{zpoints} gives details of the zero points adopted for the
CSP bandpasses.

The $\Delta m_{15}(B)$ values derived from the SNOOPy fits are given in 
Table \ref{tab:sne2}.\footnote[13]{In the case of SN~2004dt, the 
photometry began too late after maximum to obtain a reliable fit with 
SNOOPy.}  From the values of $t_{\rm max}$, we find that 28 of the 
35 SNe have pre-maximum coverage, and for 15 of these, the observations 
began at least 5 days before $B$ maximum.  In the near-infrared, 17 of the 
SNe have coverage beginning at or before maximum light, which occurs 
3--5 days prior to $B$-band maximum \citep{folatelli09}.  This was attained 
for only 4 of the 17 SNe observed during the first campaign, but was 
achieved for 13 of the 18 SNe observed in the second and third campaigns.  
This huge improvement is due to the availability of RetroCam beginning in 
the second campaign.  Since its commissioning, RetroCam has become our 
primary source of near-infrared data, and has allowed the CSP to fulfill 
its goal of obtaining excellent near-infrared coverage for the majority of 
the monitored SNe.

\section{CONCLUSIONS}
\label{sec:conc}

In this paper, we have presented the first data release of low-redshift SNe~Ia 
observed by the CSP.  The combination of pre-maximum observations, extended 
coverage, dense sampling, and high signal-to-noise ratio data produces 
a dataset of unprecedented quality.  These results will make possible 
detailed studies of the photometric properties of SNe~Ia and serve as a rich 
resource for comparison with theoretical models.  When completed, the CSP 
dataset will constitute a fundamental reference for precise determinations 
of cosmological parameters such as our own program to measure the 
$i'$-band Hubble diagram of SNe~Ia to $z \approx 0.7$ \citep{freedman09}.
The inclusion of the near-infrared data is an important addition since 
SNe~Ia are nearly perfect standard candles at these wavelengths 
\citep{krisciunas04}.  In addition, the broad wavelength range covered by 
the optical and near-infrared observations provides strong leverage for 
estimating dust extinction in the SN~Ia host galaxies \citep{folatelli09}.

\acknowledgments 

We thank James Hughes for supporting our network of computers, 
and the technical staff of Las Campanas Observatory for its help during
many observing nights. C.C. acknowledges Jos\'e Luis Prieto and 
Rodrigo Fern\'andez for their continuous support and useful discussions. 
This material is based upon work supported by the National Science 
Foundation (NSF) under grant AST--0306969. 
We also acknowledge support from {\it Hubble Space Telescope} grant 
GO-09860.07-A from the Space Telescope Science Institute, which is operated by
the Association of Universities for Research in Astronomy, Inc., under
NASA contract NAS 5-26555. M.H. acknowledges support provided by NASA 
through Hubble Fellowship grant HST-HF-01139.01-A, by Fondecyt through 
grant 1060808, from Centro de Astrof\'\i sica FONDAP 15010003, and by the 
Center of Excellence in Astrophysics and Associated Technologies (PFB 06).
G.F., M.H., and F.S. acknowledge support from the Millennium Center for
Supernova Science through grant P06-045-F funded by ``Programa Bicentenario
de Ciencia y Tecnolog\'ia de CONICYT'' and ``Programa Iniciativa Cient\'ifica
Milenio de MIDEPLAN.''  N.B.S. acknowledges the support of the 
Mitchell/Heep/Munnerlyn Chair in Astronomy at Texas A\&M University, and 
support though the Dean of the College of Sciences. 
A.V.F.'s supernova research has been funded by NSF
grants AST-0607485 and AST-0908886, as well as by the TABASGO Foundation.
KAIT and its ongoing operation were made possible by
donations from Sun Microsystems, Inc., the Hewlett-Packard Company,
AutoScope Corporation, Lick Observatory, the NSF, the University of
California, the Sylvia \& Jim Katzman Foundation, and the TABASGO
Foundation. 

\appendix

\section{OPTICAL FILTER BANDPASSES}
\label{afilters}

In order to test the CSP optical bandpass response functions given in
H06, and to improve these if possible, we employ an approach similar 
to that of \citet{stritzinger05}.  We start by multiplying the SEDs of 
the \citet{stritzinger05} spectrophotometric atlas of 102 stars 
with the Johnson $B$ and $V$ passbands from \citet{bessell90} and 
the $u'g'r'i'$ passbands from \citet{smith02} to calculate synthetic 
magnitudes.  We derive color terms and zero 
points by fitting the transformation equations 

\begin{displaymath}
B_{\rm obs}~-~B_{\rm syn}~=~ct_B \times (B_{\rm syn}~-~V_{\rm syn})~+~zp_B,
\end{displaymath}
\begin{displaymath}
V_{\rm obs}~-~V_{\rm syn}~=~ct_V \times (B_{\rm syn}~-~V_{\rm syn})~+~zp_V,
\end{displaymath}
\begin{displaymath}
u'_{\rm obs}~-~u'_{\rm syn}~=~ct_{u'} \times (u'_{\rm syn}~-~g'_{\rm syn})~+~zp_{u'},
\end{displaymath}
\begin{displaymath}
g'_{\rm obs}~-~g'_{\rm syn}~=~ct_{g'} \times (g'_{\rm syn}~-~r'_{\rm syn})~+~zp_{g'},
\end{displaymath}
\begin{displaymath}
r'_{\rm obs}~-~r'_{\rm syn}~=~ct_{r'} \times (r'_{\rm syn}~-~i'_{\rm syn})~+~zp_{r'},~{\rm and}
\end{displaymath}
\begin{displaymath}
i'_{\rm obs}~-~i'_{\rm syn}~=~ct_{i'} \times (r'_{\rm syn}~-~i'_{\rm syn})~+~zp_{i'},
\end{displaymath}

\noindent
where the ``obs'' subscript denotes the magnitudes from the \citet{landolt92} 
and \citet{smith02} catalogs.  If the color terms, $ct_X$, differ from zero, 
the bandpasses 
are shifted in wavelength without changing their shapes.  Magnitudes and 
color terms are recalculated, and the process is repeated until the color
terms are consistent with zero.

In $B$ and $V$, we used the 91 stars contained in both the 
spectrophotometric atlas and the photometric catalog of \citet{landolt92} 
which are not listed as possible variables. 
The derived filter shifts of $12.4 \pm 3$~\AA~in $B$ and $8.2 \pm
4$~\AA~in $V$ are similar to the values of 8.5~\AA~in $B$ and 6~\AA~in
$V$ found by \citet{stritzinger05}. The small differences 
may be due to the inclusion of stars noted as variable or possibly variable 
by \citet{stritzinger05} which were not included in our analysis. 

The same analysis was performed for the $u'g'r'i'$ bandpasses.  The
\citet{smith02} $u'$ filter had to be slightly modified: at wavelengths 
shorter than 3100~\AA, the transmission was set to zero because the 
\citet{stritzinger05} spectra begin at 3050~\AA.  This modification
produces photon losses of $< 6$\%.  Unfortunately, only 16 of the 102 stars
in the \citet{stritzinger05} spectrophotometric atlas are included in the
\citet{smith02} catalog.  From this limited sample, the color terms for 
the $g'$ and $r'$ filters
were found to be consistent with zero without applying any shifts.  
The color terms of the $u'$ and $i'$ filters differed from zero by 
$\sim2\sigma$, but given the small sample size, we did not
consider this conclusive.  Hence, we opted not to apply any shifts to
the \citet{smith02} $u'g'r'i'$ bandpasses.

Next, we used these revised {\em standard} bandpasses to calculate synthetic 
magnitudes in the $BV$ and $u'g'r'i'$ systems for the 91 non-variable stars 
in the \citet{stritzinger05} spectrophotometric atlas.  Final 
adjustment of the zero points for the each bandpass were derived by 
calculating synthetic magnitudes in $BV$ for $\alpha$~Lyrae and $u'g'r'i'$ 
for BD$+17^{\circ}4708$, and matching these to their standard magnitudes
as given by \citet{bessell98}, \citet{johnson53}, \citet{fukugita96}, and
\citet{smith02}.  Likewise, synthetic magnitudes in the CSP {\em natural}
system were 
calculated using the $BV$ and $ugri$ bandpasses given in H06.
Values for the color terms and zero points were derived from least-squares fits
to the transformation equations

\begin{displaymath}
B_{\rm std}~-~B_{\rm nat}~=~ct_b \times (B_{\rm nat}~-~V_{\rm nat})~+~zp_b,
\end{displaymath}
\begin{displaymath}
V_{\rm std}~-~V_{\rm nat}~=~ct_v \times (V_{\rm nat}~-~i_{\rm nat})~+~zp_v,
\end{displaymath}
\begin{displaymath}
u'_{\rm std}~-~u_{\rm nat}~=~ct_u \times (u_{\rm nat}~-~g_{\rm nat})~+~zp_u,
\end{displaymath}
\begin{displaymath}
g'_{\rm std}~-~g_{\rm nat}~=~ct_g \times (g_{\rm nat}~-~r_{\rm nat})~+~zp_g,
\end{displaymath}
\begin{displaymath}
r'_{\rm std}~-~r_{\rm nat}~=~ct_r \times (r_{\rm nat}~-~i_{\rm nat})~+~zp_r,~{\rm and}
\end{displaymath}
\begin{displaymath}
i'_{\rm std}~-~i_{\rm nat}~=~ct_i \times (r_{\rm nat}~-~i_{\rm nat})~+~zp_i,
\end{displaymath}

\noindent
where the ``std'' subscript denotes the synthetic magnitudes in the 
{\em standard} system, and the ``nat'' subscript signifies synthetic 
magnitudes calculated in the {\em natural} bandpasses.  The natural 
bandpasses were shifted in wavelength,
maintaining their shapes, until the calculated color terms matched the 
observed ones listed in Figure~4 of H06.  As illustrated in
Figures~\ref{fig:fcterms}, excellent 
results were achieved for the $BVgri$ filters by applying the following
small shifts: $B = 21.8 \pm 9.7$~\AA, $V = 28.6 \pm 17.5$~\AA, 
$g = -14.2 \pm 14.1$~\AA, $r = -5.0 \pm 17.0$~\AA, and $i = -10.1 \pm 28.6$~\AA.
The quoted errors for these shifts were derived from matching the formal errors
of the observed color terms (see Figure~\ref{fig:colorc}).

In the case of the CSP $u$ filter, Figure~\ref{fig:fcterms}
shows that the observed color term is not at all close to that 
predicted from the bandpass given in H06. Indeed, even after applying 
quite large shifts we were unable to match the observed color term.  
Reasonable agreement could only be achieved after applying a cutoff 
to the bandpass on the blue side beginning at $\sim3800$~\AA\ and falling 
to zero at $\sim3300$~\AA.  This modified $u$ filter bandpass, which must
be considered very preliminary at this point, is
plotted in Figure~\ref{fig:filters}, and the results of the
synthetic color-term calculation are shown in 
Figure~\ref{fig:fcterms}.
We have verified that the transmission
of the CSP $u$ filter is consistent with the measurements of
the manufacturer, and so this blue cutoff in the bandpass 
must be produced by the Swope CCD imager, or the telescope 
optics.  We are continuing to investigate this discrepancy
so as to provide a more reliable response function for the
CSP $u$ bandpass.

\section{$YJH$ MEASUREMENTS OF FEIGE~16}
\label{feige16}

According to \citet{landolt83, landolt92}, the A0V star Feige~16 
\citep{feige58} has $UBVRI$ colors close to zero. It therefore affords us 
the opportunity to check the $Y$-band photometric system presented by 
\citet{hillenbrand02} and in Appendix~C of H06.  Fifteen measurements
of Feige~16 were conducted on six nights with RetroCam.  
The photometry was computed with respect to several (four to 
six) \citet{persson98} solar-analog standards on each night.  We discarded
one $J$ and four $H$ measurements as they clearly lay well
outside the cluster of values that were averaged with equal weight.
Table \ref{tab:feige16} contains the results, together with the values 
given by \citet{leggett06}, transformed according to their prescription, 
from their Mauna Kea Observatory (MKO) system to the LCO system.  We have added 
0.015 mag (1$\sigma$ of the mean) in quadrature to our uncertainties to 
account for small unknown systematics. For example, although we are 
treating our $JH$ data as if they were on the system defined by 
\citet{persson98}, the fact is that there could be small differences due 
to the different cameras being used then and now (the filter bandpasses are 
identical).

The LCO and (transformed) MKO $JH$ measurements agree well, within the
stated combined uncertainties.  The important result for the calibration of 
the $Y$ and $J$ bands is that the colors $Y-J = -0.009$ mag and $Y-H = 0.013$ mag 
are satisfyingly close to zero.  This confirms that our $Y$-band measurements 
are indeed on the \citet{elias82} system where $\alpha$~Lyrae has 0.00 color.
It also confirms the $Y-J$ colors for the \citet{persson98} standard stars
predicted from model atmospheres presented in Appendix~C of H06.
	
The $V$ magnitude of Feige~16 is 12.400 \citep{landolt83, landolt92}. 
Assuming that all the near-infrared colors are actually zero, the average of 
all the $YJHK_s$ measurements (on the LCO system) is 12.355 mag, indicating 
that the star may be slightly reddened. If so, $E(B-V) = 0.015$ mag.

\section{PHOTOMETRIC ZERO POINTS}
\label{zpoints}

The photometric zero point for a filter $X$ with a bandpass response function
$S_{X}$ can be defined as\[
ZP_{X}=2.5\log\left[\int_0^{\infty}
  S_{X}\left(\lambda\right)\, F_{0}\left(\lambda\right) \frac{\lambda}{hc}\, d\lambda\right]+m_{0},\]
where $F_{0}$ is the spectral energy distribution (SED)
of a fiducial source and $m_{0}$ is the
observed magnitude of this source through filter $X$.  For the purpose of 
computing photometric zero points for the CSP optical and near-infrared 
bandpasses, we have adopted the SED of Vega 
($\alpha$ Lyr) as presented by \citet{bohlin04}. In the following subsections, 
we outline how the zero points were derived for each bandpass. 
Table \ref{tab:zpts} shows the final zero points adopted.

\subsection{$BV$ Photometry}

The CSP uses the \citet{landolt92} standard stars.
To derive zero points for our $BV$ filters, we use the SED of Vega as 
defined by \citet{bohlin04}\footnote{This SED can be found at CALSPEC:
  ftp://ftp.stsci.edu/cdbs/current\_calspec/alpha\_lyr\_stis\_004.ascii
.} 
and the magnitudes adopted by 
\citet{fukugita96}:  $B_{\rm Vega} = 0.03$ and $V_{\rm Vega} = 0.03$
mag.  We convert these magnitudes to the LCO natural system using the
color terms from \citet{hamuy06}.  For the V-band transformation, an
$i^\prime$ magnitude is needed for Vega.  For consistency, we use the
\citet{bohlin04} SED for Vega and the $i^\prime$ filter function from
\citet{fukugita96} to compute a synthetic 
$i^\prime_{\rm Vega} = 0.382$ mag.  

\subsection{$ugri$ Photometry}

The CSP uses the \citet{smith02} standard stars for the $ugri$
photometry, which are tied to the SED of BD$+17^{\circ}4708$.  
In order to maintain 
consistency with our choice of the \citet{bohlin04} SED for Vega, 
we use the SED of BD$+17^{\circ}4708$ as given by \citet{bohlin04b}
\footnote{This SED can be found at CALSPEC:
  ftp://ftp.stsci.edu/cdbs/current\_calspec/bd\_17d4708\_stisnic\_002.ascii
.}. For the assumed $ugri$ magnitudes of BD$+17^{\circ}4708$, we use
the values listed in  
\citet{smith02} for the USNO 1.0-m photometric telescope.  These magnitudes 
are then converted to the LCO natural system using the color terms from 
\citet{hamuy06}.

\subsection{Near-Infrared Photometry}

Our $YJHK$ photometry uses the standards of \citet{persson98}, which
are tied to the \citet{elias82} standards. The \citet{elias82}
standards are, in turn, tied to Vega. We are therefore on a Vega
system and use the \citet{bohlin04} SED. We assume $J_{\rm Vega}=-0.001$,
$H_{\rm Vega}=0.000$, and $K_{\rm Vega}=-0.001$ mag \citep{cohen99},
as well as $Y_{\rm Vega}=0.000$ mag \citep{hamuy06}.

\clearpage

\figcaption[]{A mosaic of $V$-band CCD images of 35 SNe~Ia observed by the CSP.
 \label{fig:fcharts}}   

\figcaption[]{Swope $V$-band CCD images of SN~2004ef ({\em left}) and
SN~2006X ({\em right}) after a template image of the host galaxy was subtracted 
at the  location of  each SN.
 \label{fig:subtractions}}

  \figcaption[]{Day $-$3 spectrum of SN~2004gu ({\it solid line})  
  compared to day $-$2 spectra of  SN~2006gz  ({\it top panel}) and 
  SN~1999ac ({\it bottom panel}). Prominent ions are indicated.
  \label{fig:spectra}}

\figcaption[]{Extinction coefficients for the $u'g'r'i'BV$ filters as a
  function of time.  The average coefficient is plotted with a horizontal
  line and its numerical value is indicated in each panel along with the
  root-mean square (rms) scatter (in parentheses) expressed in 
  thousandths of a magnitude.
  \label{fig:extinc}}   

\figcaption[]{Color coefficients for the $u'g'r'i'BV$ filters as a
  function of time.  The average coefficient is plotted with a horizontal
  line and its numerical value is indicated in each panel along with the
  rms scatter (in parentheses) expressed in thousandths of a magnitude.
  \label{fig:colorc}}   

\figcaption[]{$uBgVriYJHK_s$ light curves of the 35 SNe~Ia in the 
  natural system of the Swope telescope.  Uncertainties in the photometry 
  are smaller than the points, unless shown.  The smooth curves show the 
  fits derived with SNOOPy.  Note that observations obtained after $\sim$85 days 
  past $B$-band maximum are not plotted.\label{fig:flcurves}}

\figcaption[]{Color transformation terms calculated from synthetic 
  photometry of 91 Landolt standard stars.  The dashed blue lines show 
  the predicted color-term slope for the CSP filter response functions
  given in H06, whereas the solid red lines represent the actual 
  observed slopes.  The dots correspond to synthetic photometry of the 
  Landolt standards using the modified CSP $ugriBV$ natural bandpasses 
  presented in Appendix~\ref{afilters} and shown in Figure~\ref{fig:filters}.
  \label{fig:fcterms}}

\figcaption[]{Comparison of revised CSP $ugriBV$ natural filter bandpasses 
 ({\em solid lines}) with the USNO40 $u'g'r'i'$ and Landolt $BV$ standard 
 bandpasses \citep{stritzinger05} ({\em dashed lines}). The USNO40 and Landolt
 curves have been adjusted to approximately the peak throughputs of the CSP 
 filters.\label{fig:filters}}

\clearpage
\newpage

% Finding charts
%\input{charts.tex}
\begin{figure}[t]

\epsscale{.54}
\plottwo{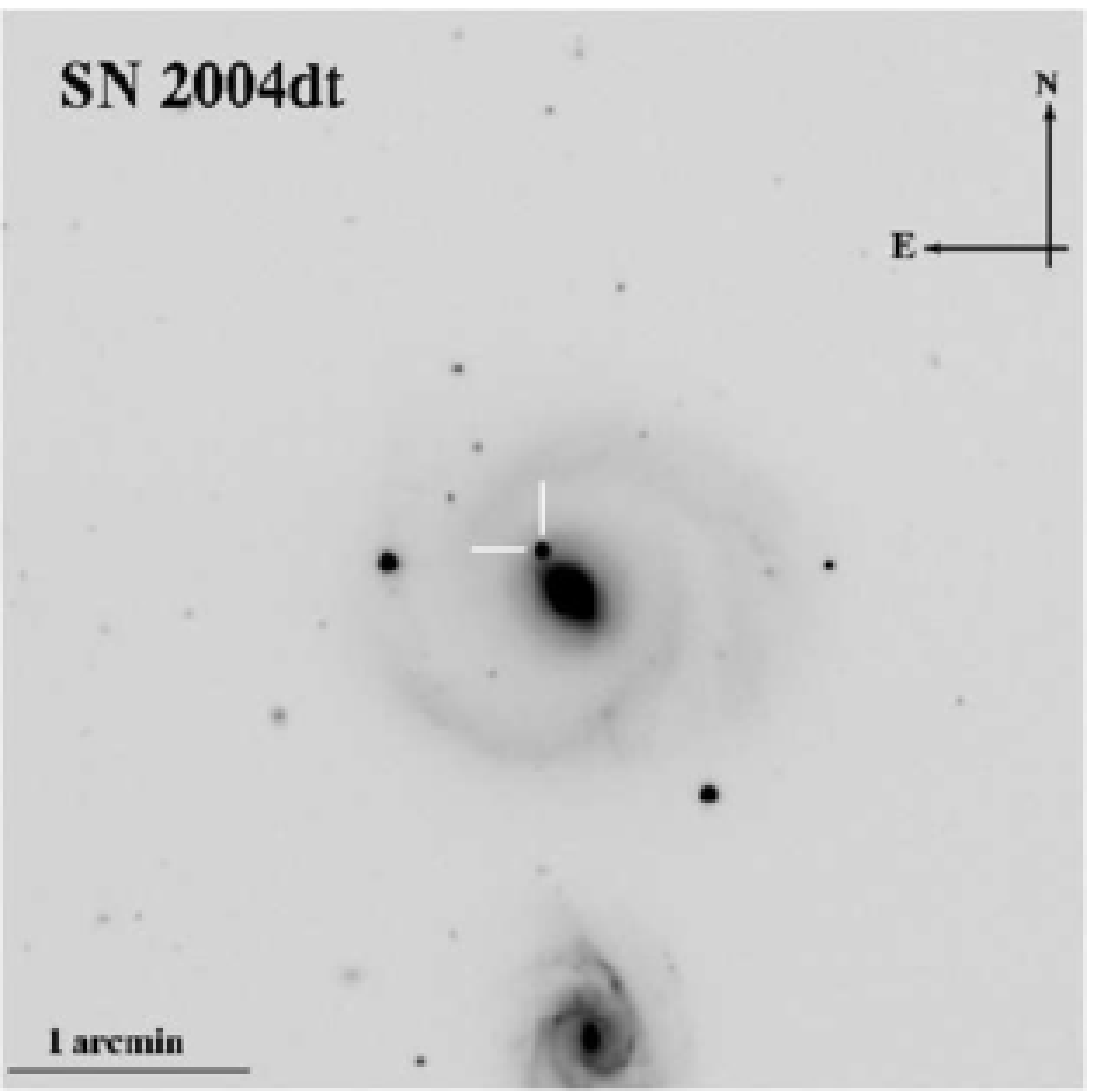}{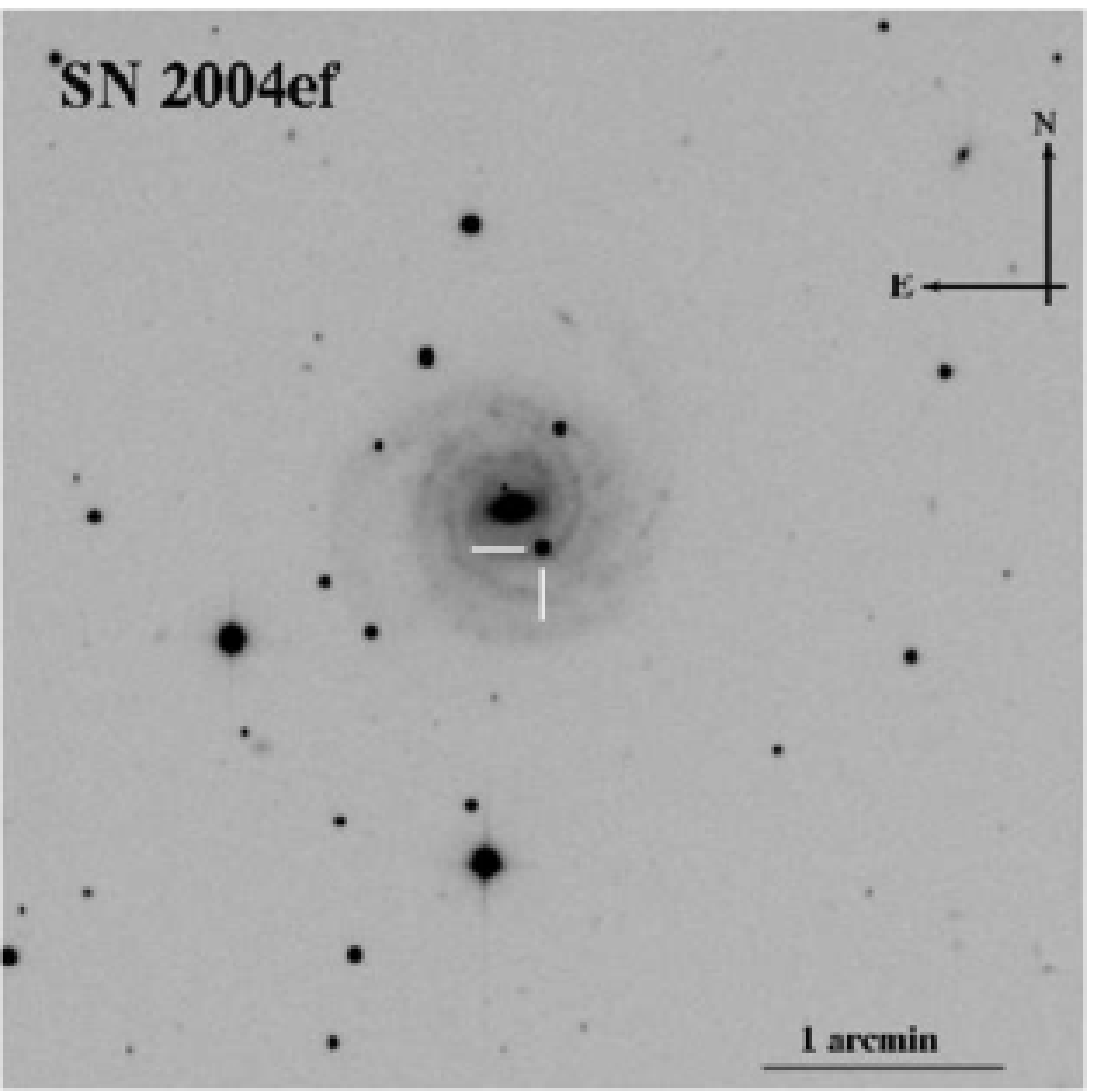}
\plottwo{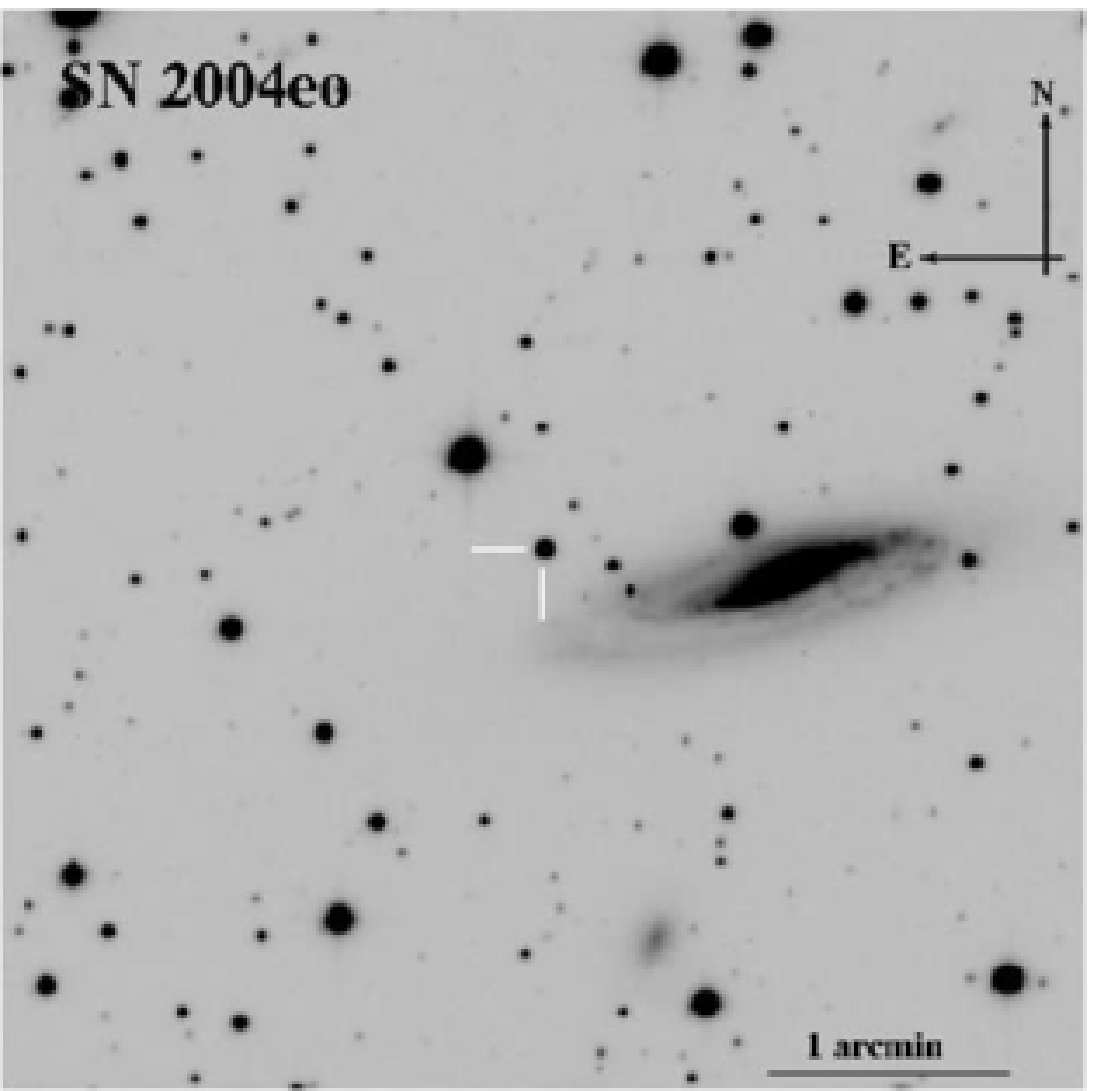}{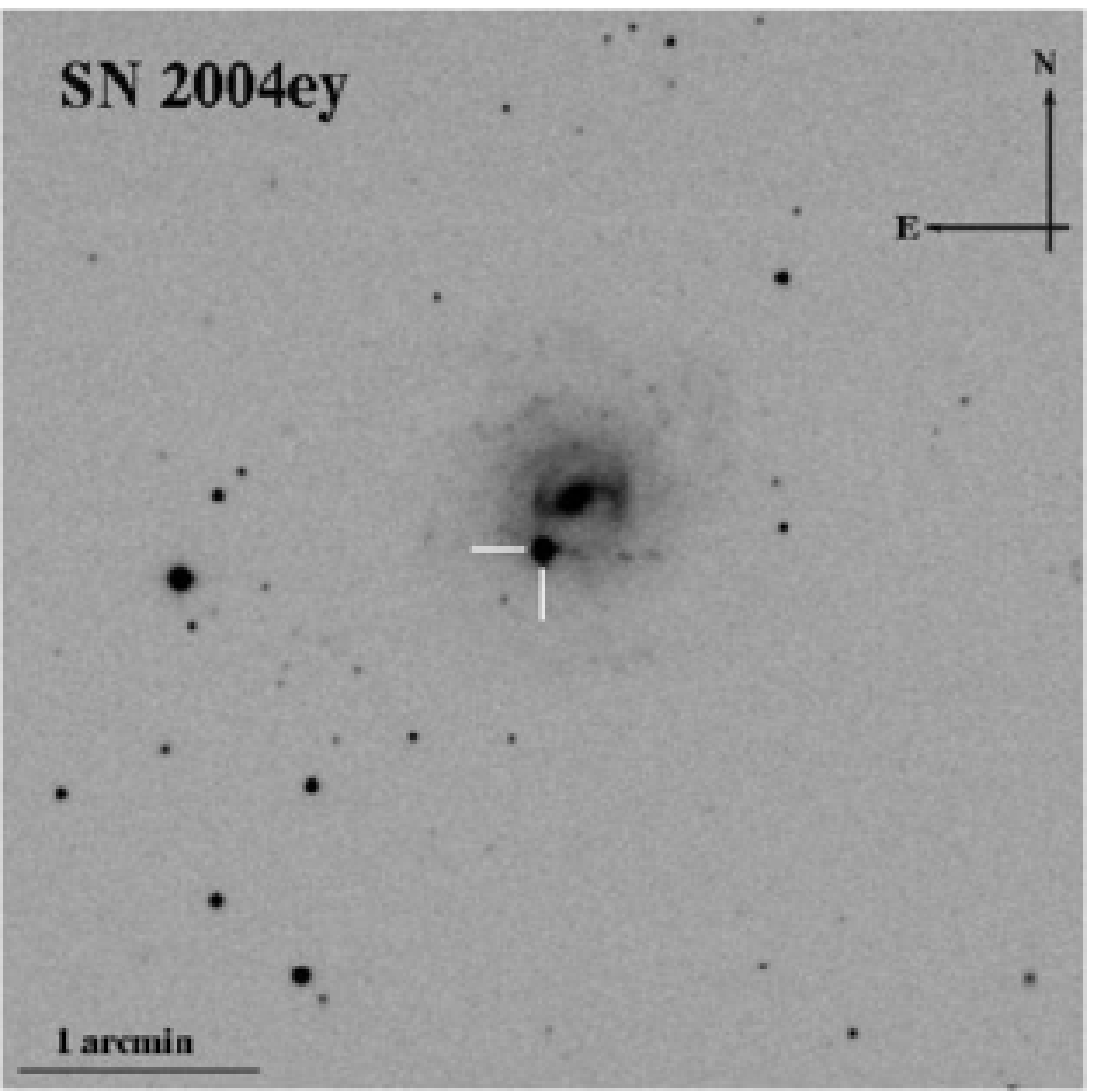}
\newline                                                                     
\plottwo{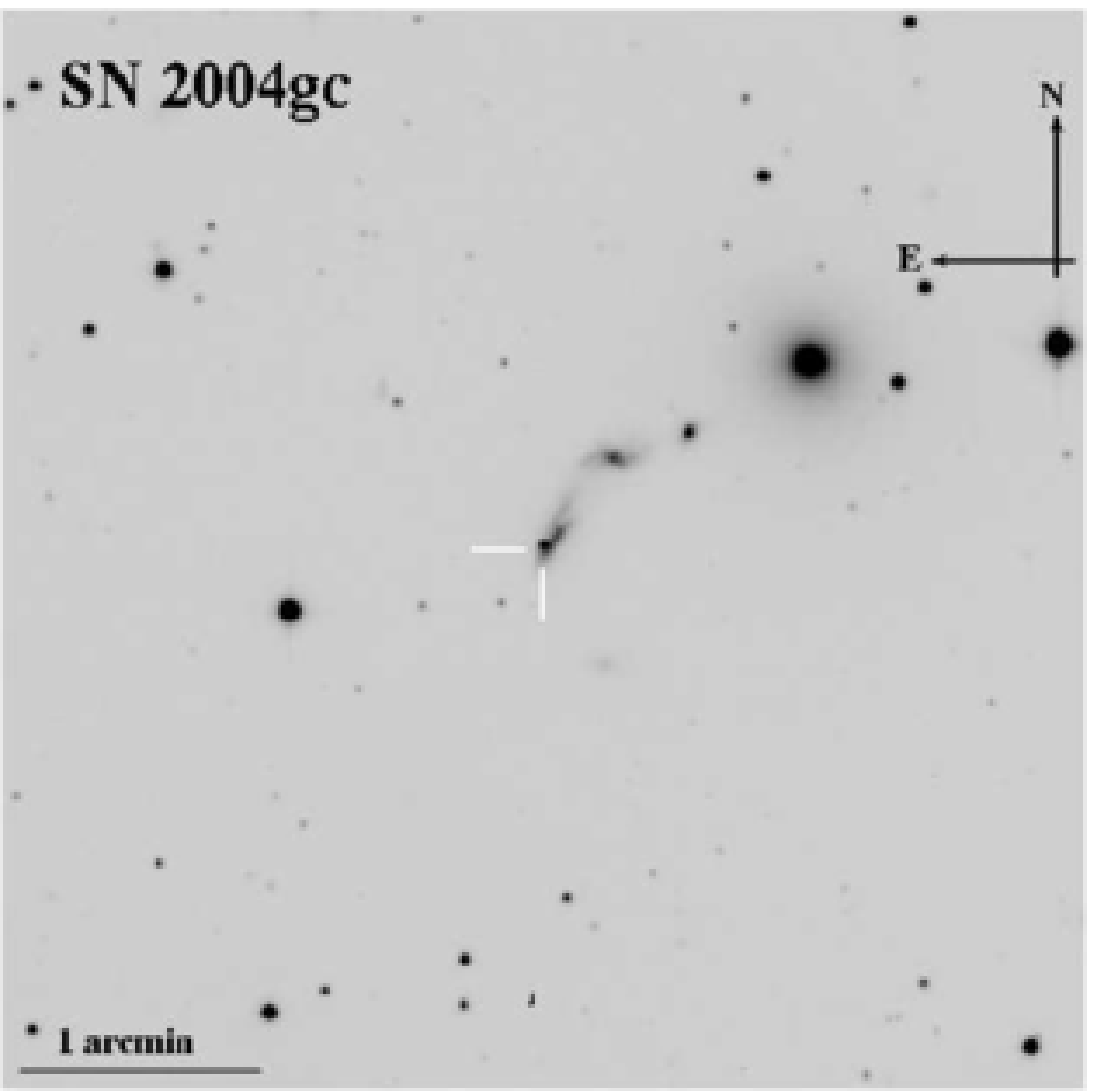}{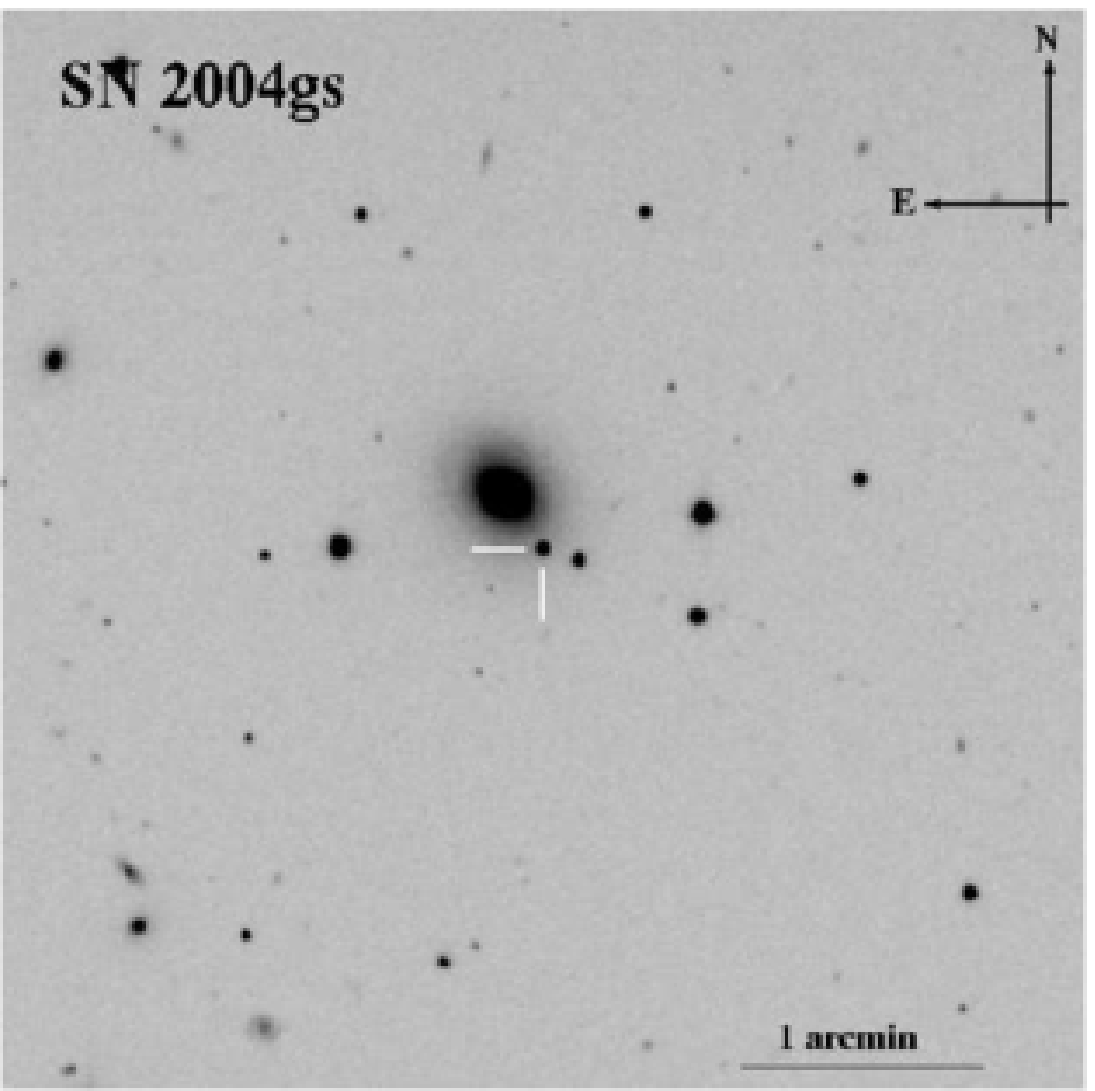}
\plottwo{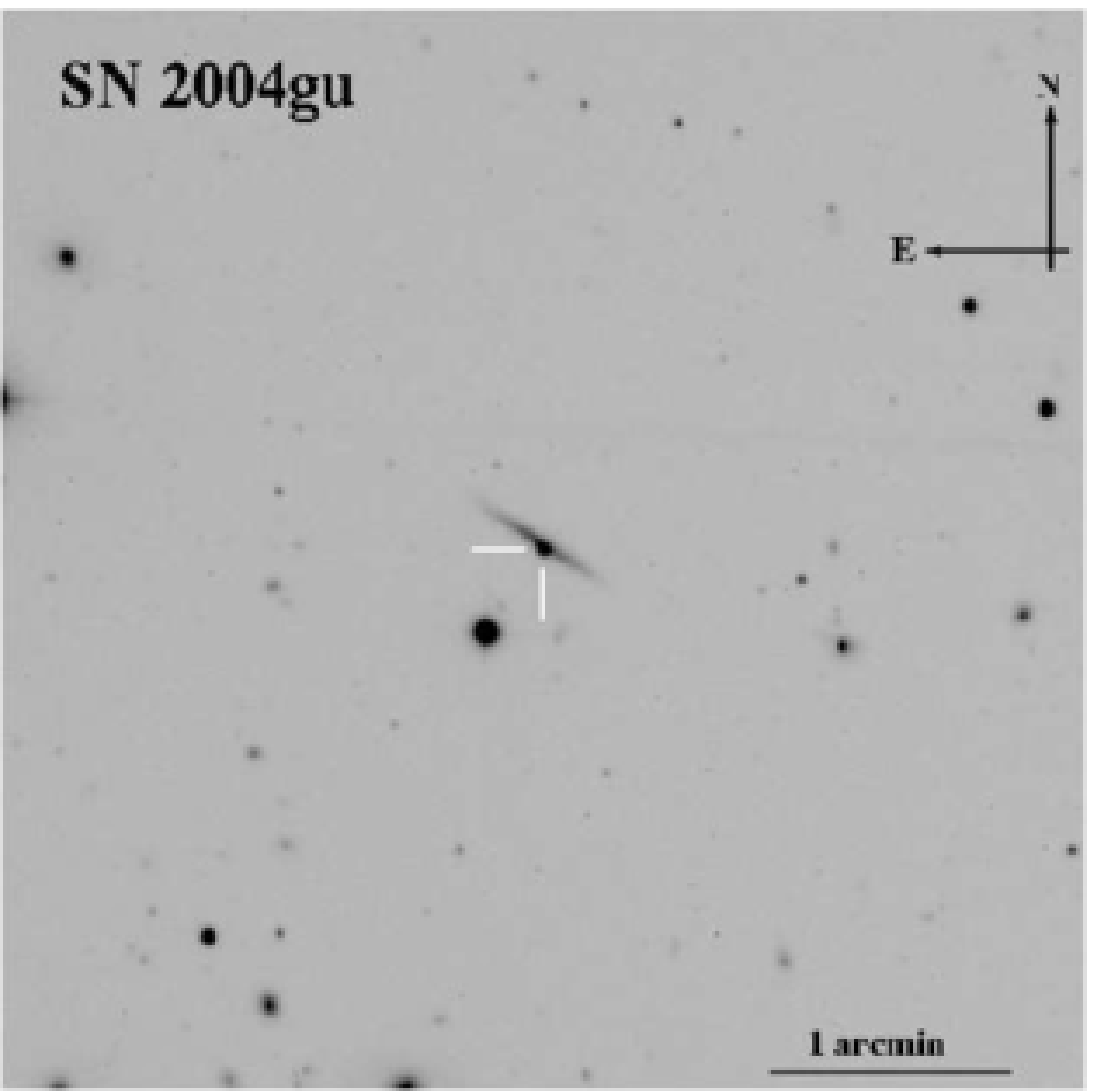}{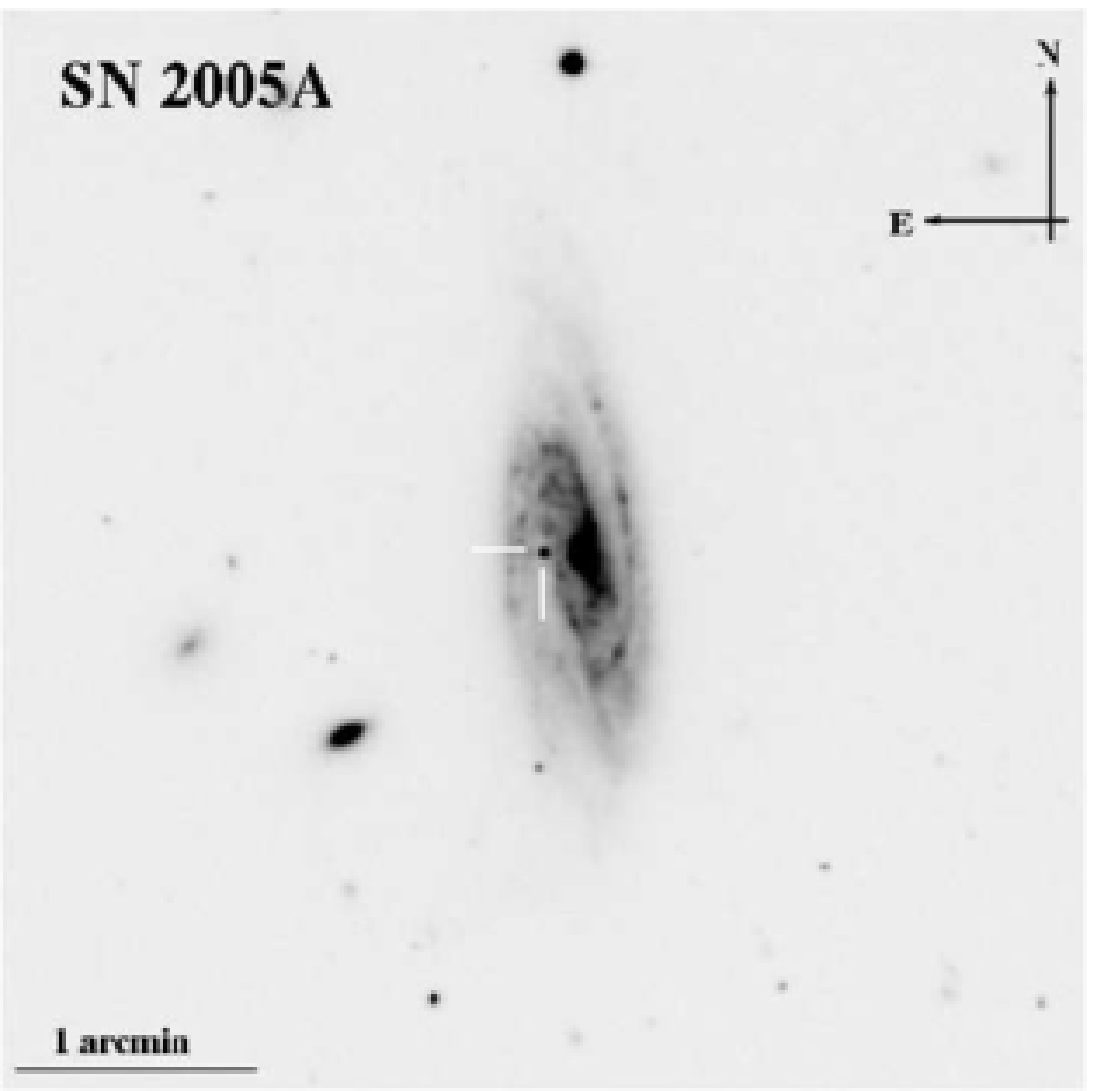}
\newline                                                                     
\plottwo{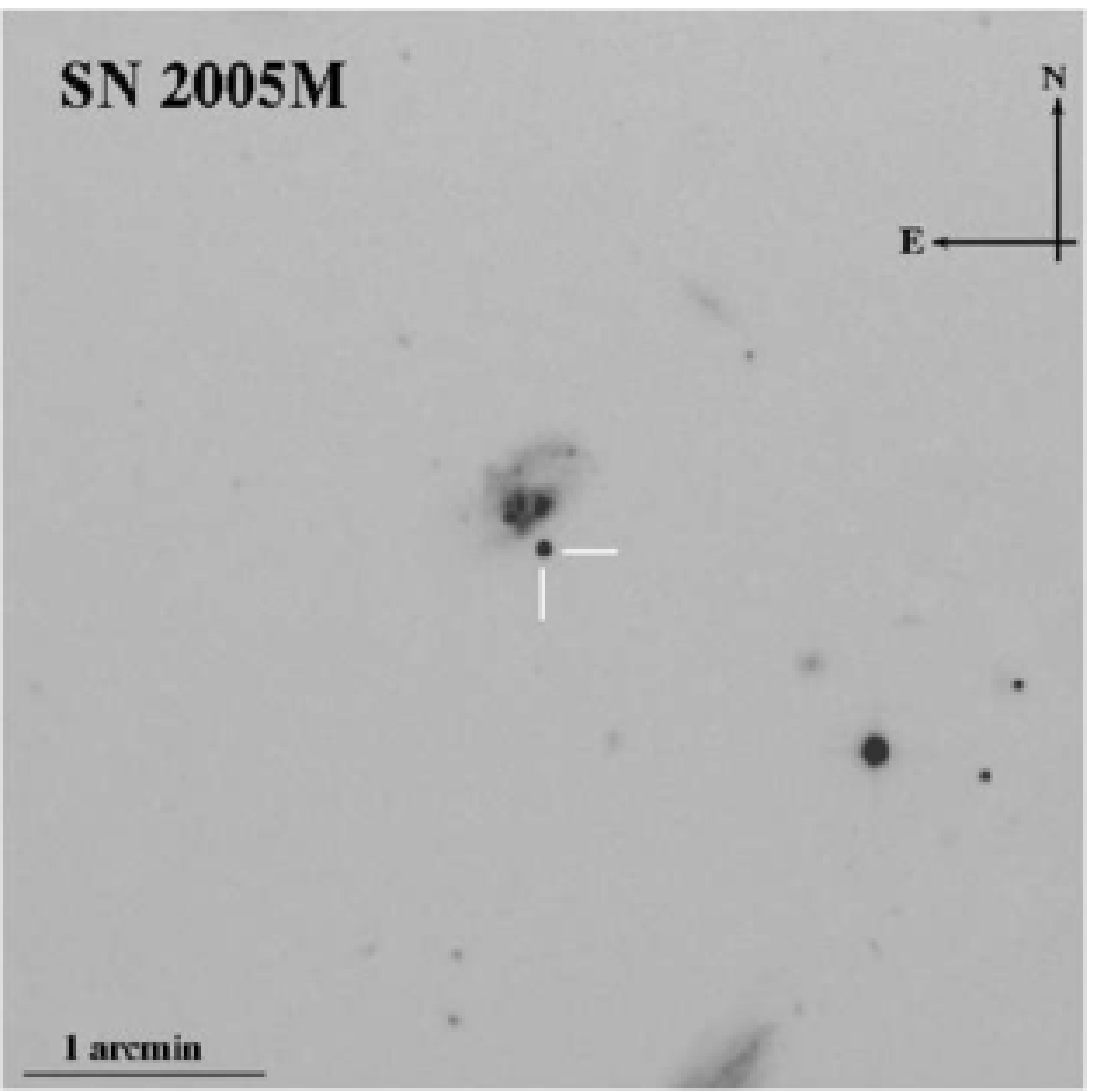}{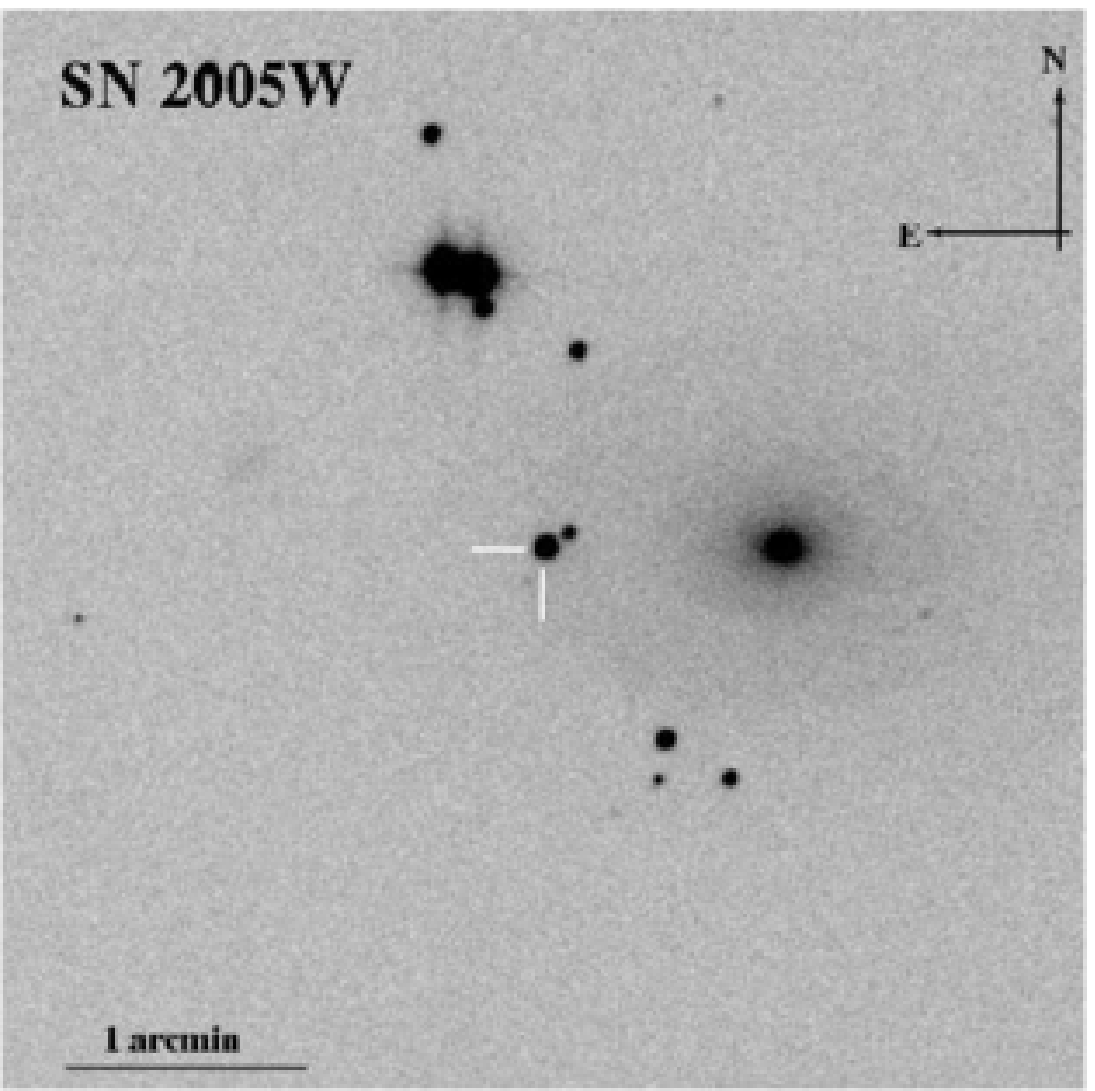}
\plottwo{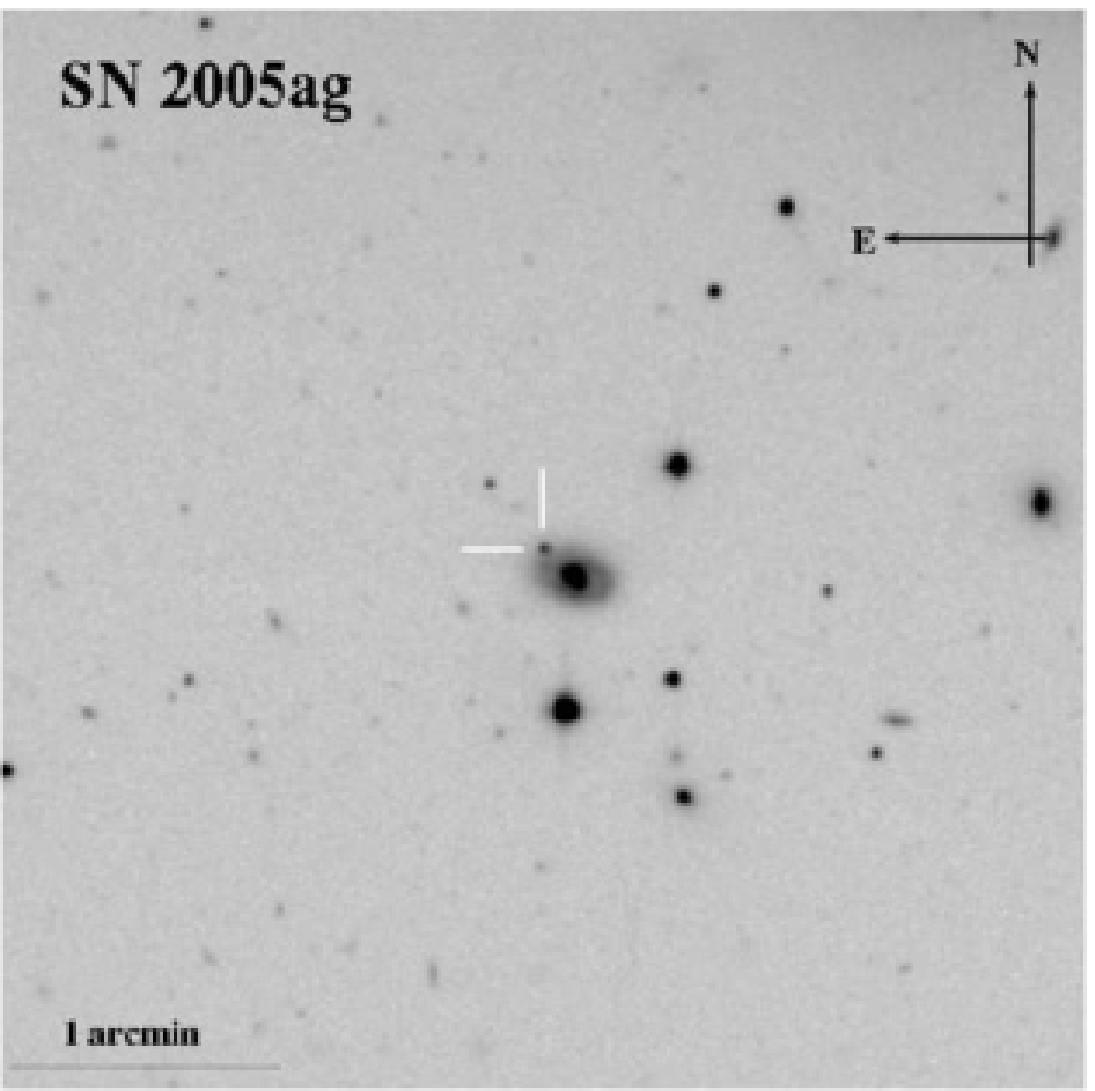}{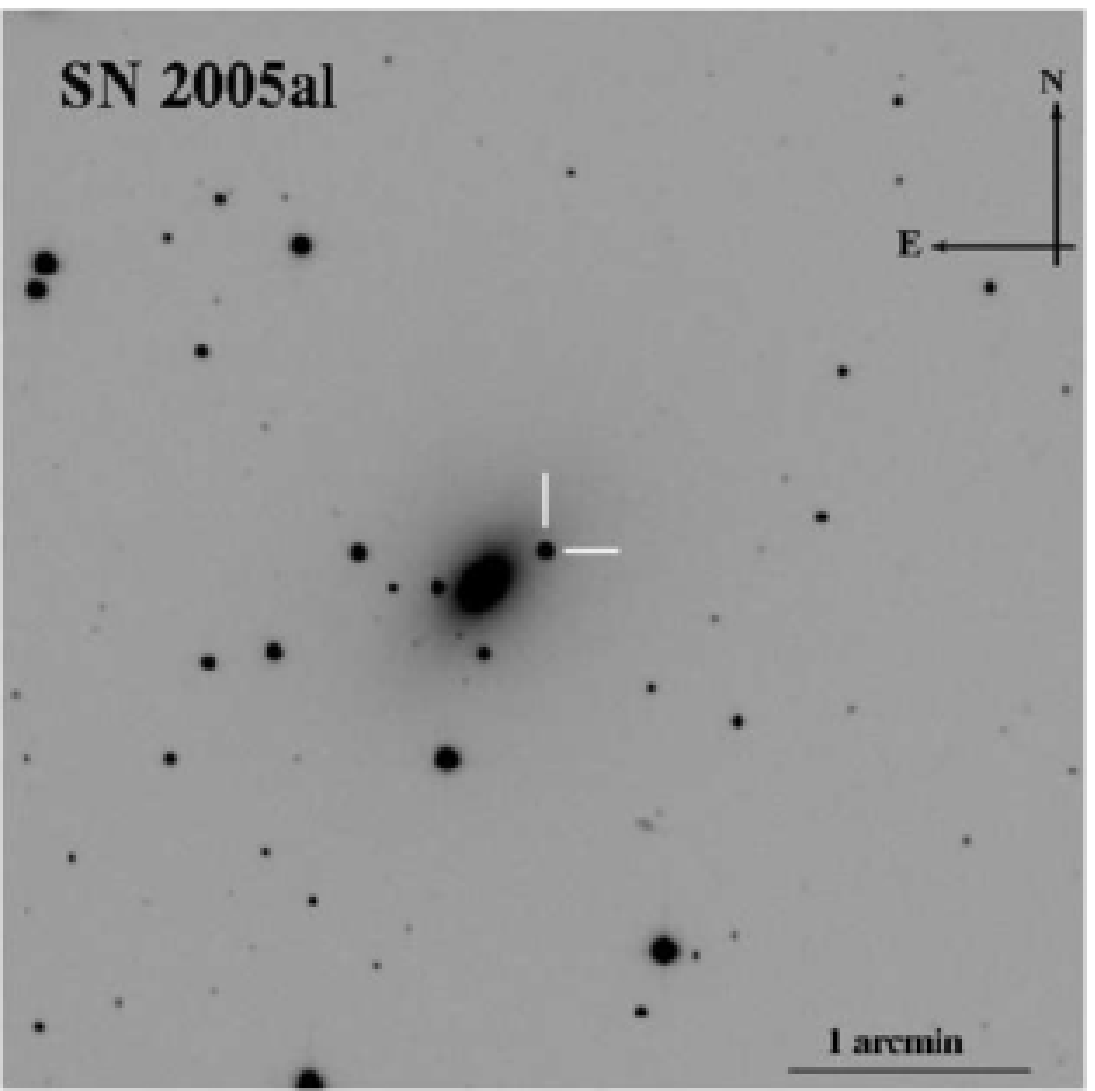}
\newline                                                                     
\plottwo{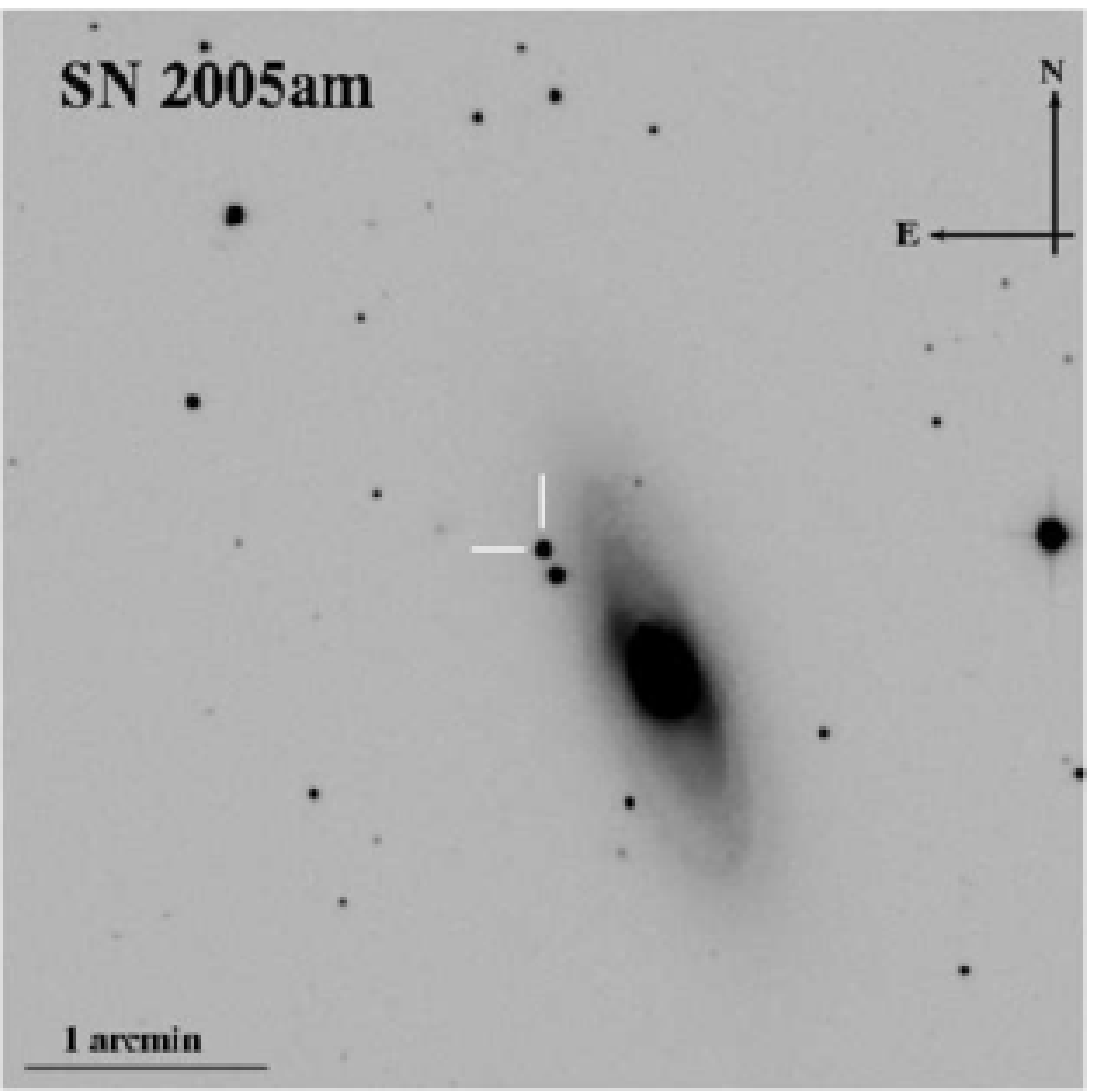}{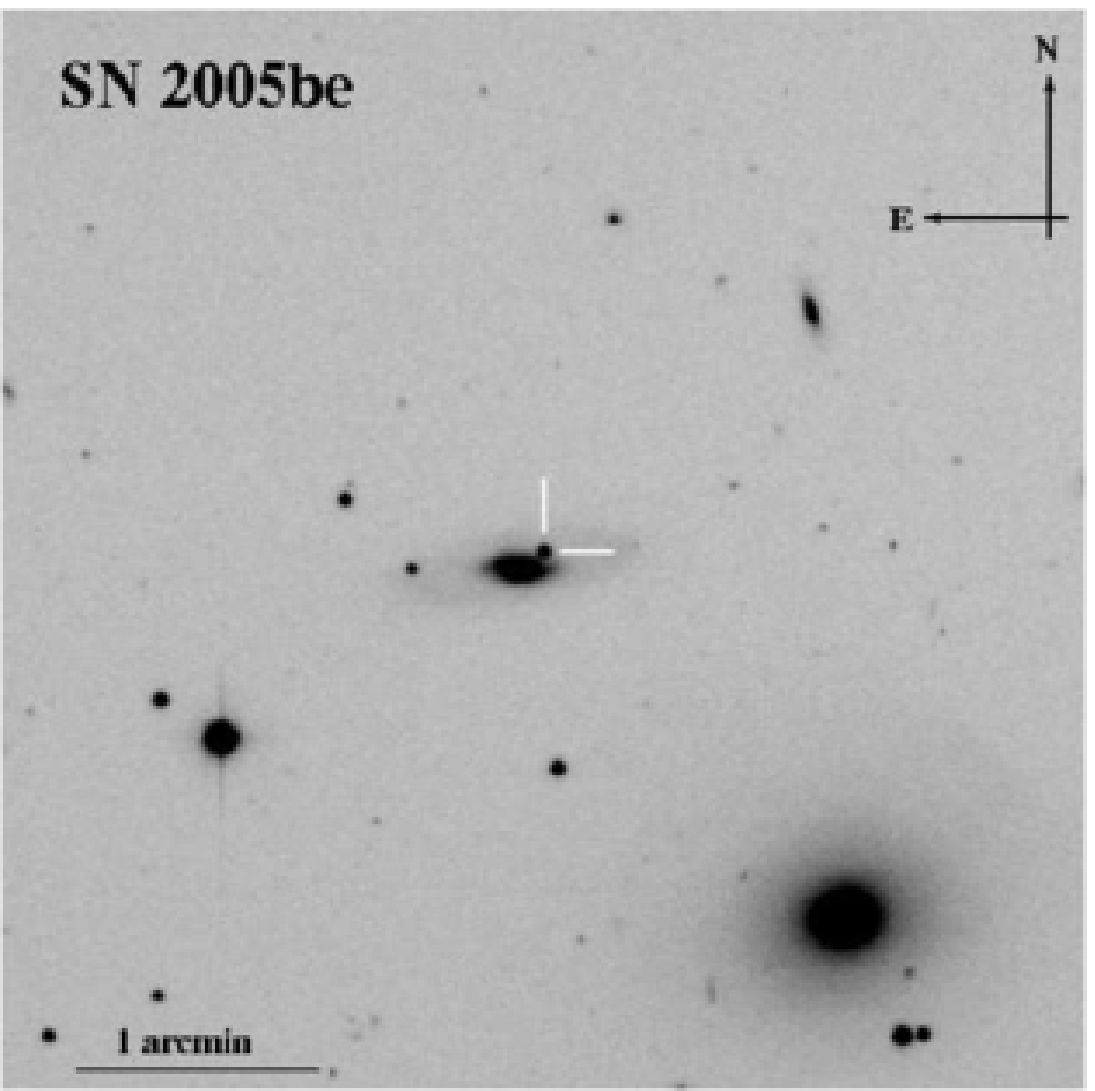}
\plottwo{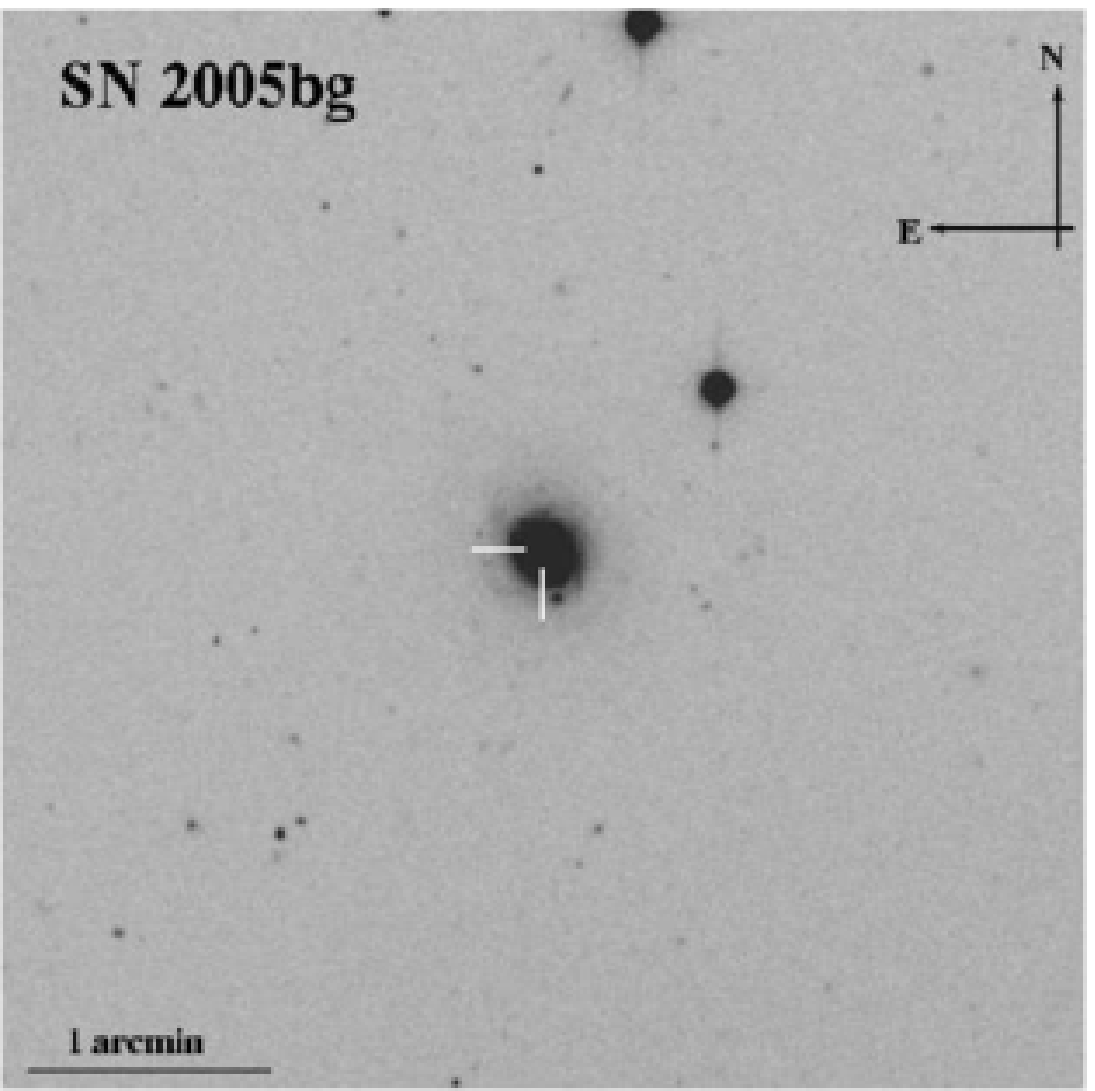}{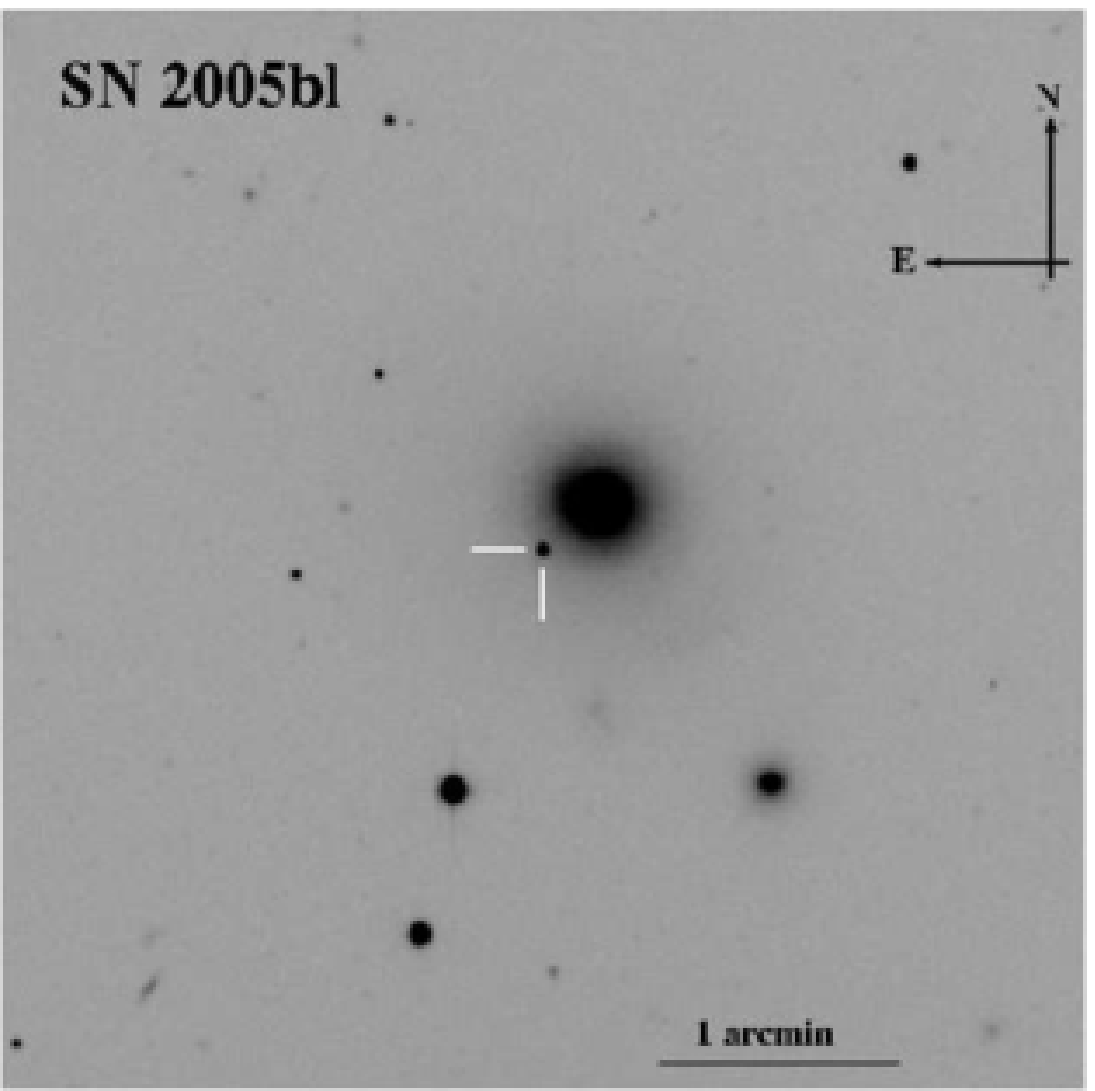}
\newline                                                                     
\plottwo{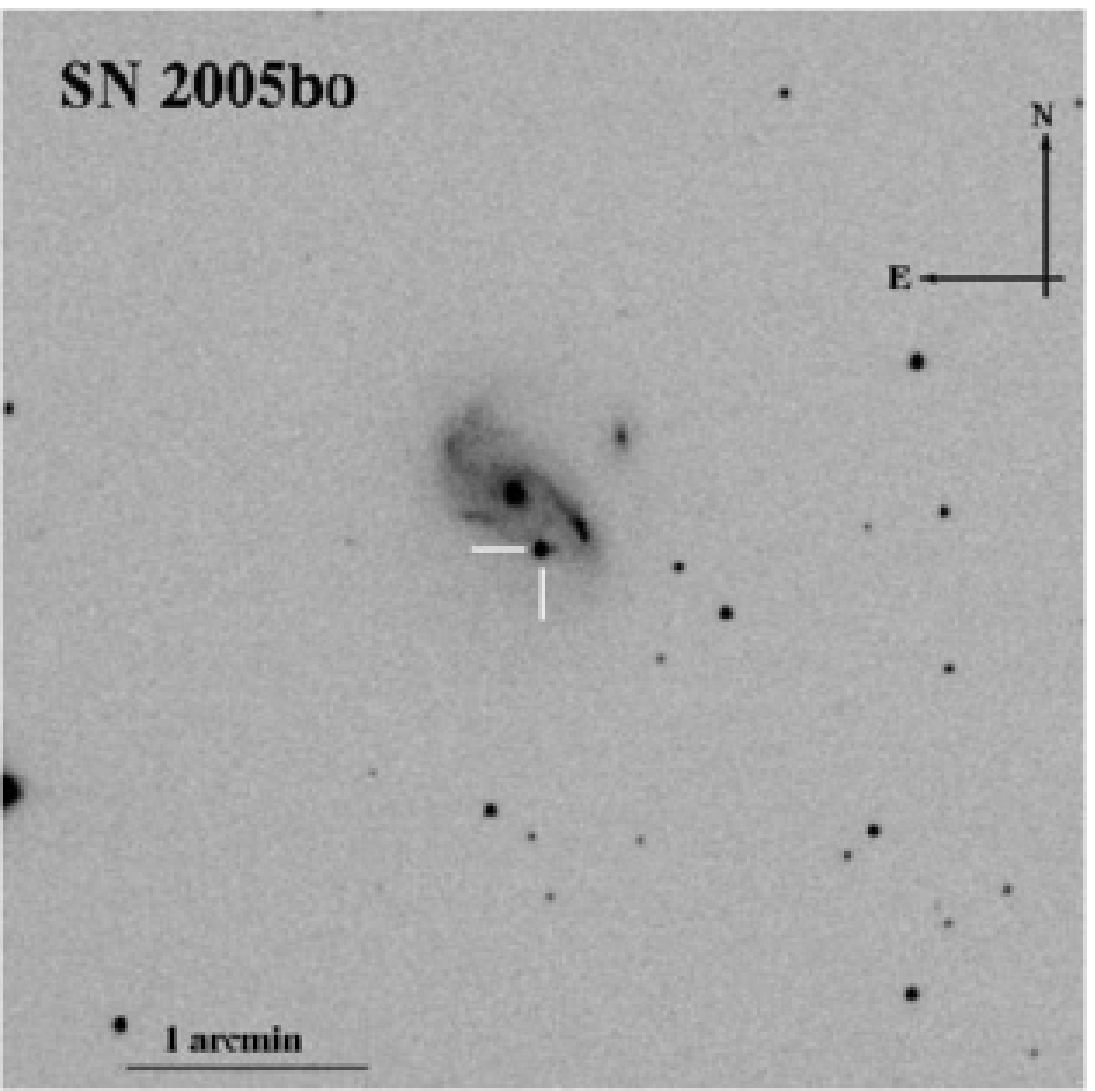}{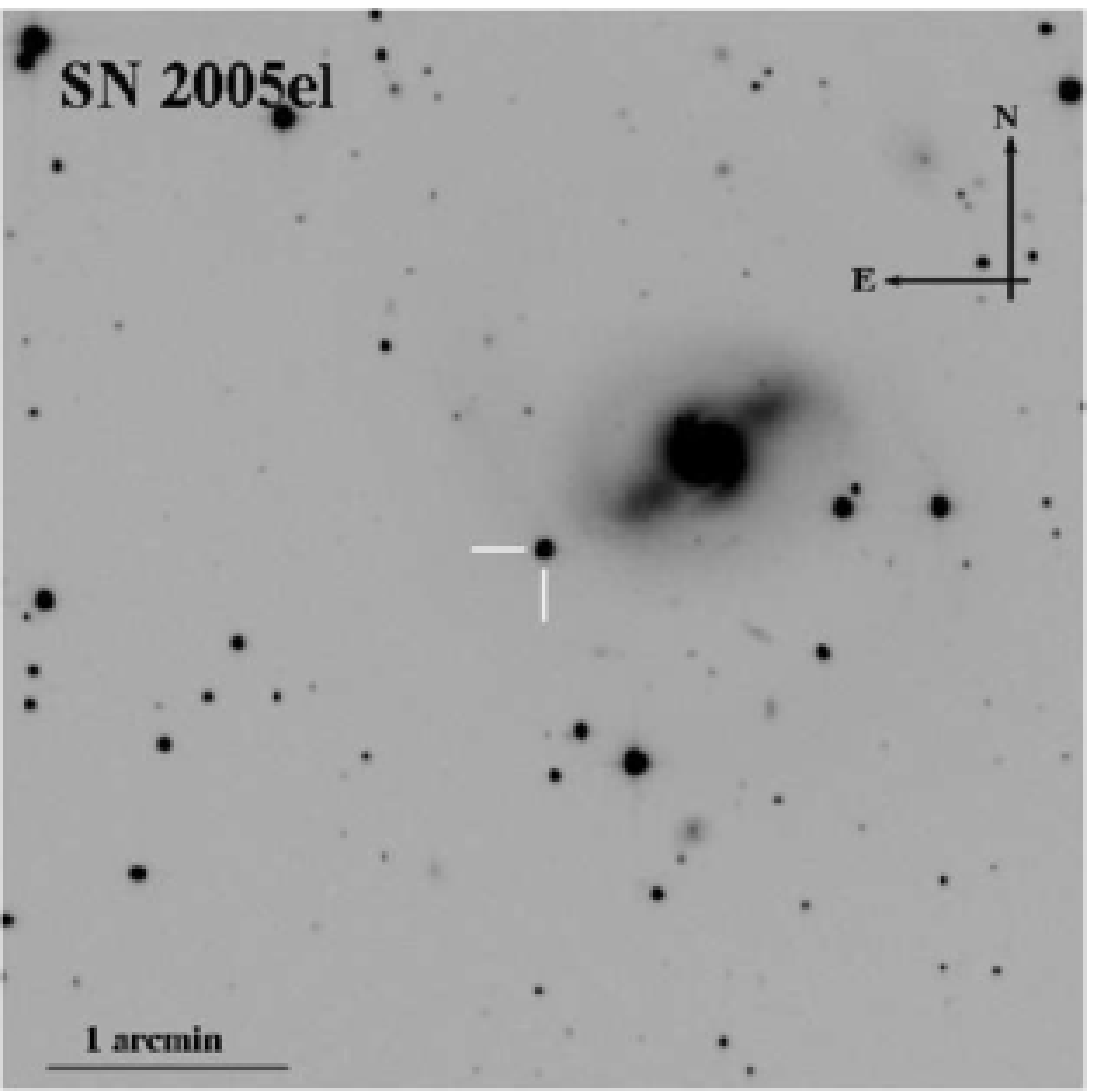}
\plottwo{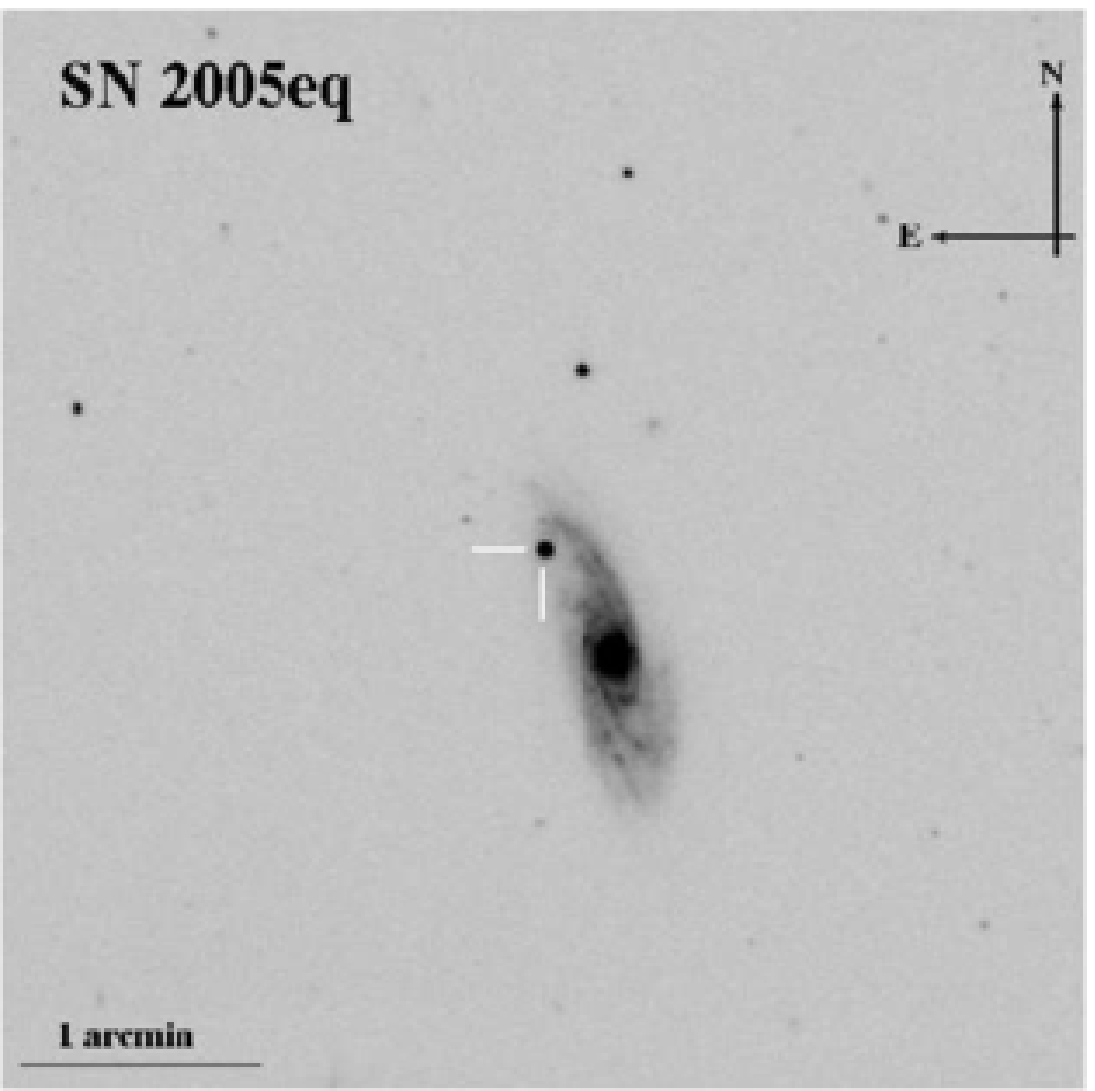}{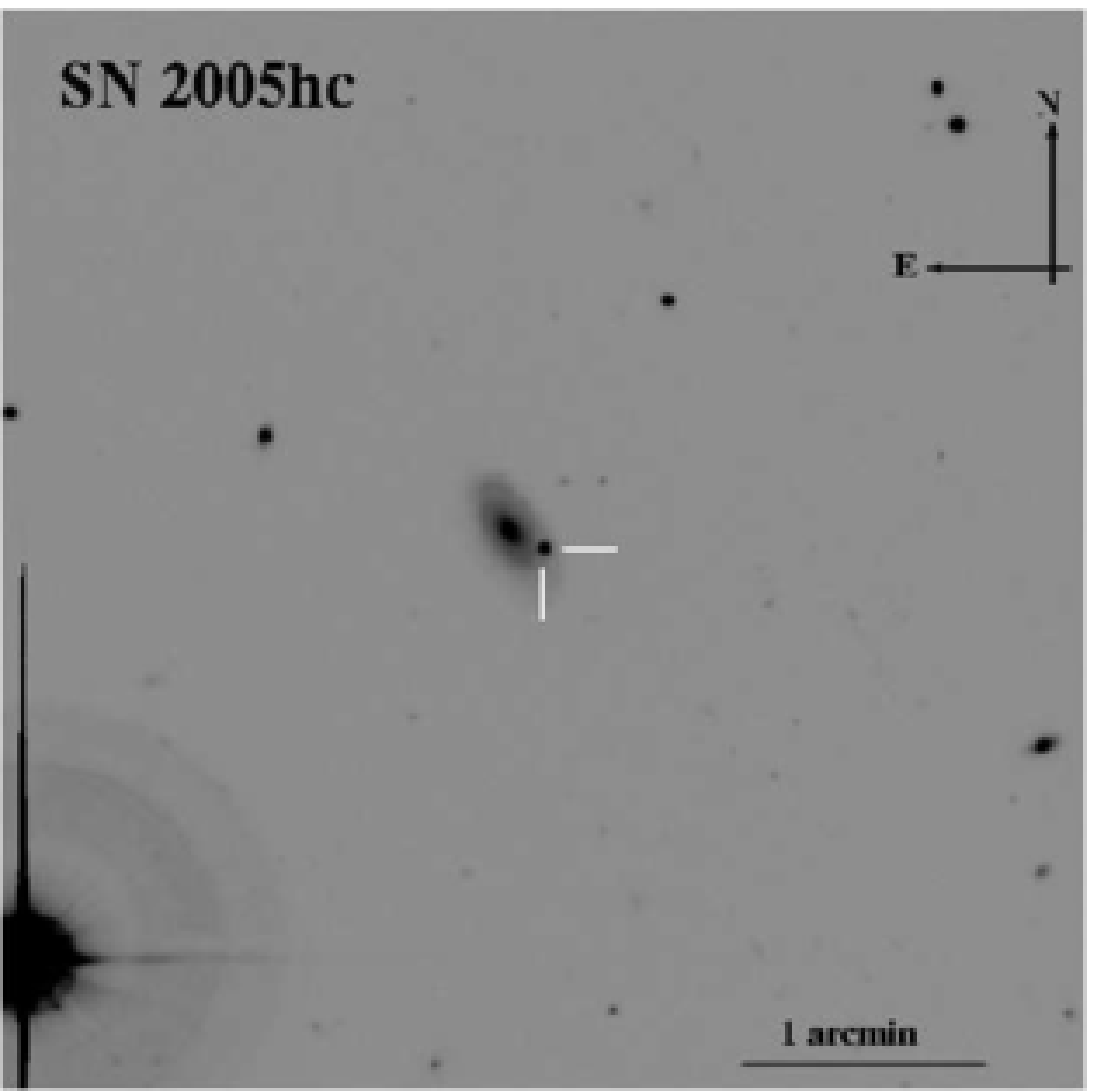}
{\center Contreras {\it et al.} Fig. \ref{fig:fcharts}}
\end{figure}
\clearpage
\newpage

\begin{figure}[t]

\plottwo{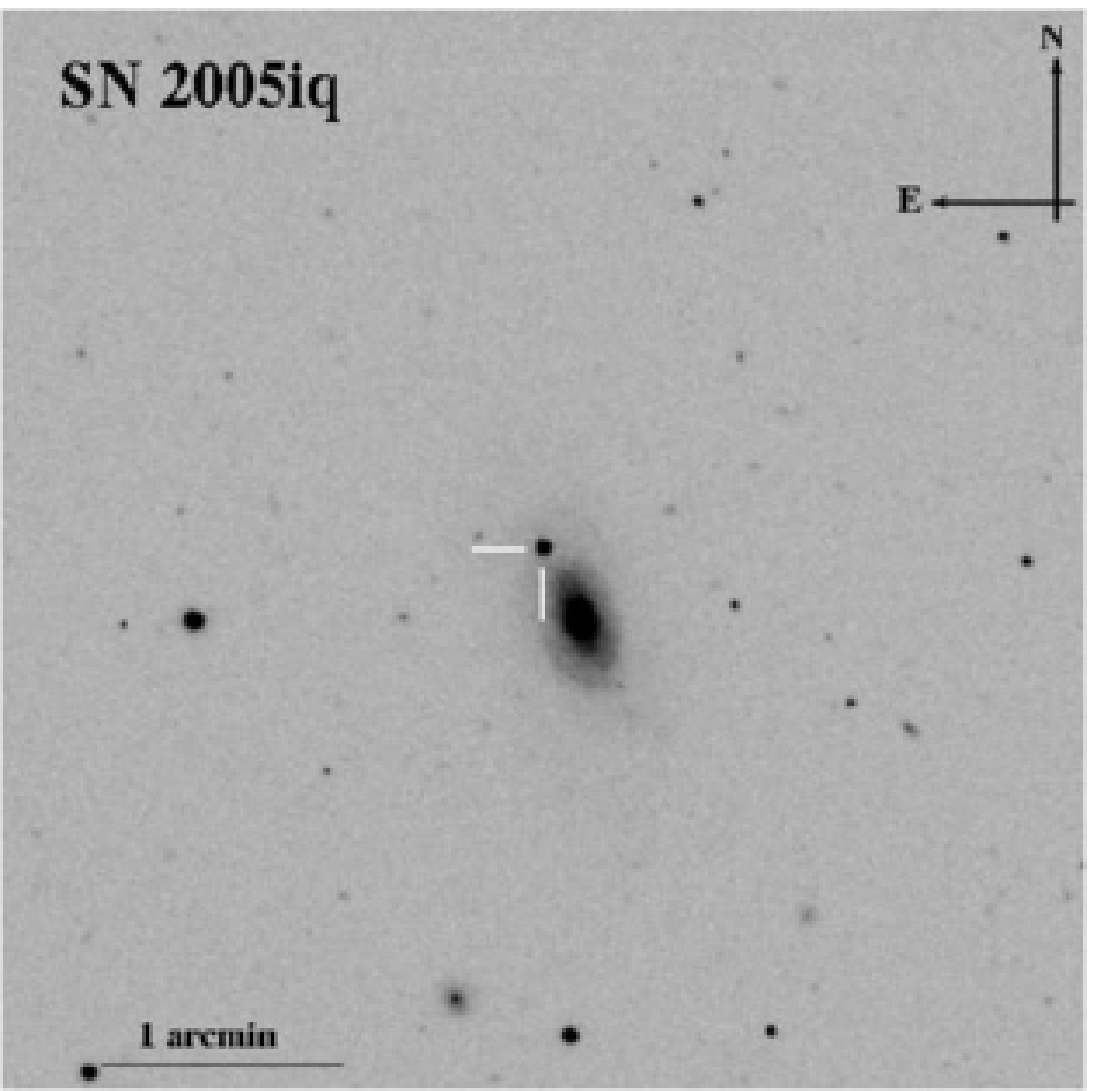}{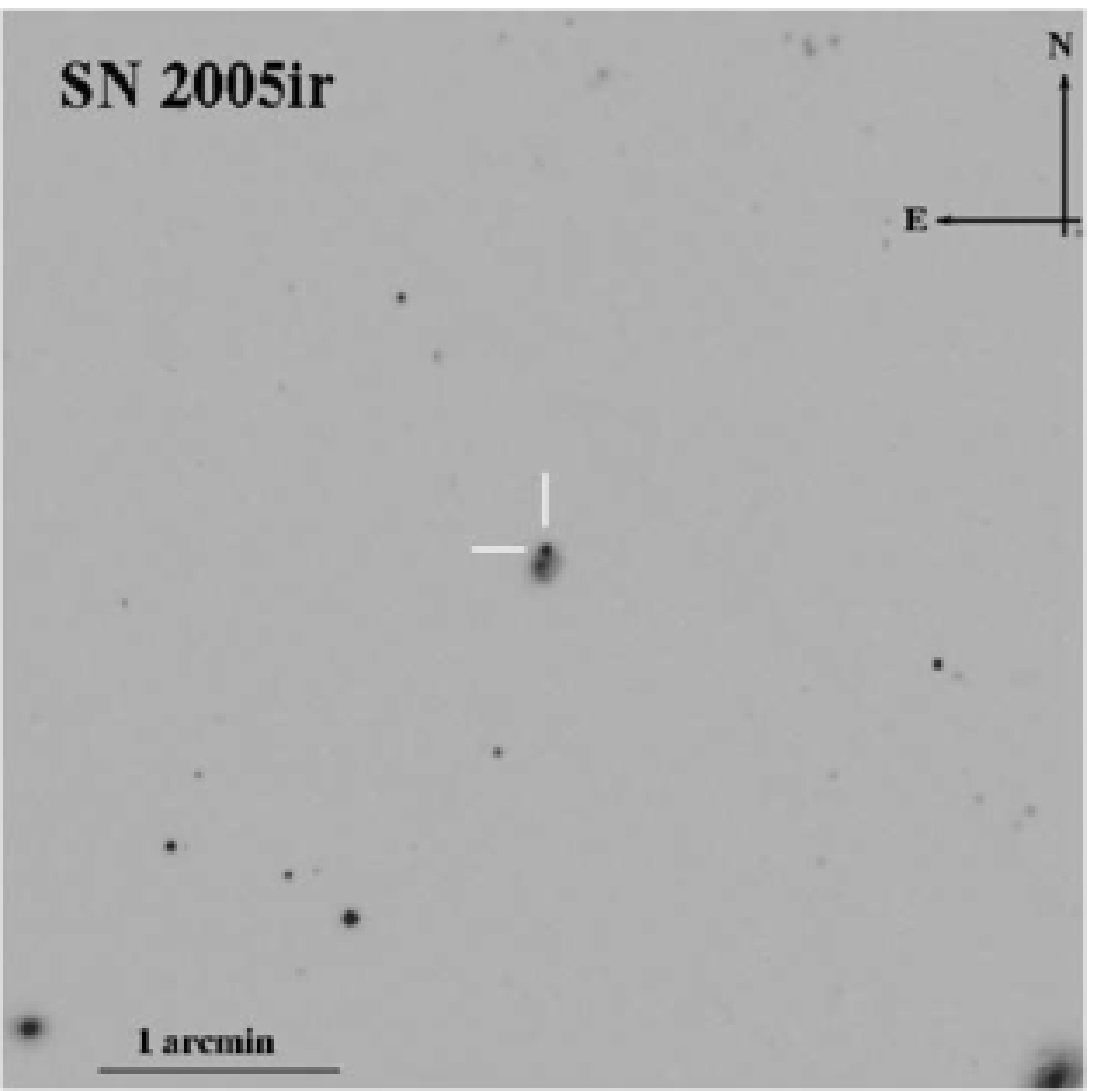}
\plottwo{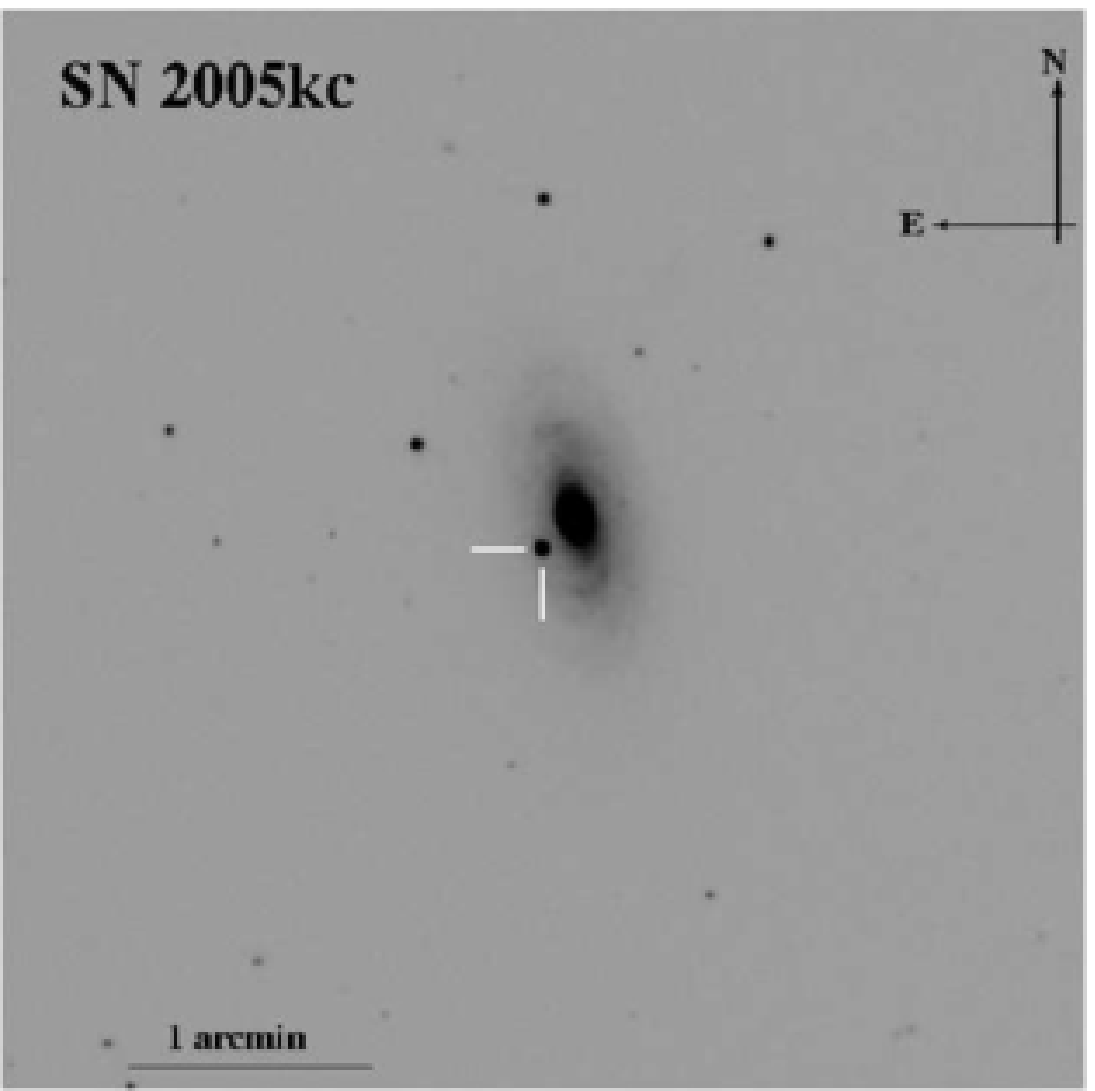}{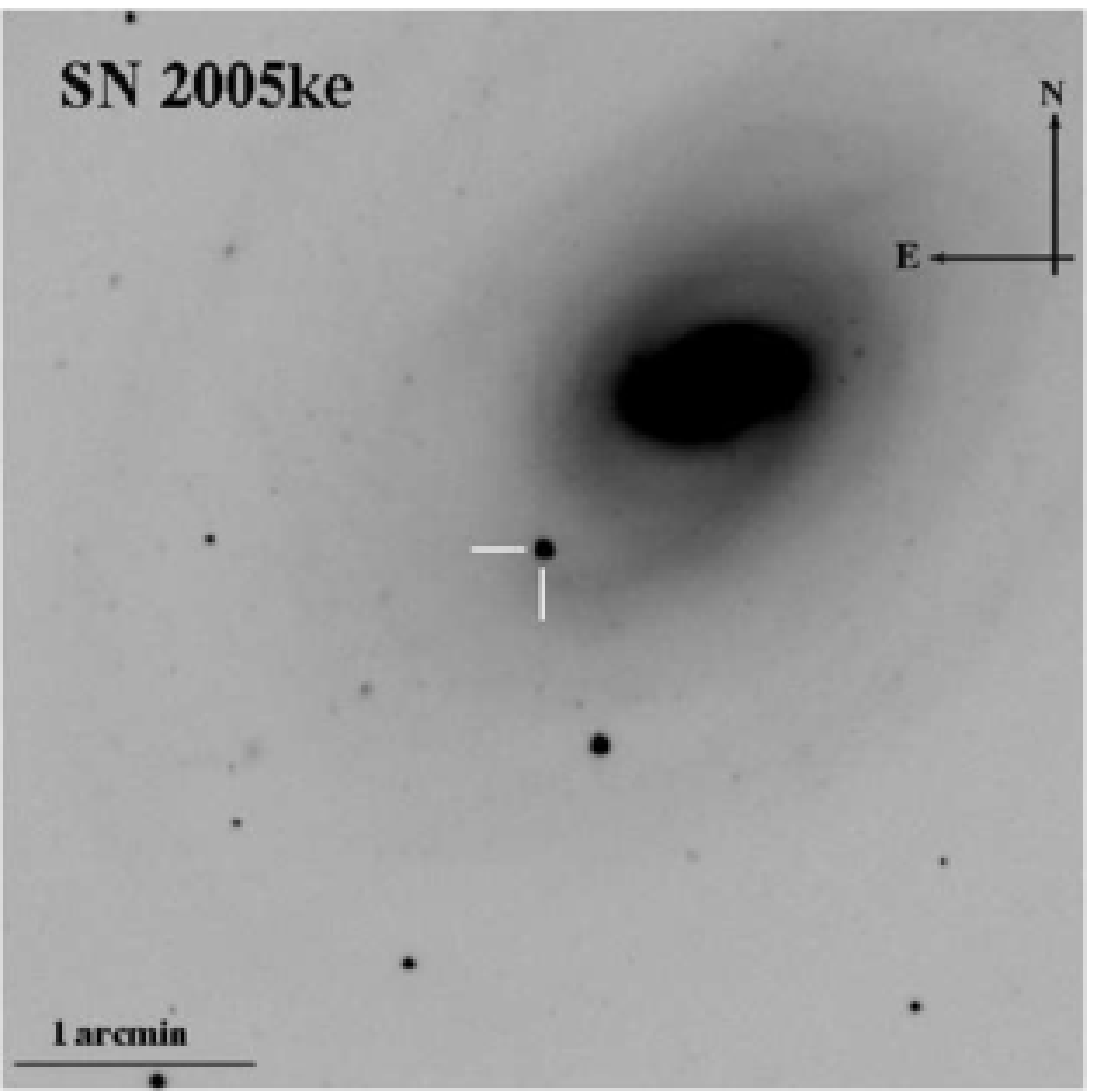}
\newline                                                                     
\plottwo{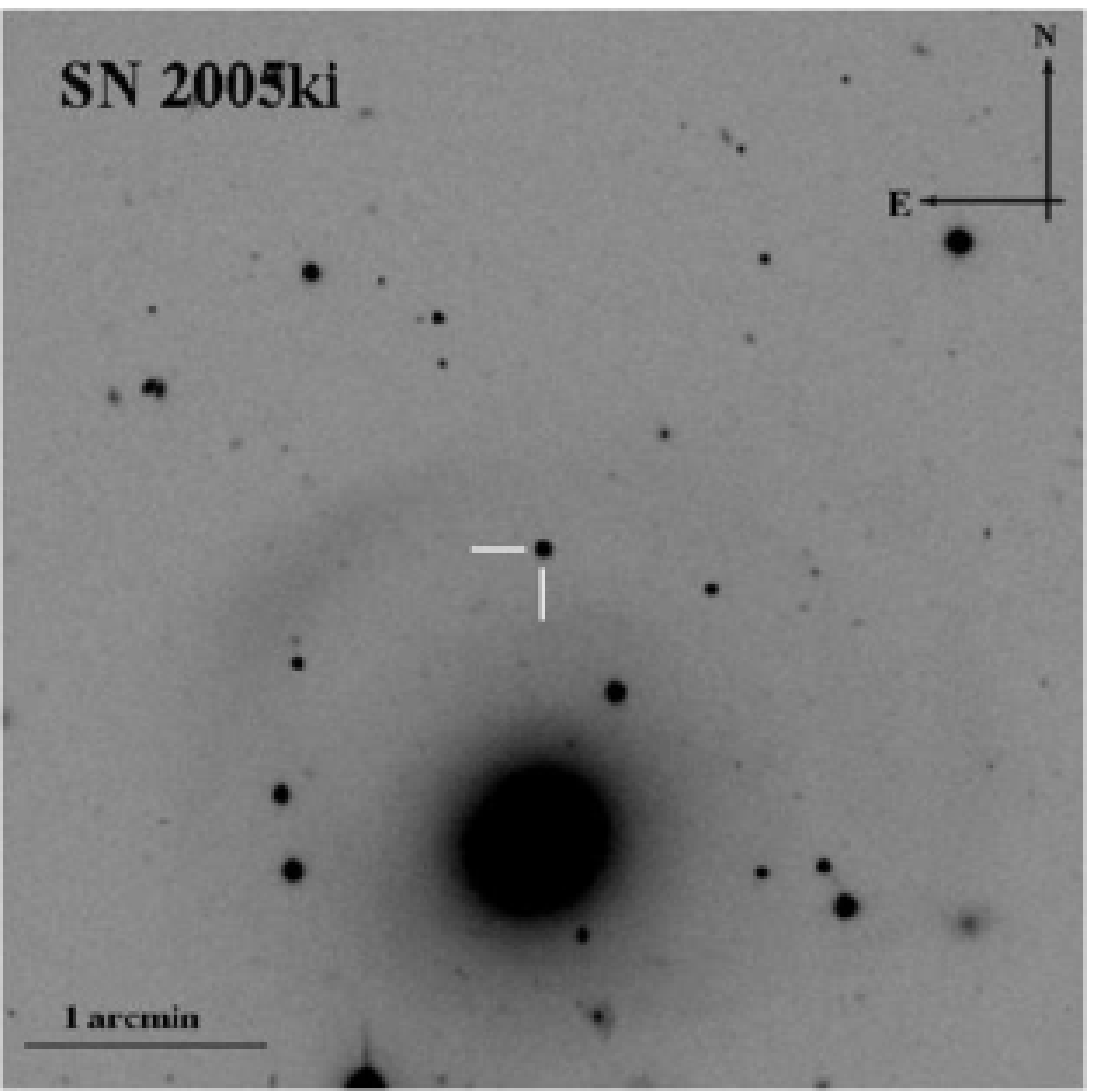}{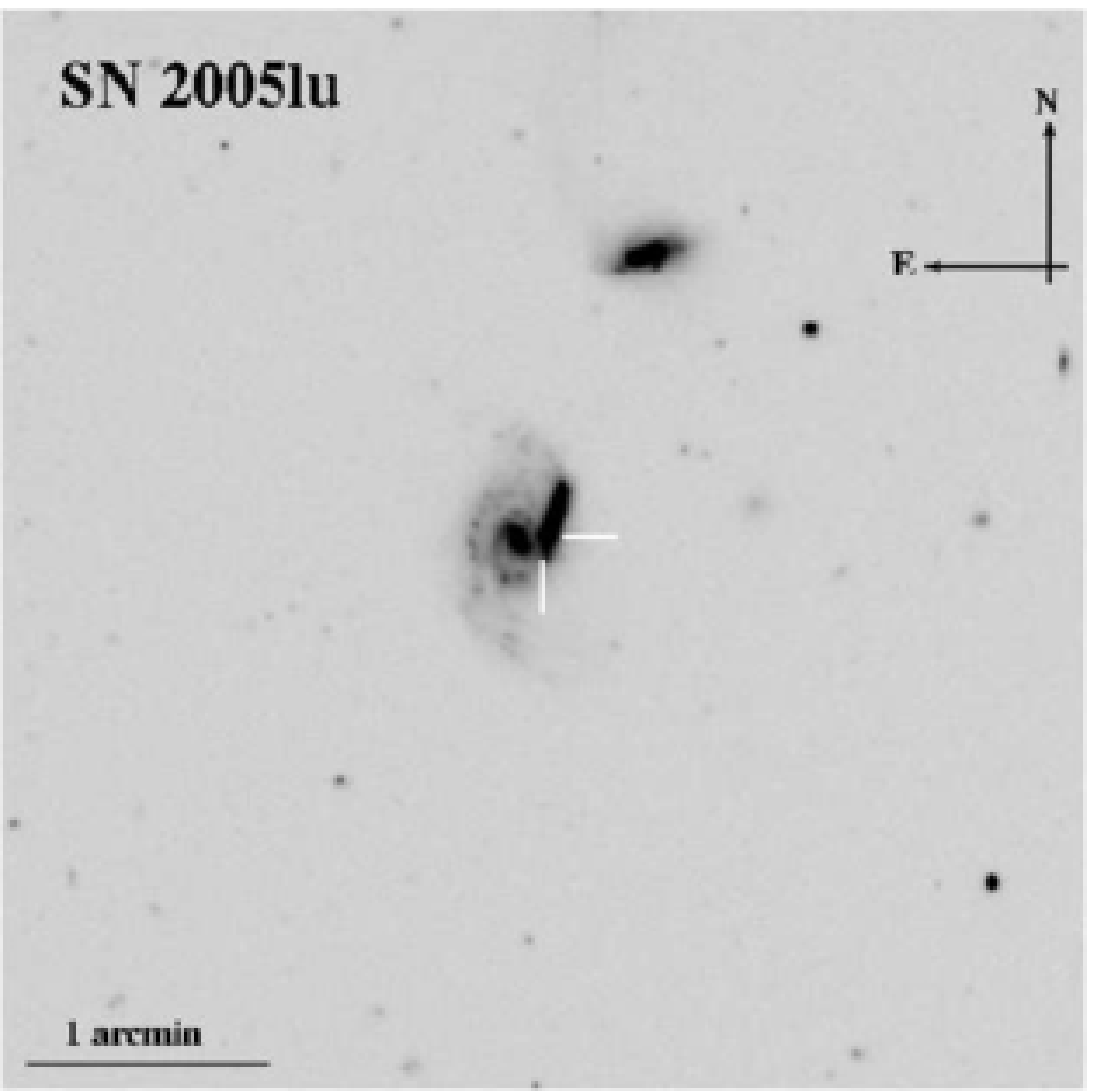}
\plottwo{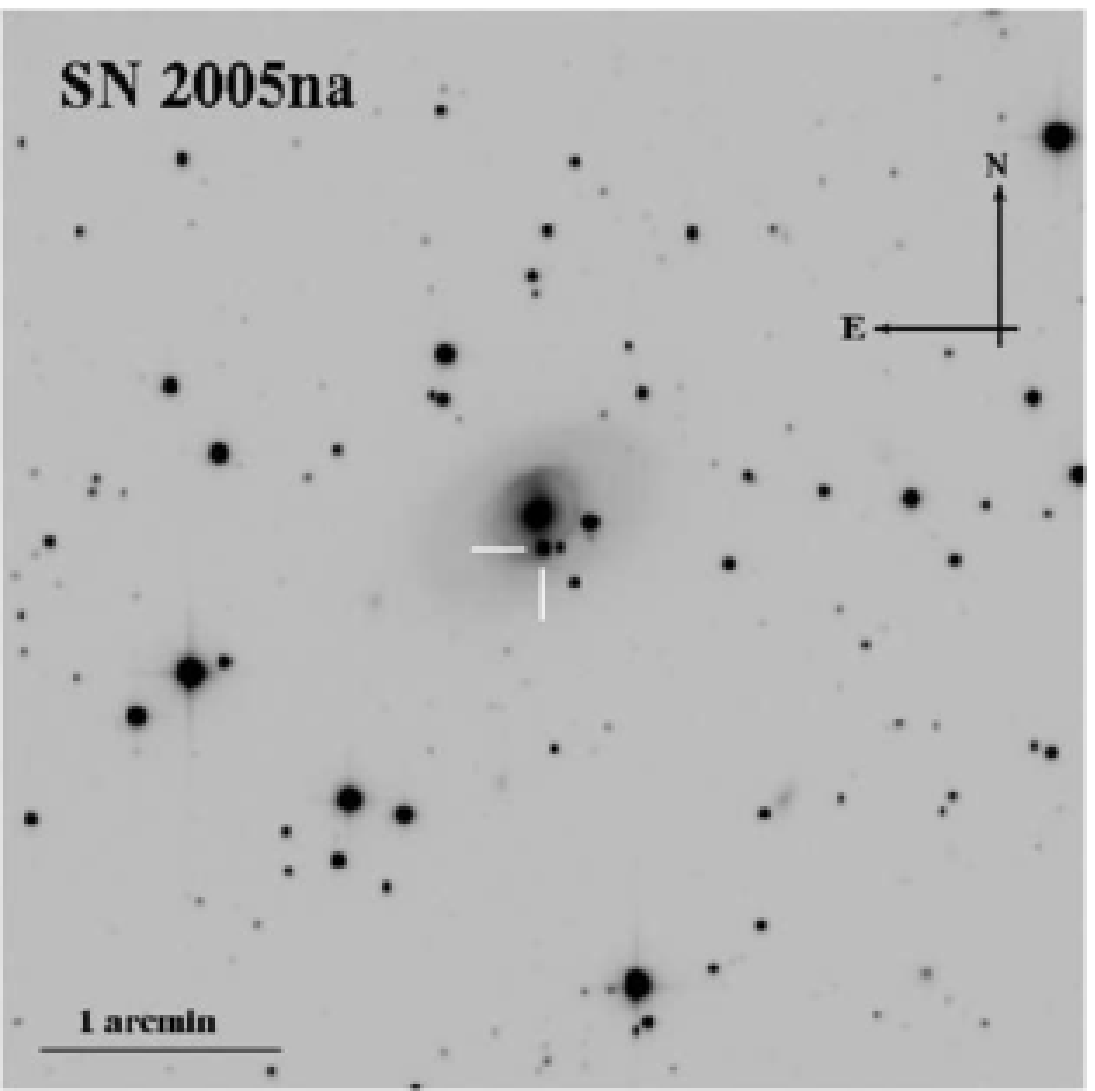}{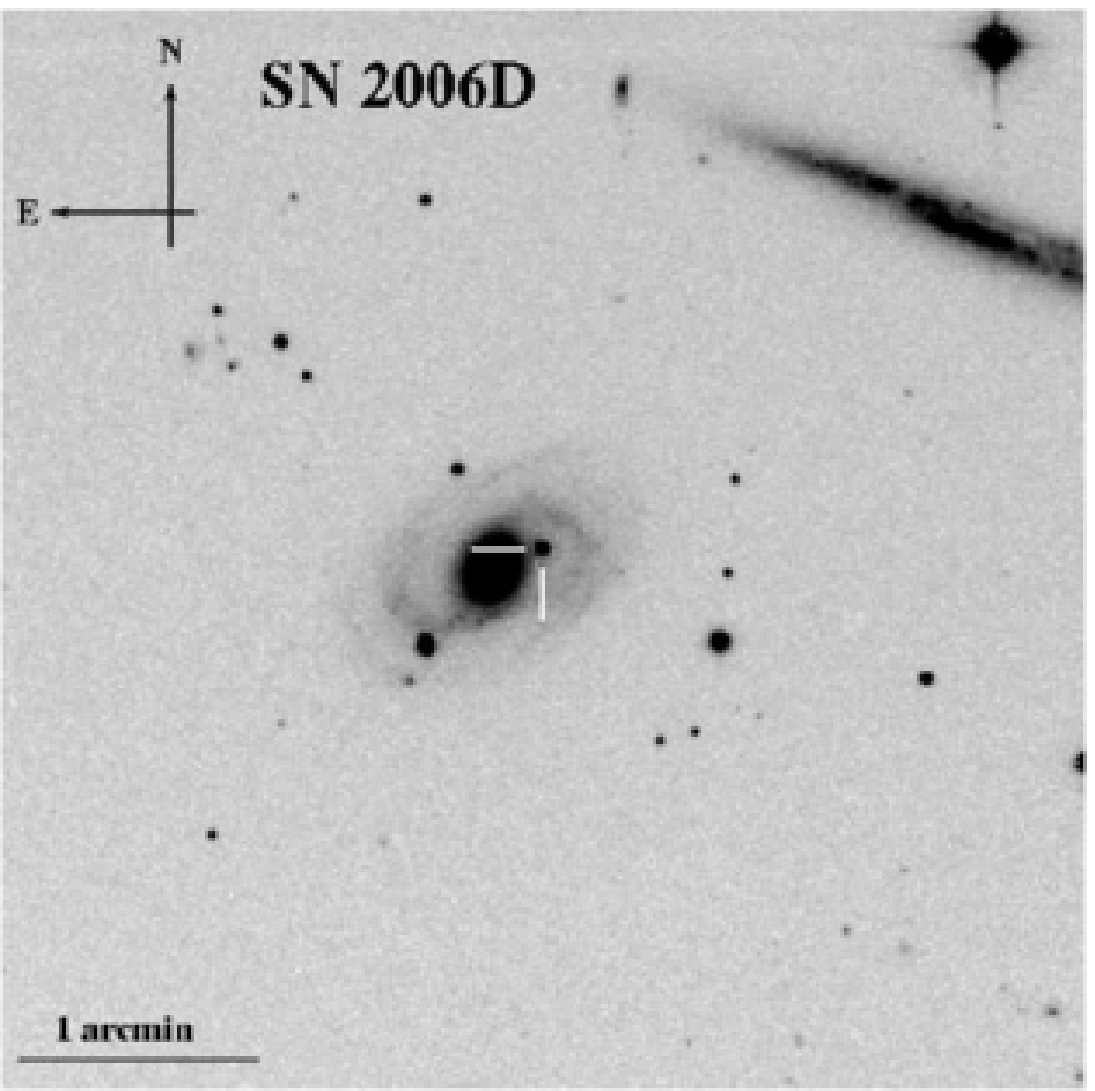}
\newline                                                                     
\plottwo{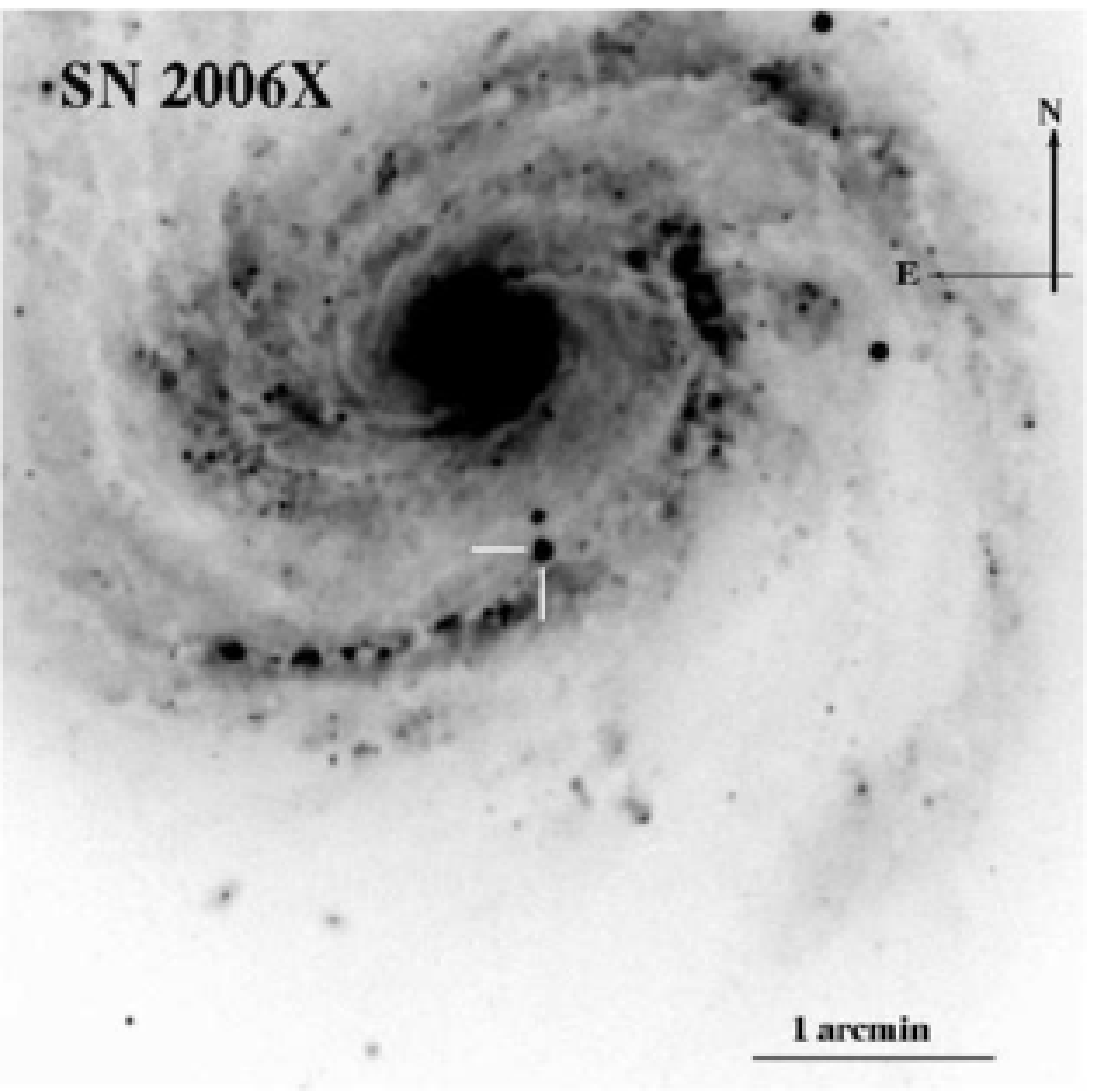}{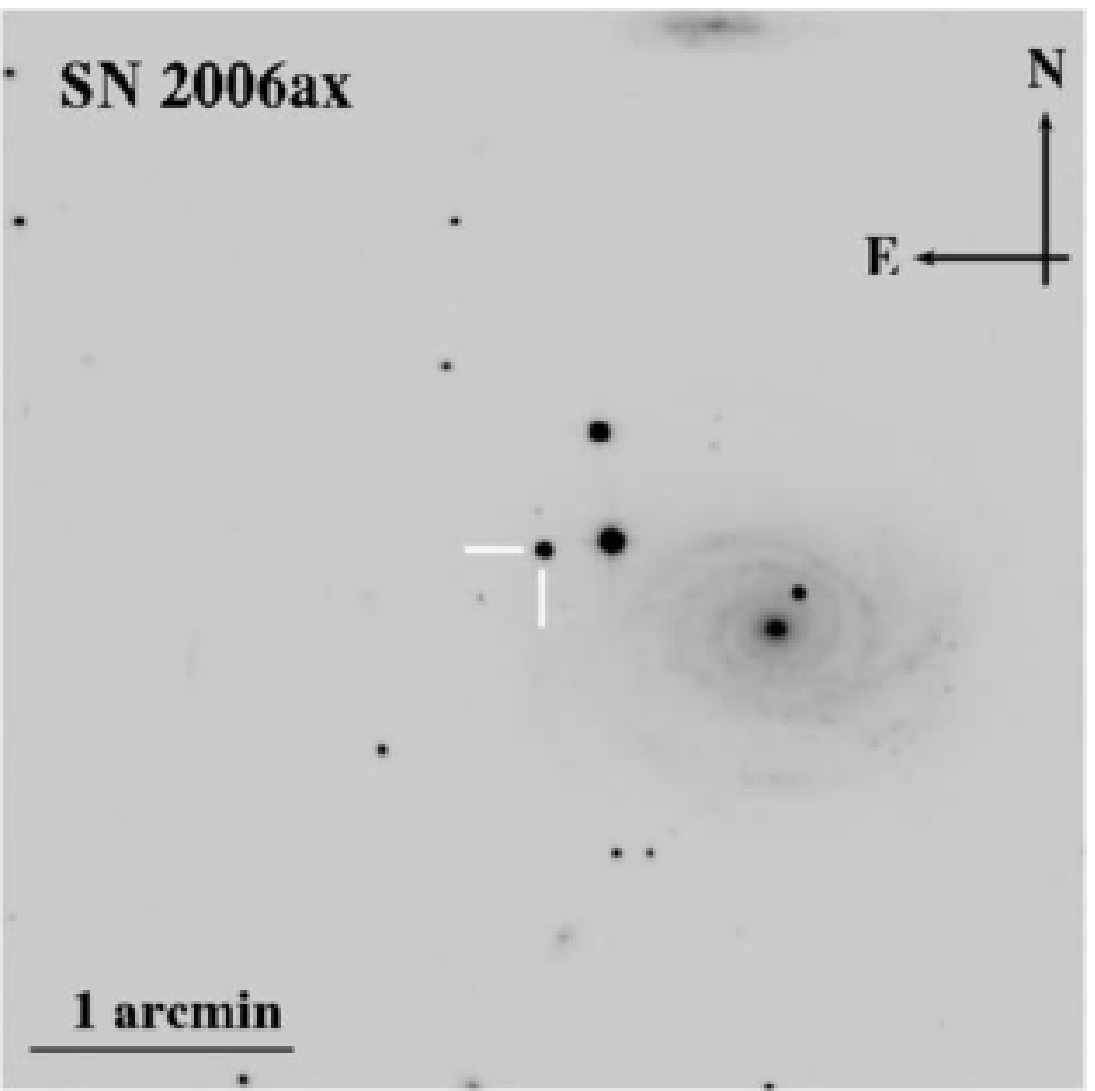}
\plottwo{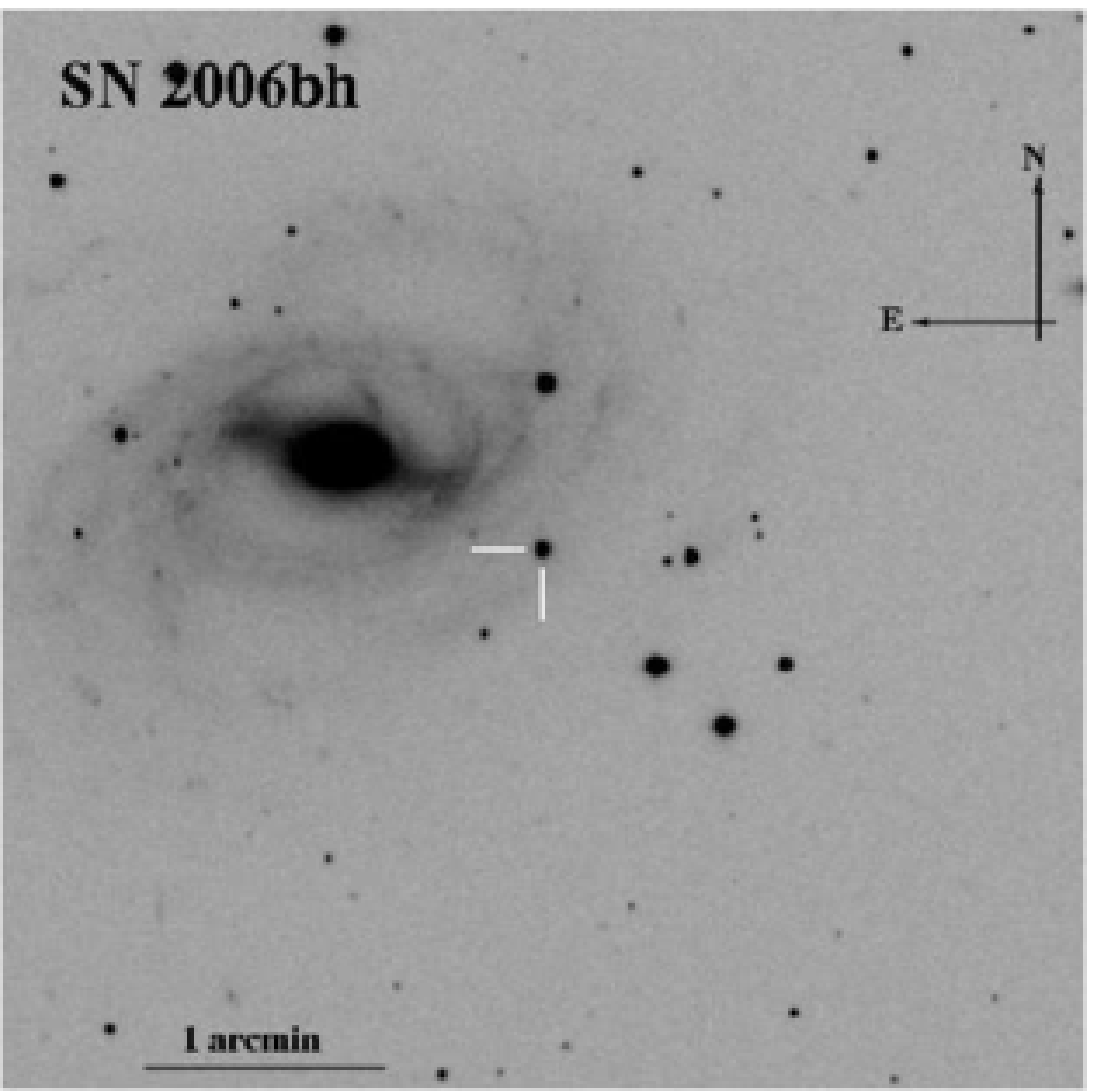}{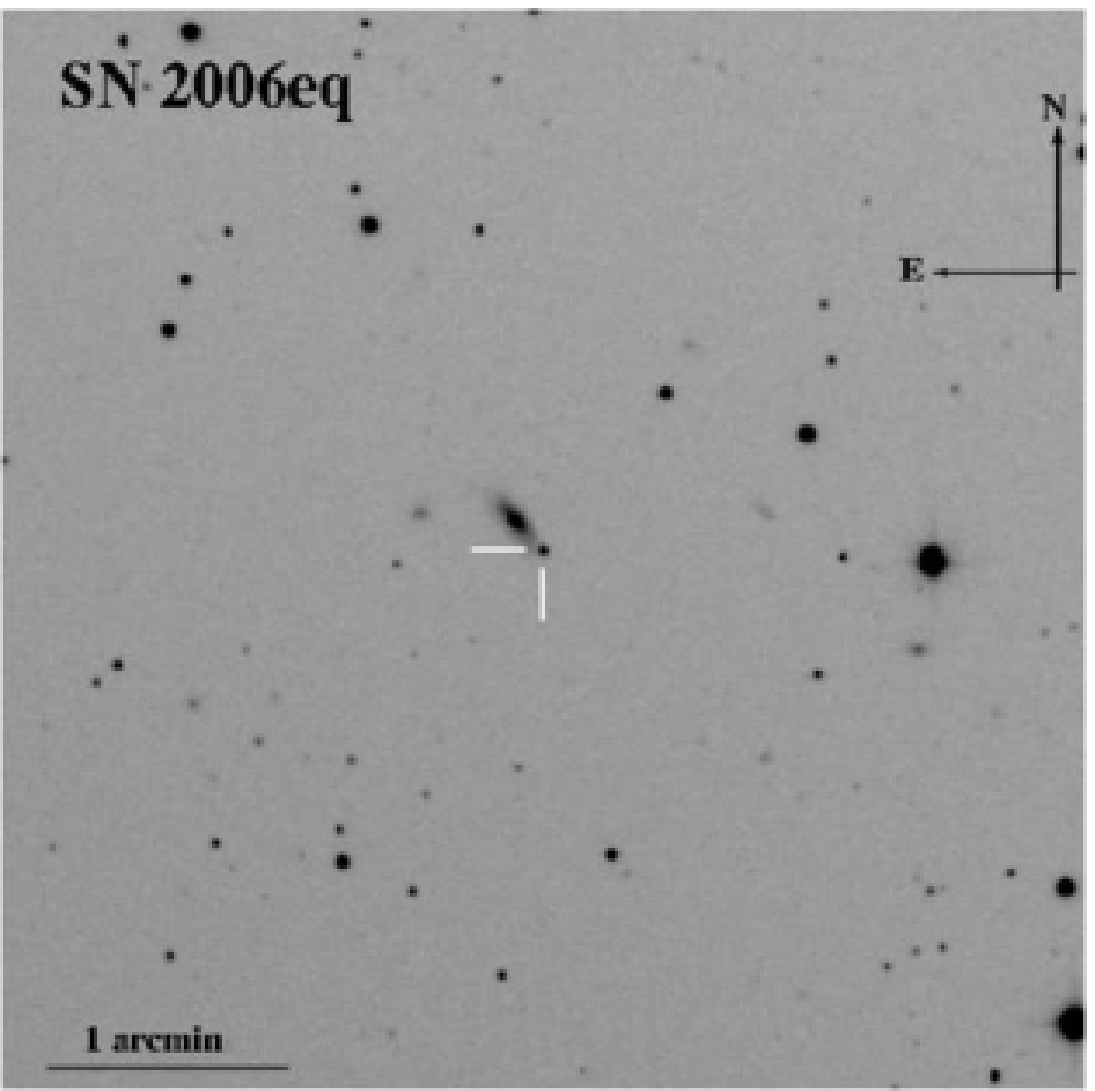}
\newline                                                                     
\plottwo{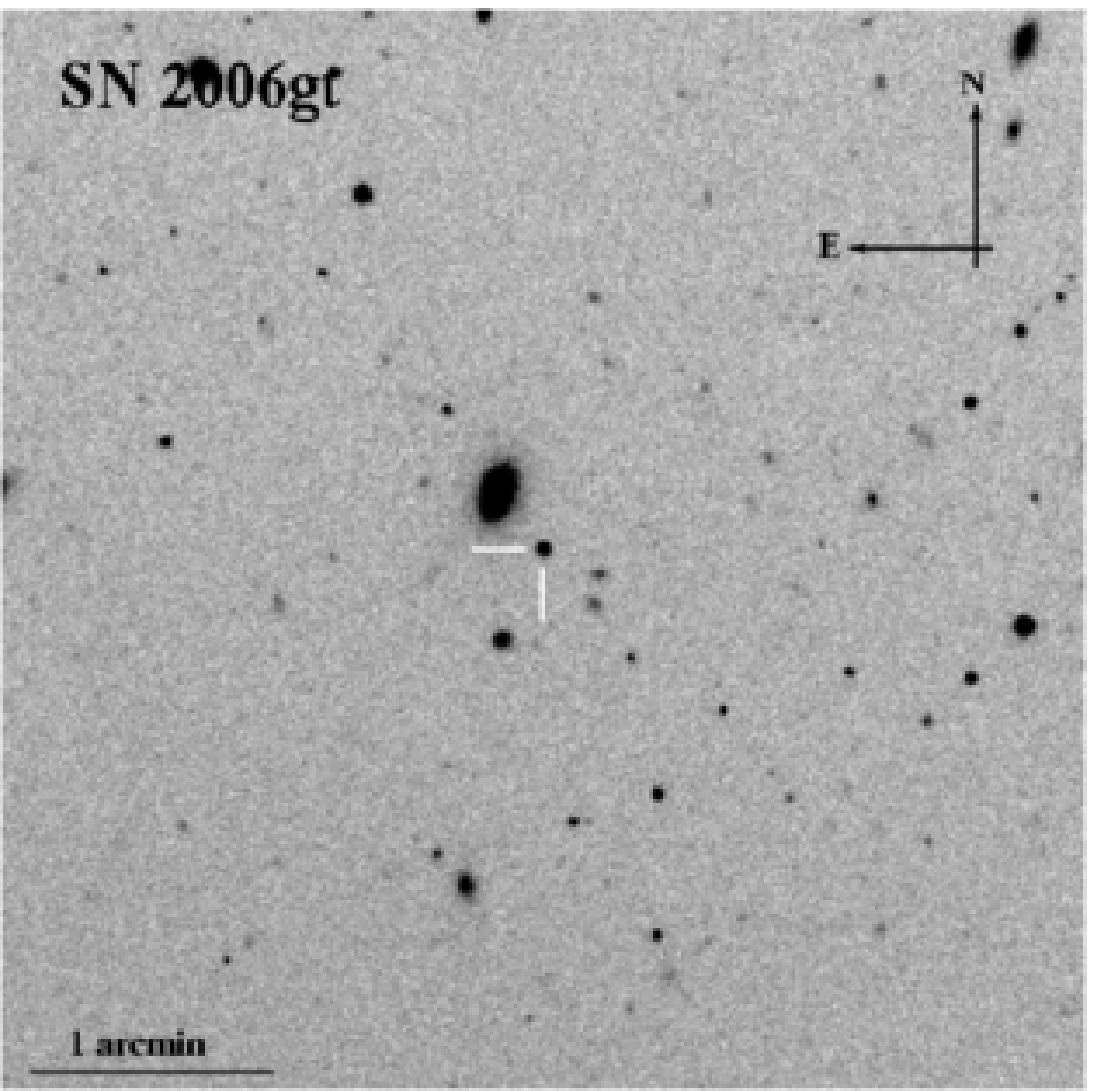}{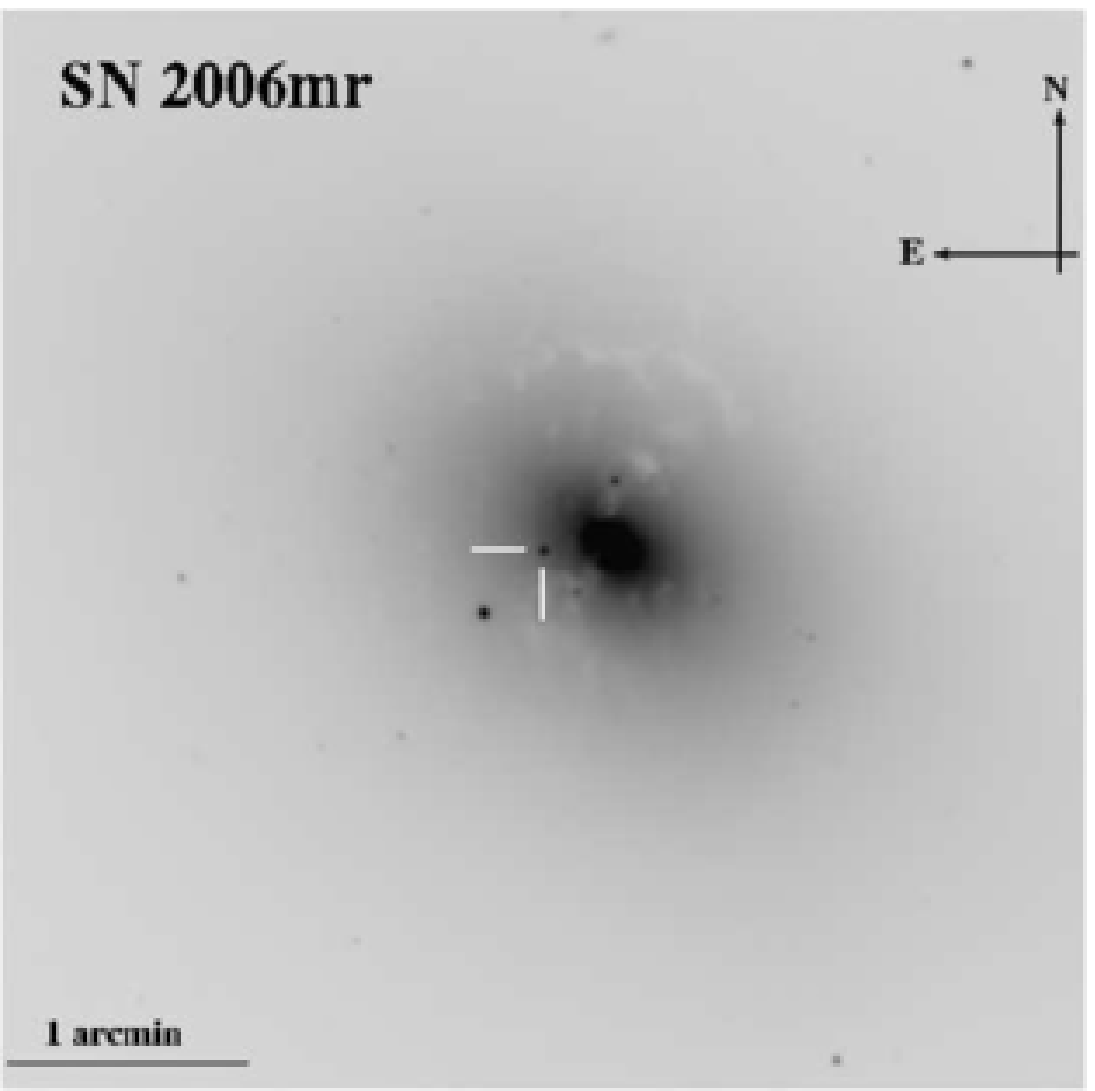}
\plottwo{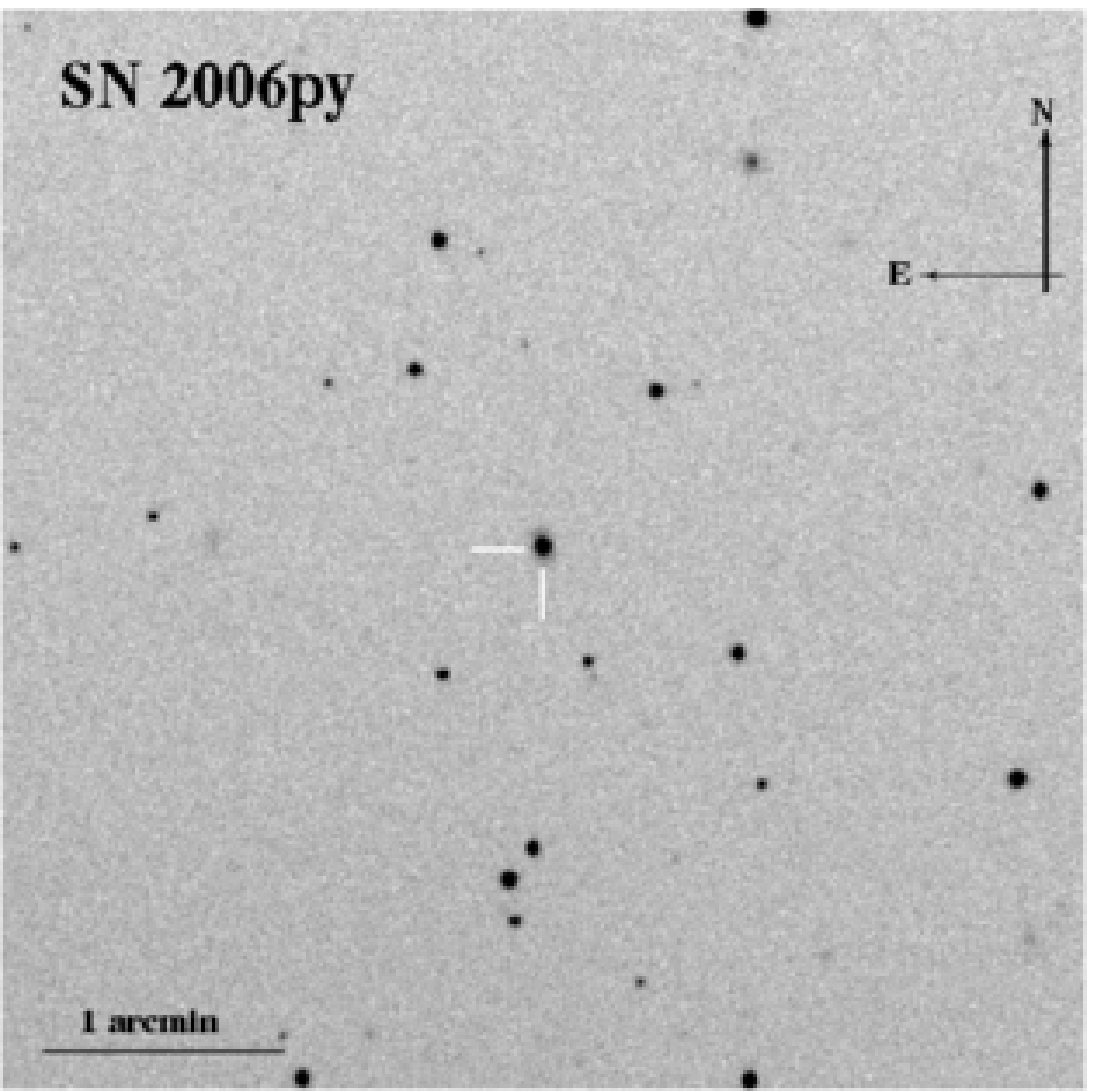}{charts/white.eps}
\newline                                                                     
{\center Contreras {\it et al.} Fig. \ref{fig:fcharts}}
\epsscale{1.}

\end{figure}
\clearpage
\newpage

\clearpage

\begin{figure}
%\epsscale{0.71}
\plottwo{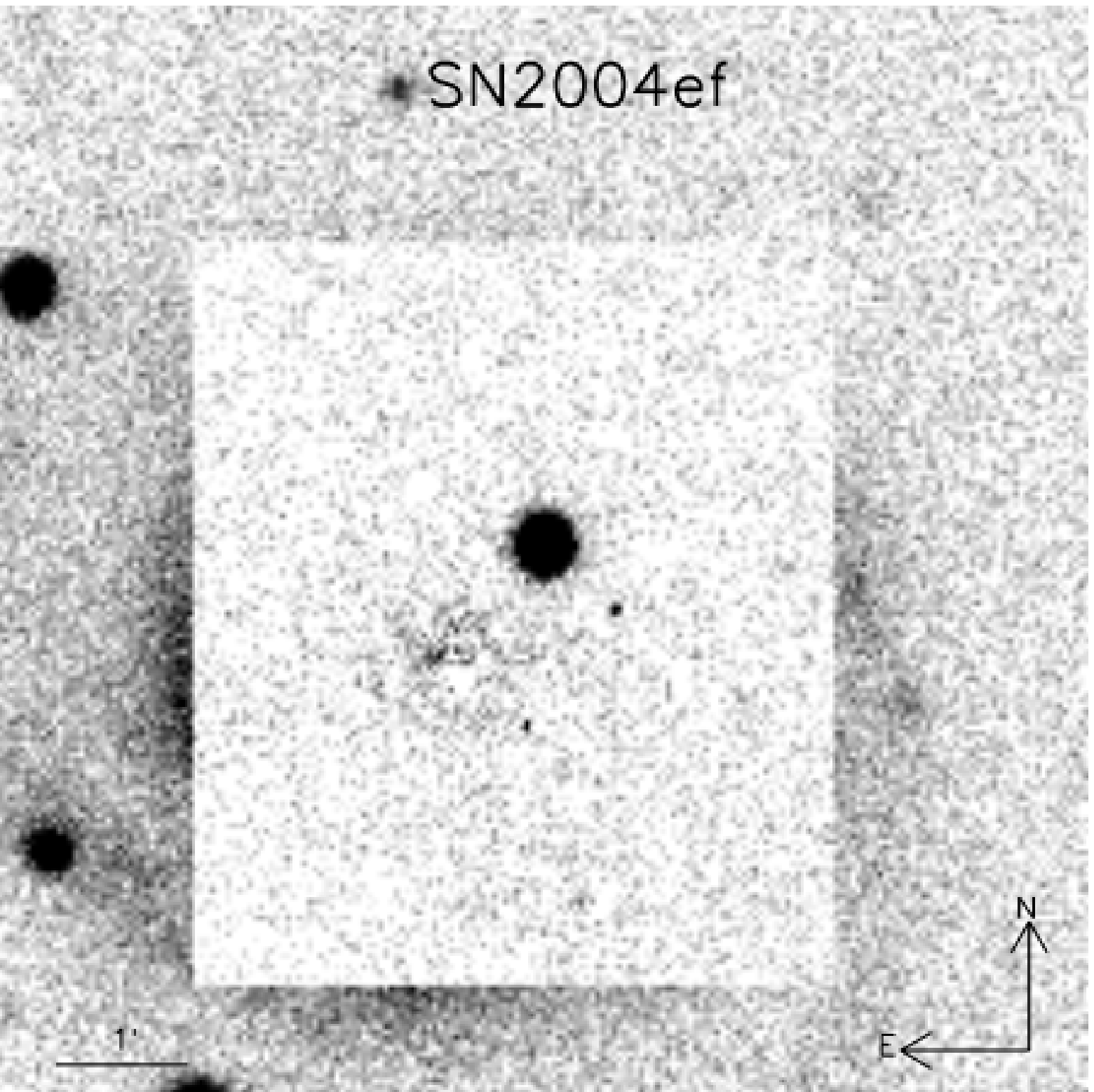}{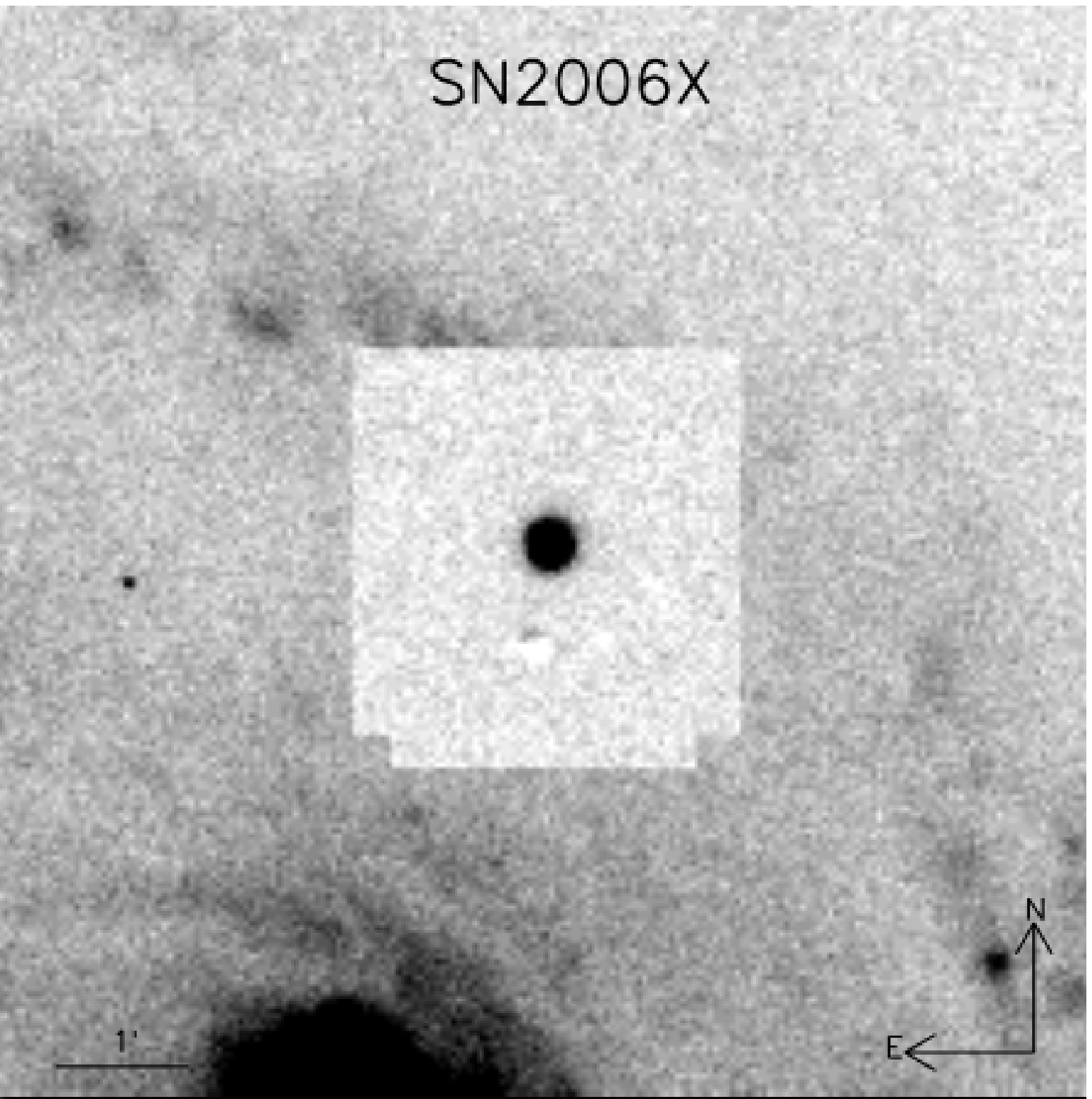}
{\center Contreras {\it et al.} Fig. \ref{fig:subtractions}}
\end{figure}

\clearpage
\begin{figure}
%\epsscale{0.71}
\plotone{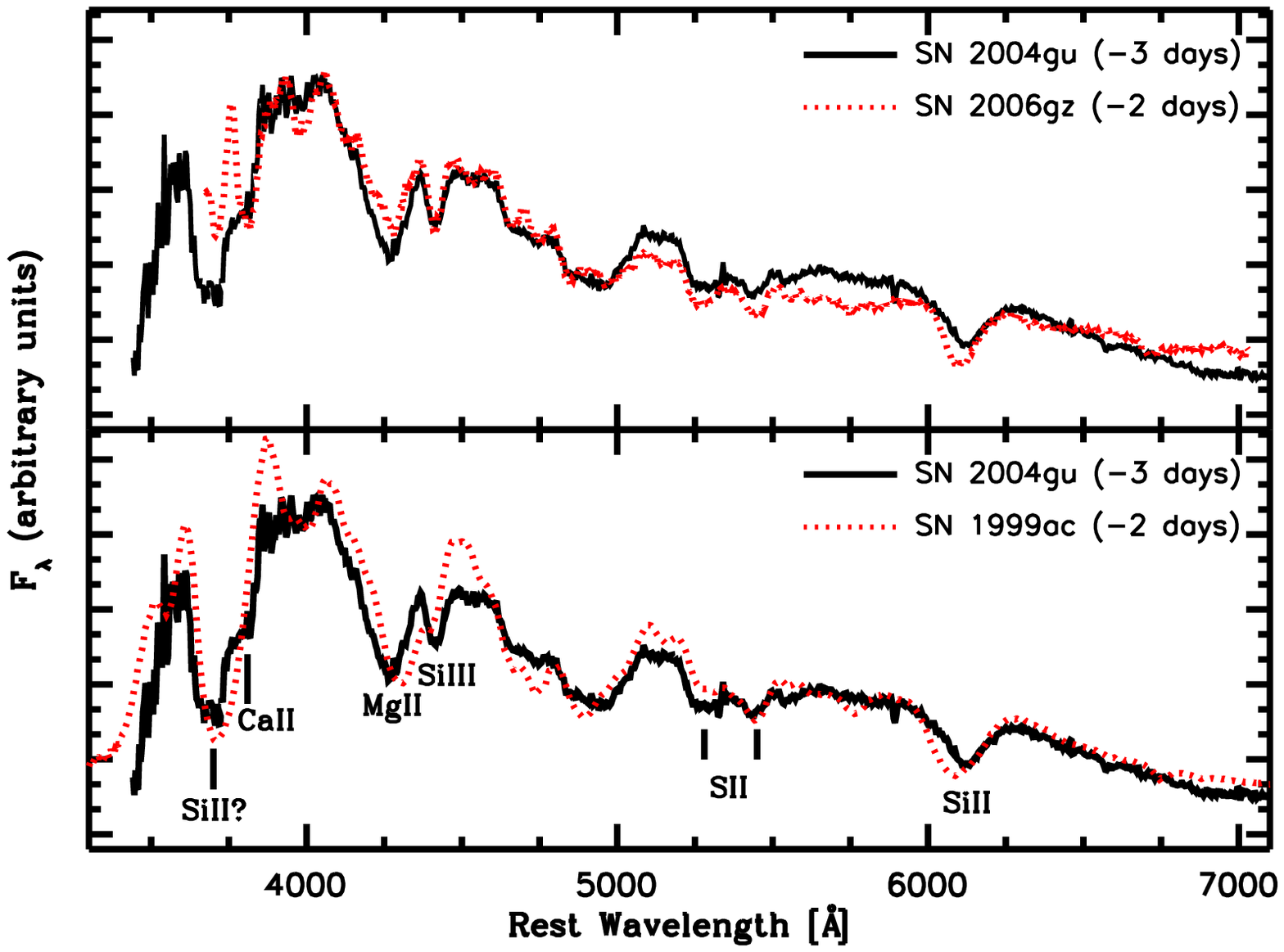}
{\center Contreras {\it et al.} Fig. \ref{fig:spectra}}
\end{figure}

\begin{figure}
\epsscale{0.71}
\plotone{opt_ir_plots/opt_extinction_coeff.eps}
{\center Contreras {\it et al.} Fig. \ref{fig:extinc}}
\end{figure}

\begin{figure}
\plotone{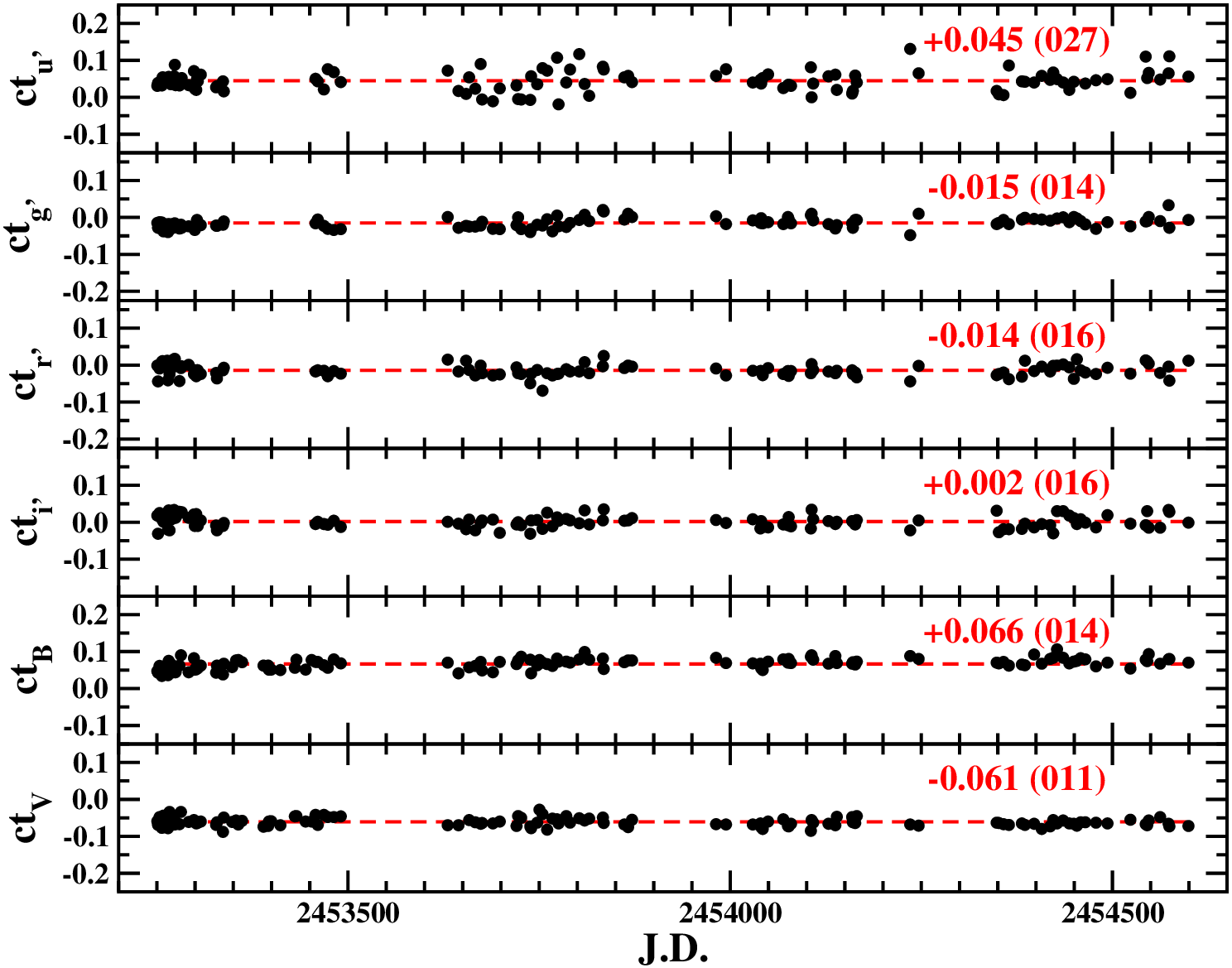}
{\center Contreras {\it et al.} Fig. \ref{fig:colorc}}
\end{figure}

\begin{figure}[t]

  \epsscale{1.}
  \plottwo{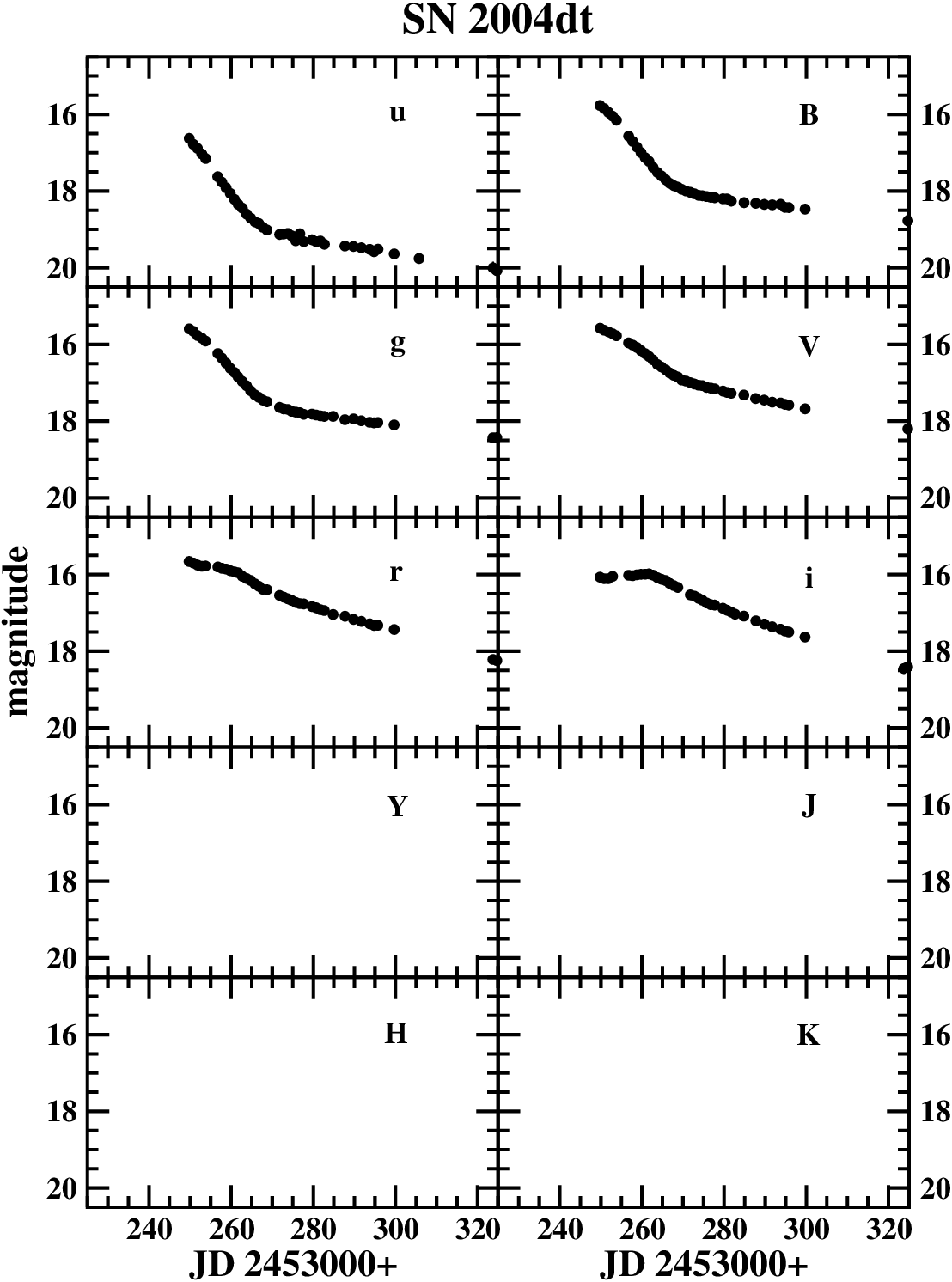}{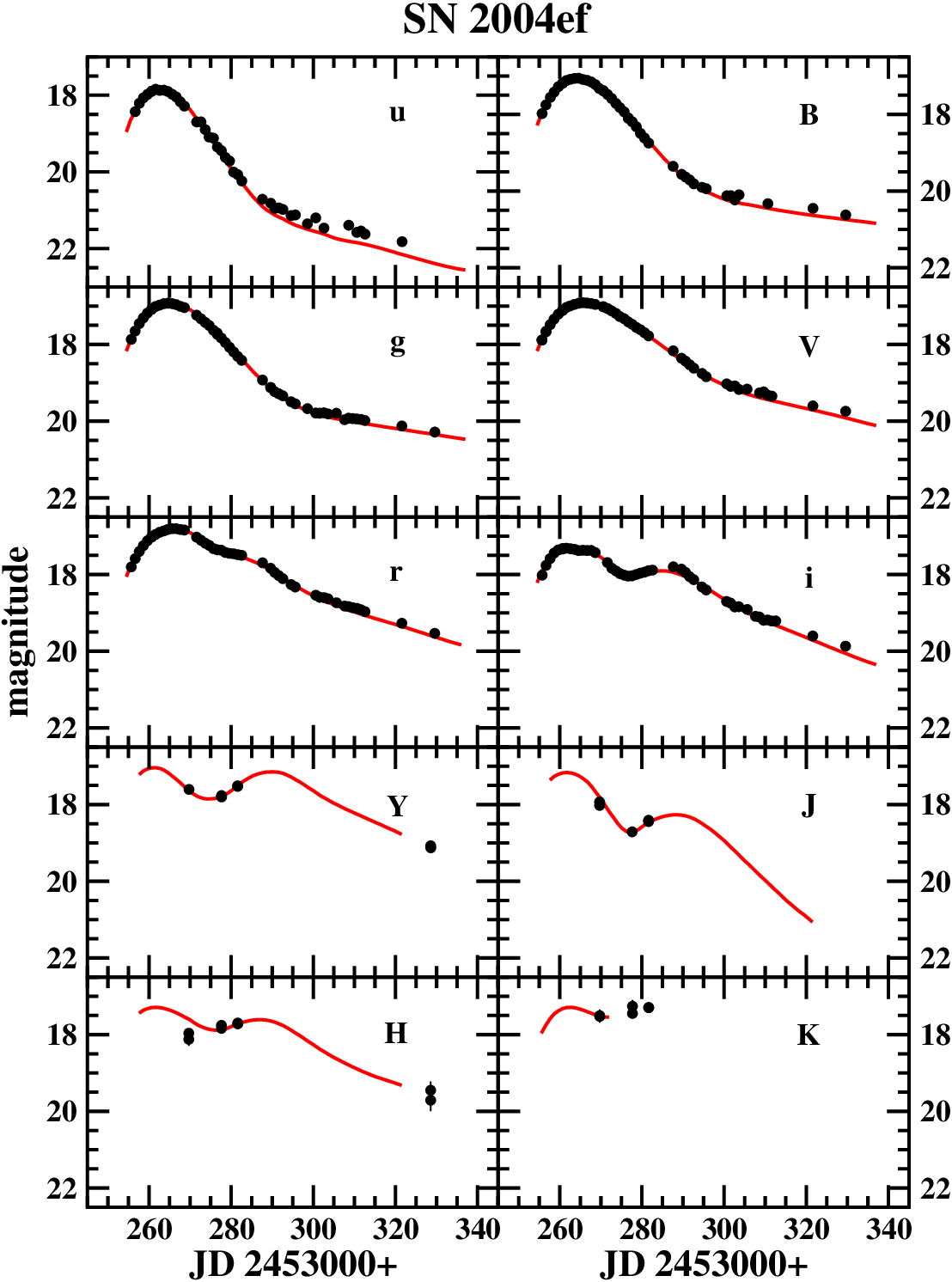}
  \newline
  \newline
  \plottwo{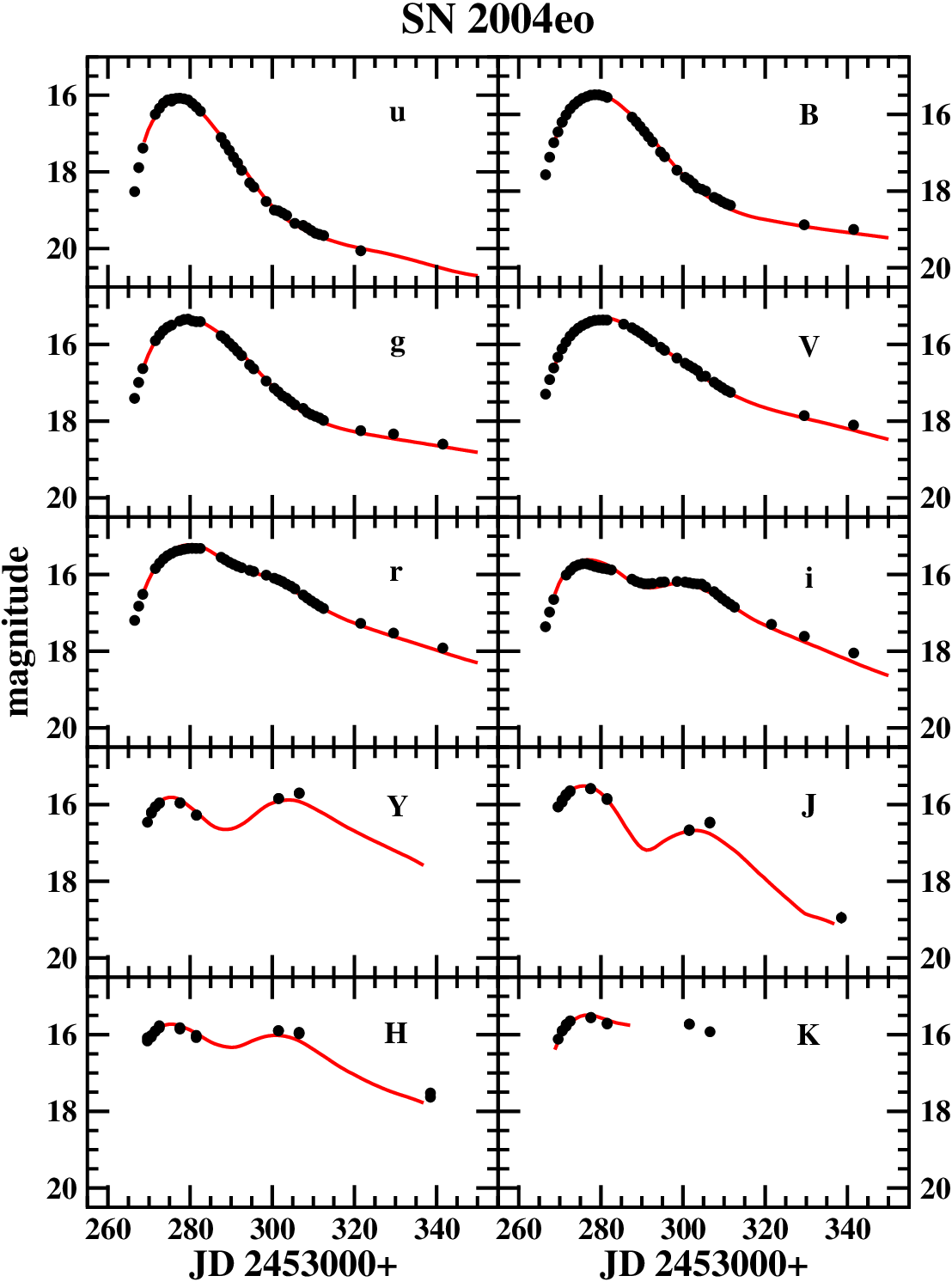}{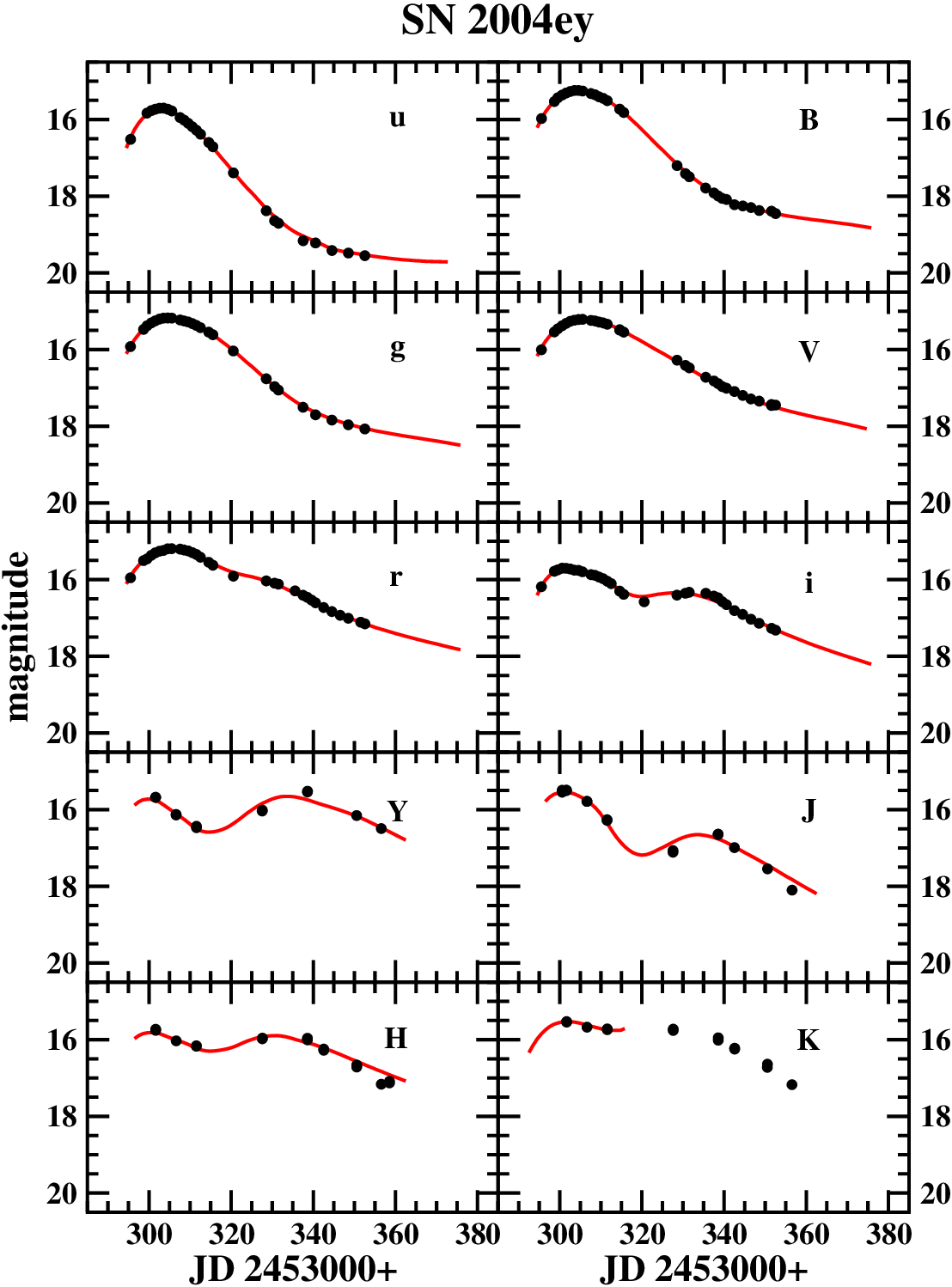}
  {\center Contreras {\it et al.} Fig. \ref{fig:flcurves}}

\end{figure}

\clearpage
\newpage

\begin{figure}[t]

  \plottwo{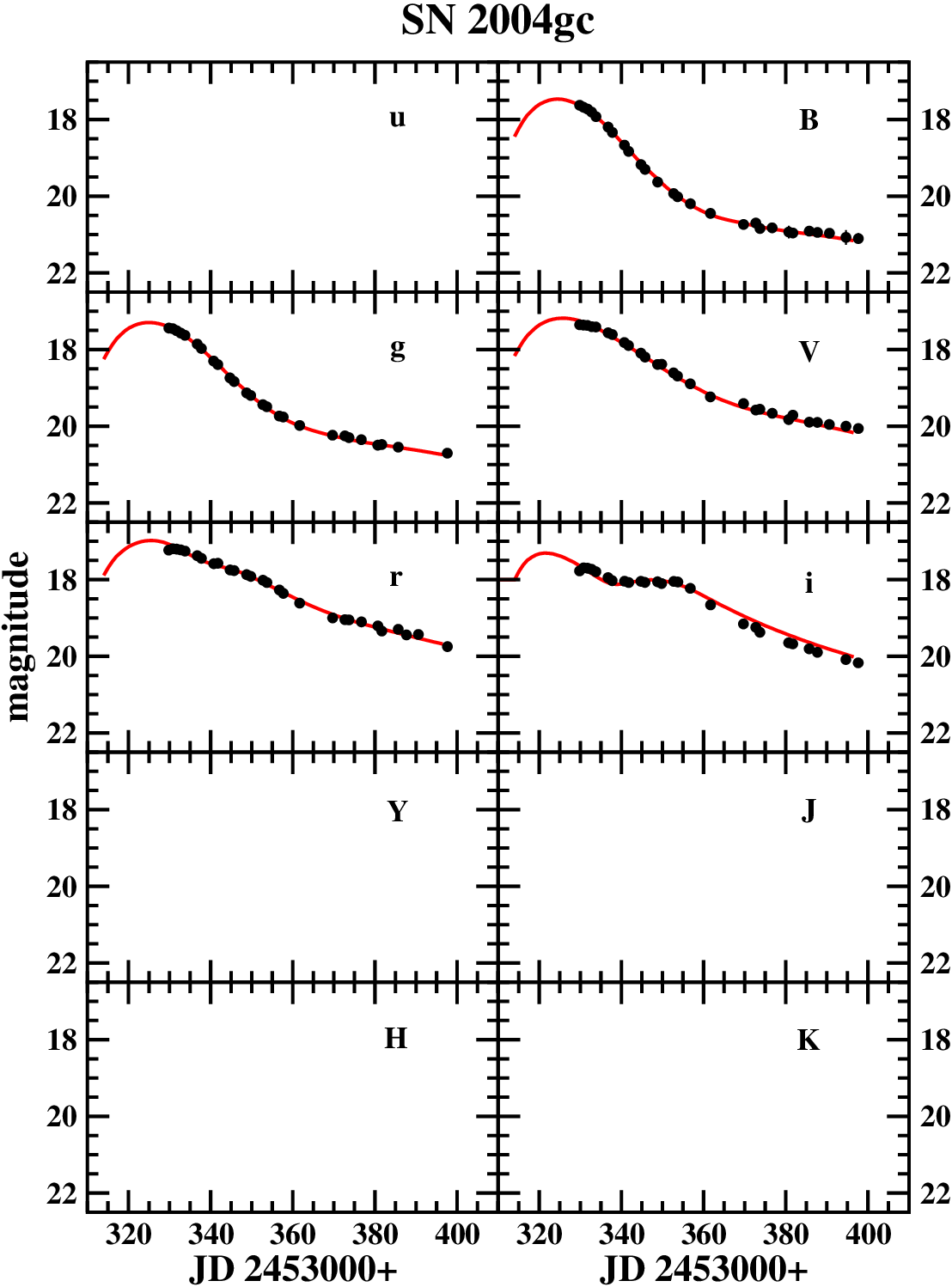}{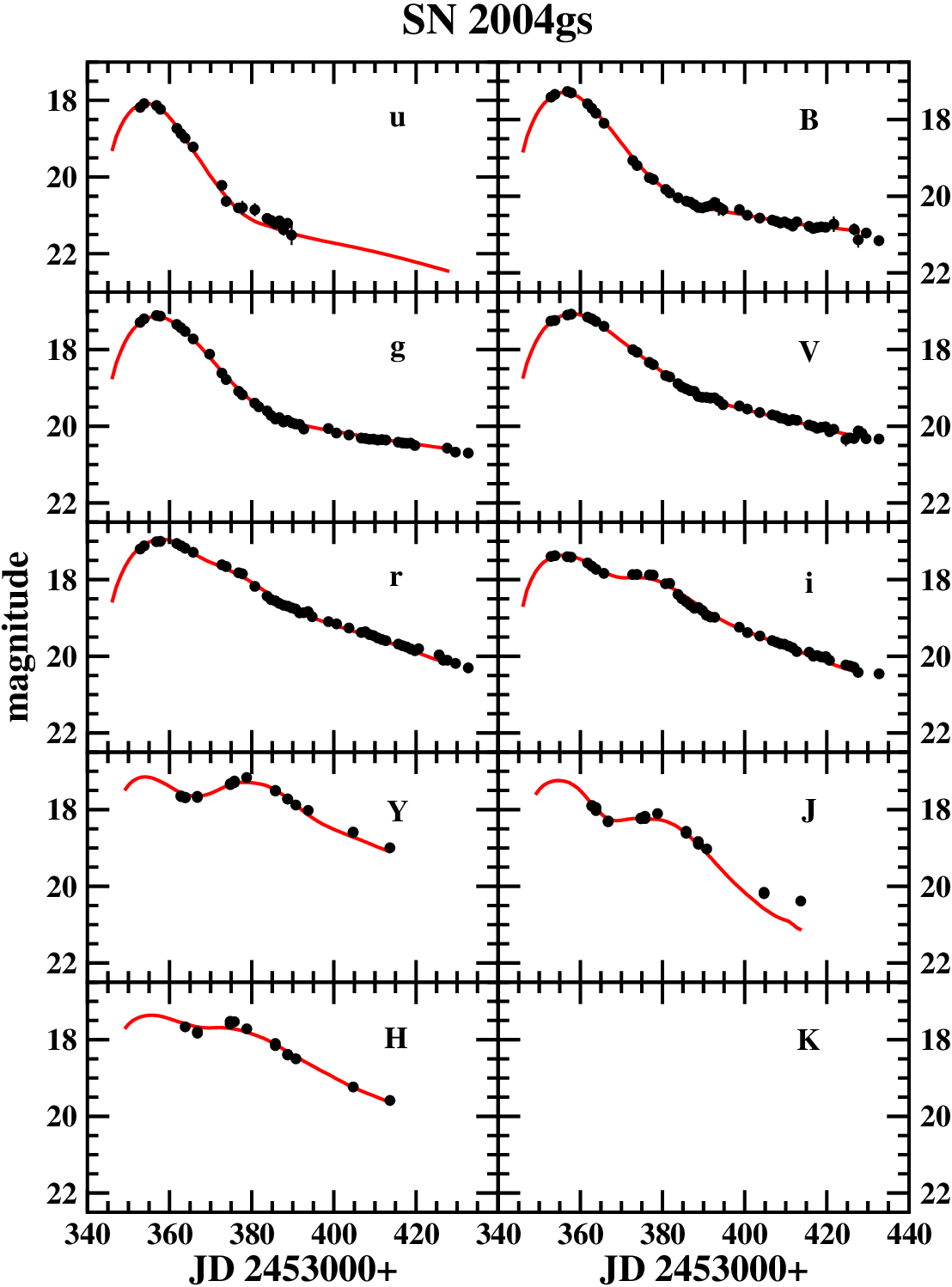}
  \newline
  \newline
  \plottwo{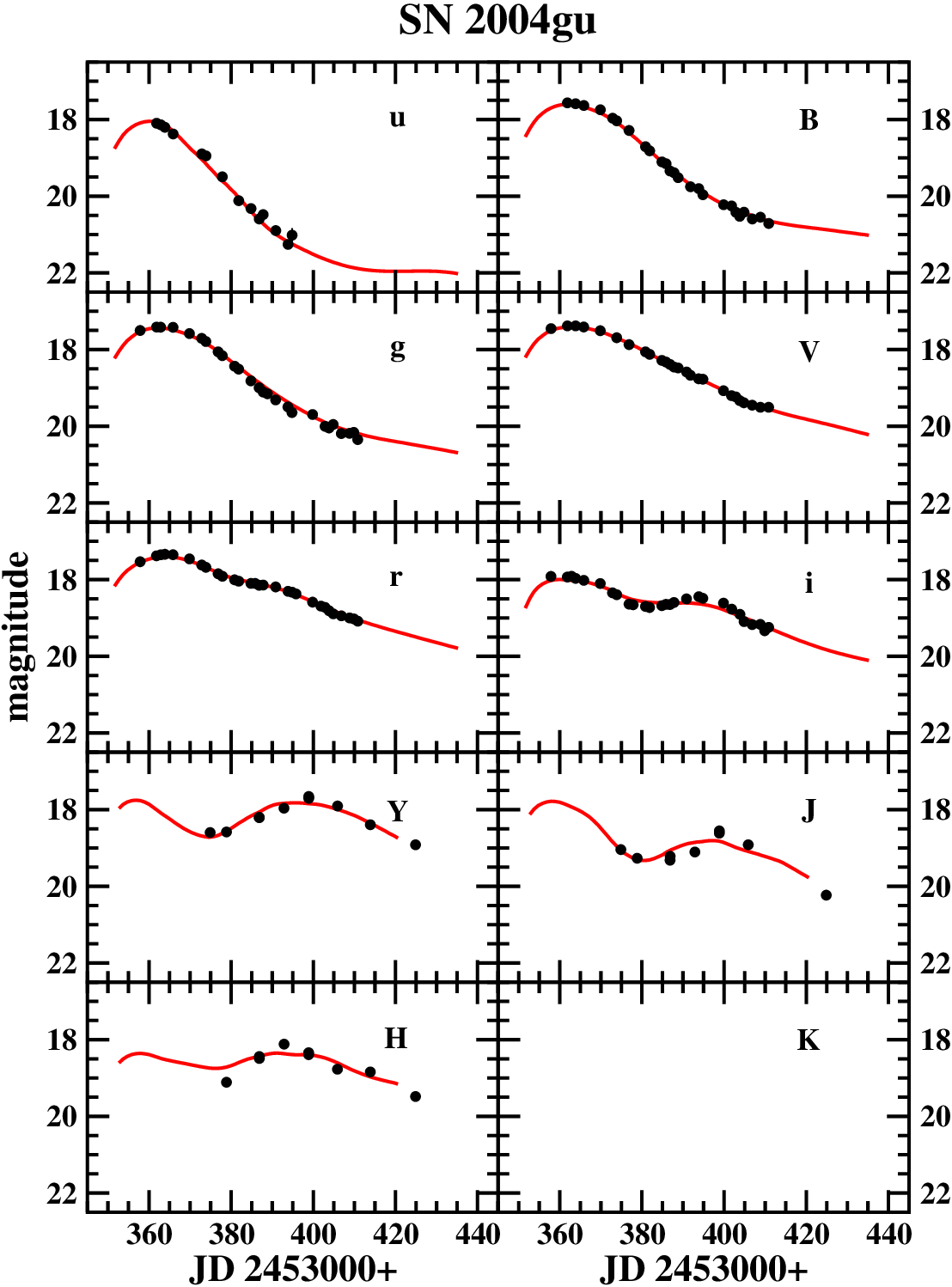}{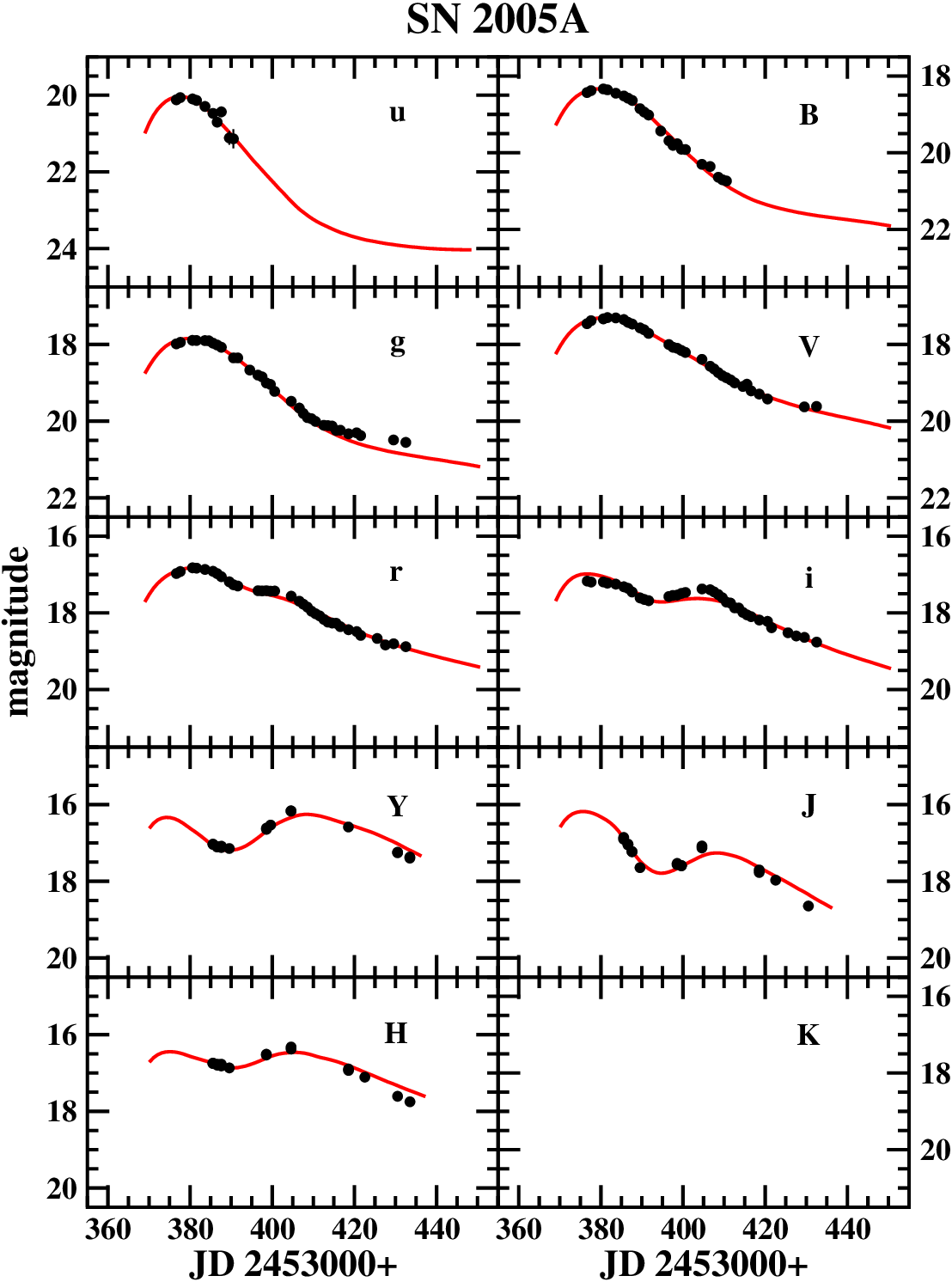}
  {\center Contreras {\it et al.} Fig. \ref{fig:flcurves}}

\end{figure}

\clearpage
\newpage

\begin{figure}[t]

  \plottwo{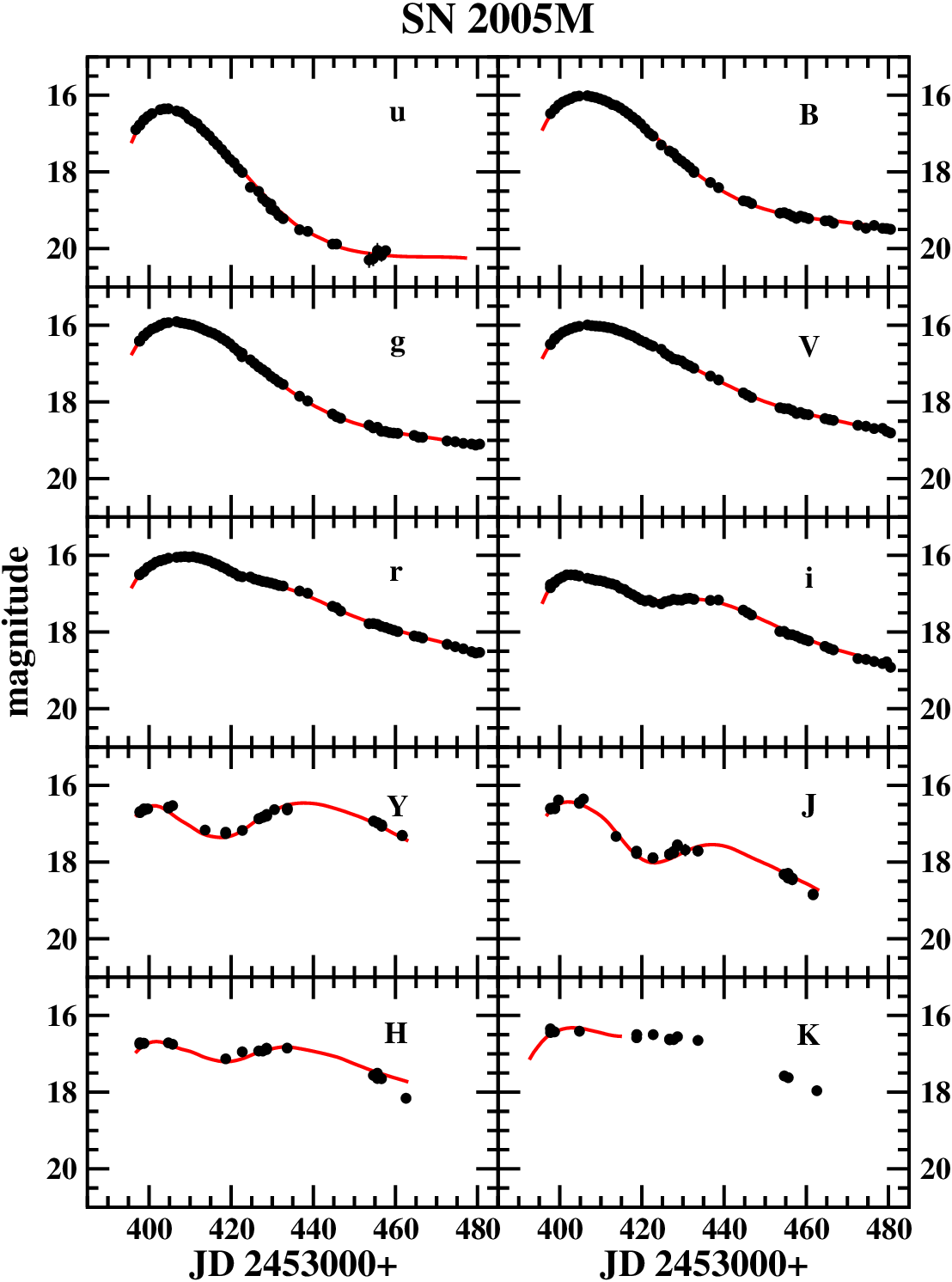}{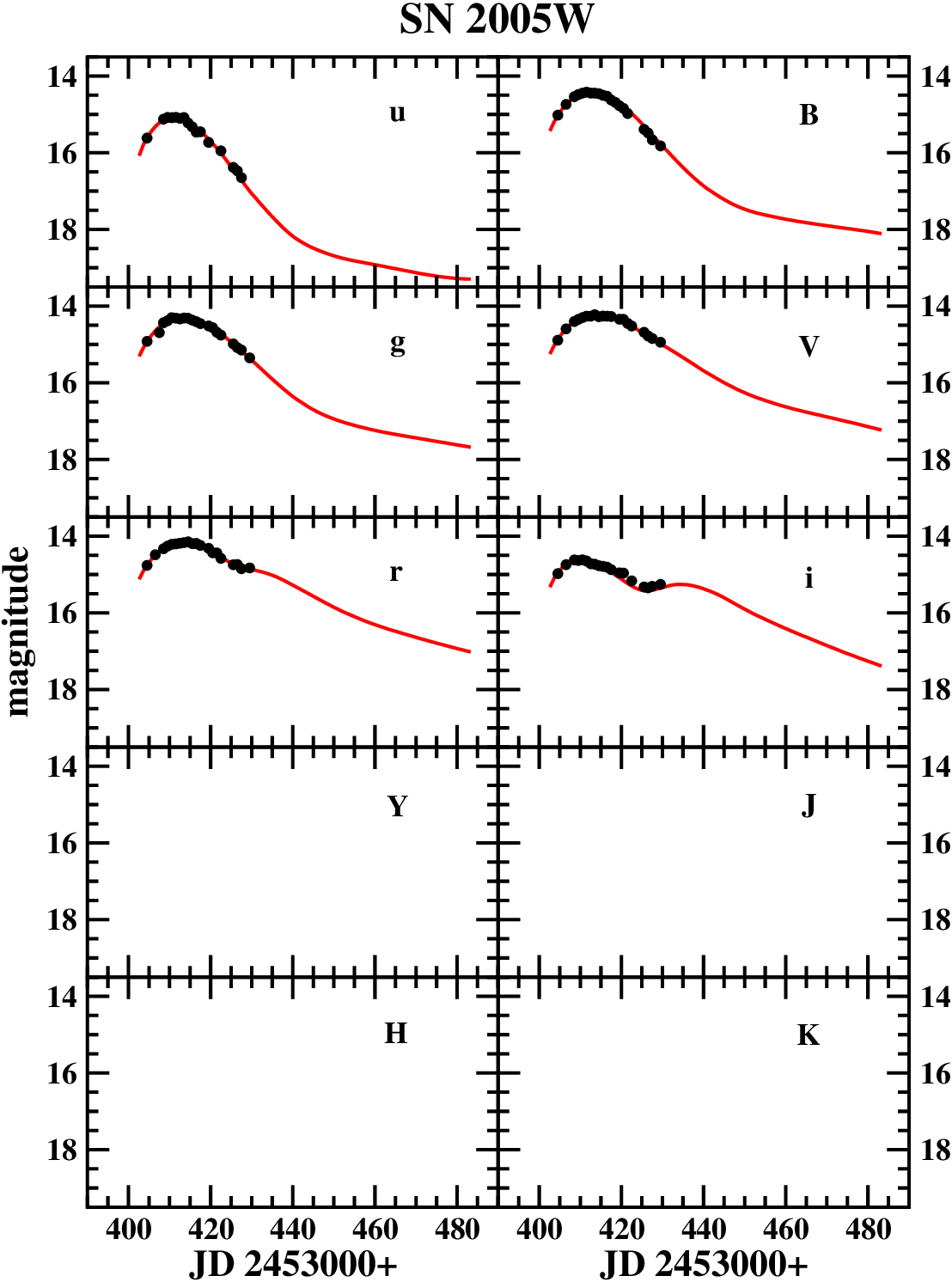}
  \newline
  \newline
  \plottwo{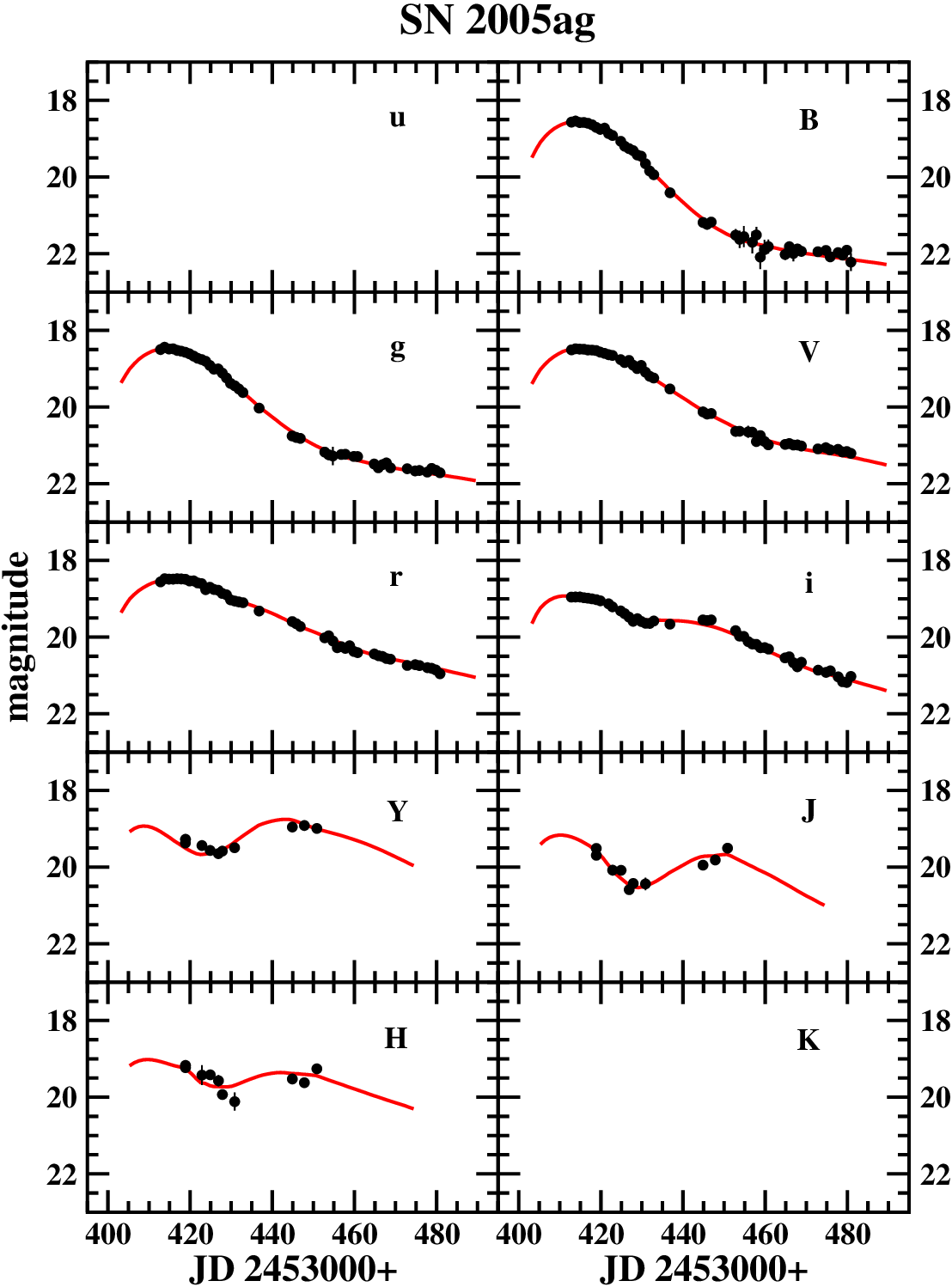}{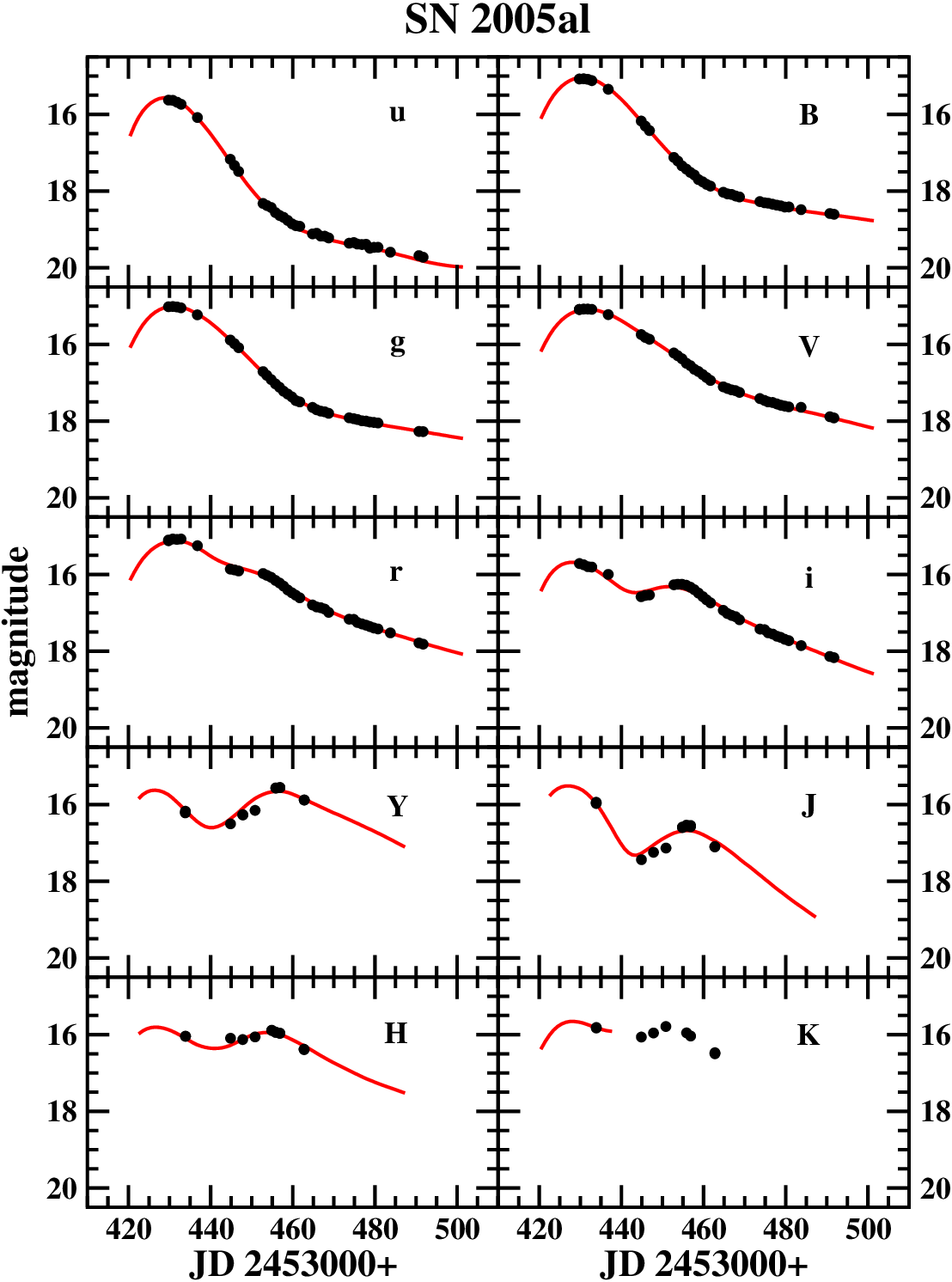}
  {\center Contreras {\it et al.} Fig. \ref{fig:flcurves}}

\end{figure}

\clearpage
\newpage

\begin{figure}[t]

  \plottwo{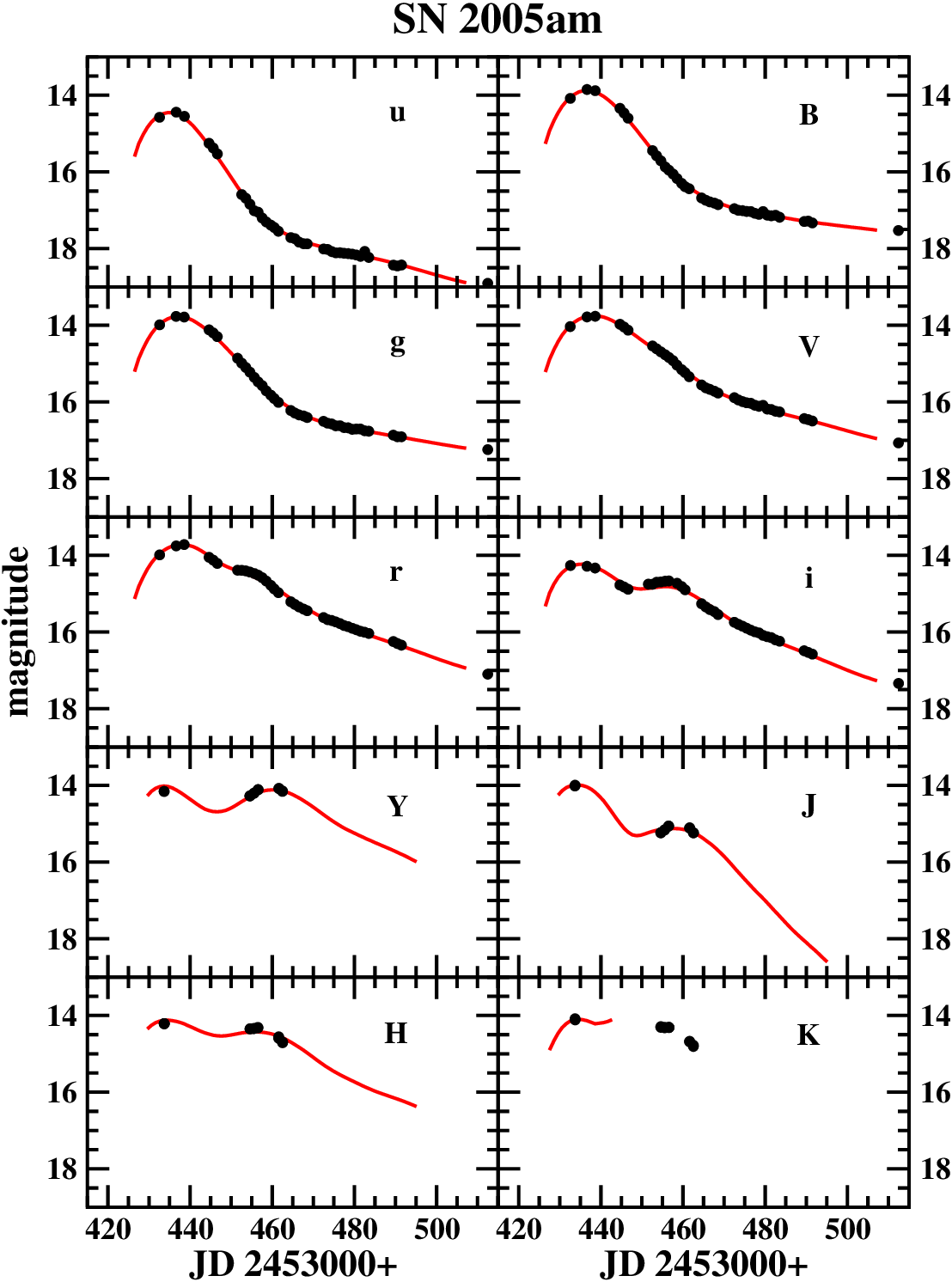}{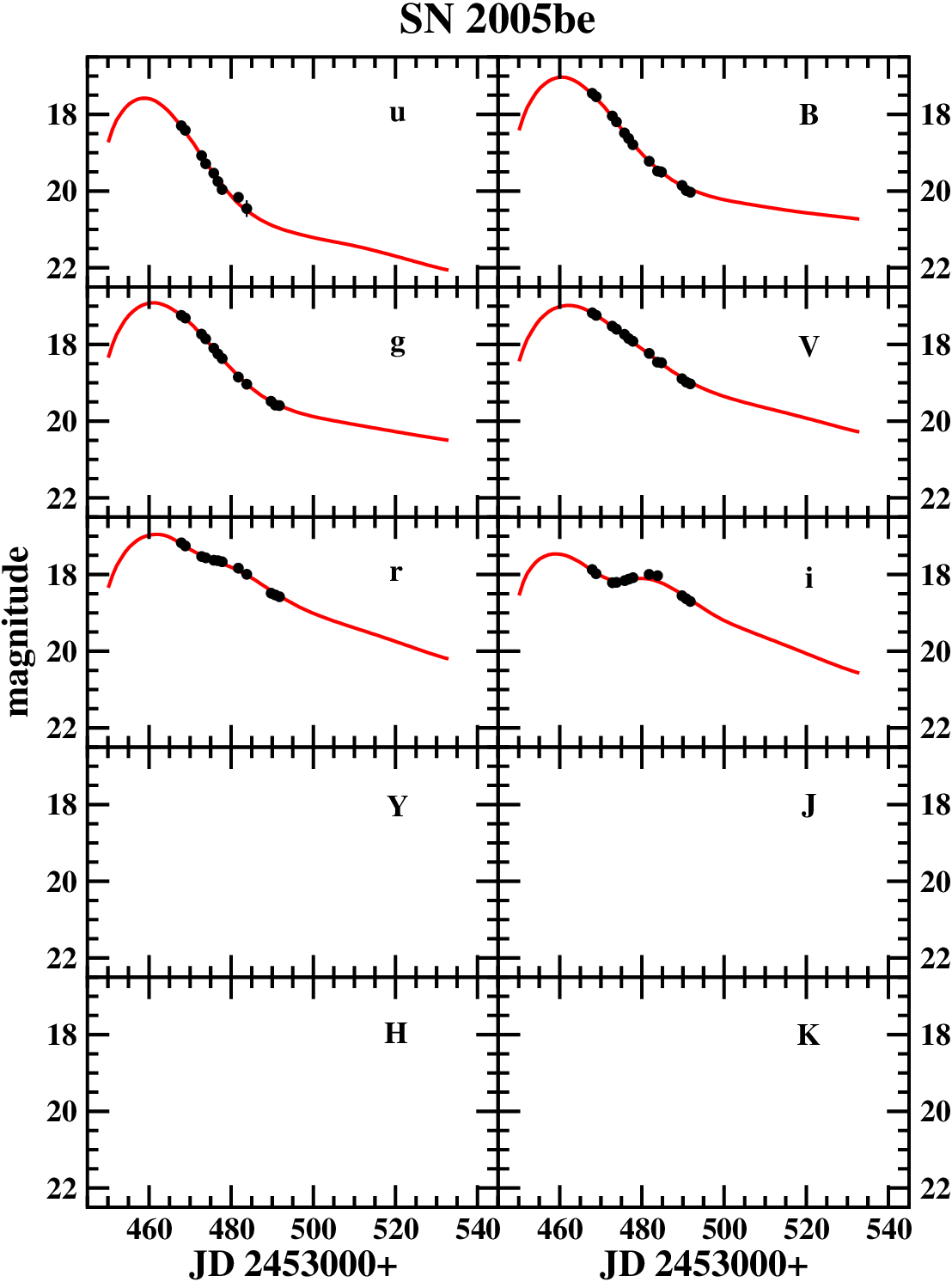}
  \newline
  \newline
  \plottwo{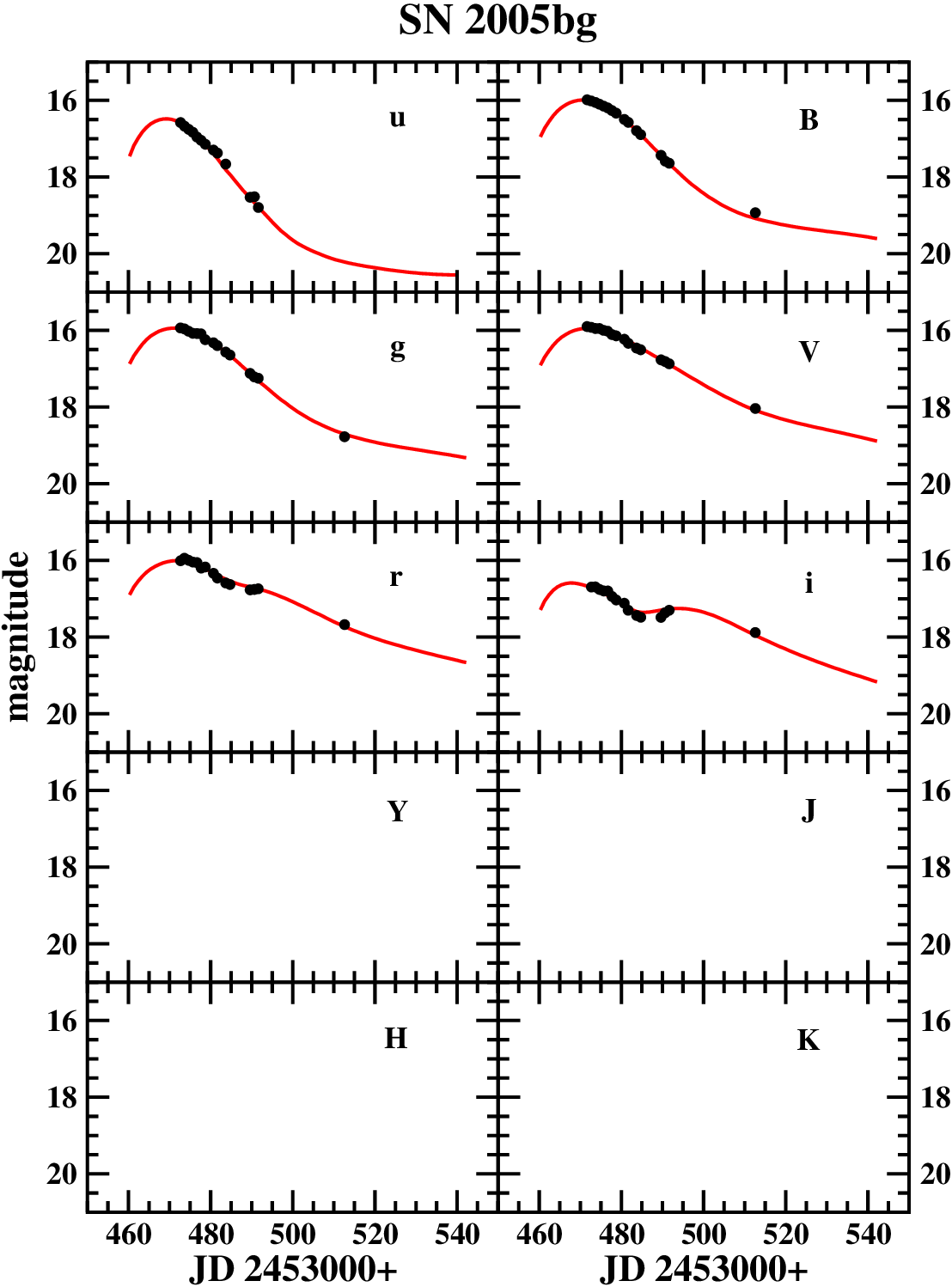}{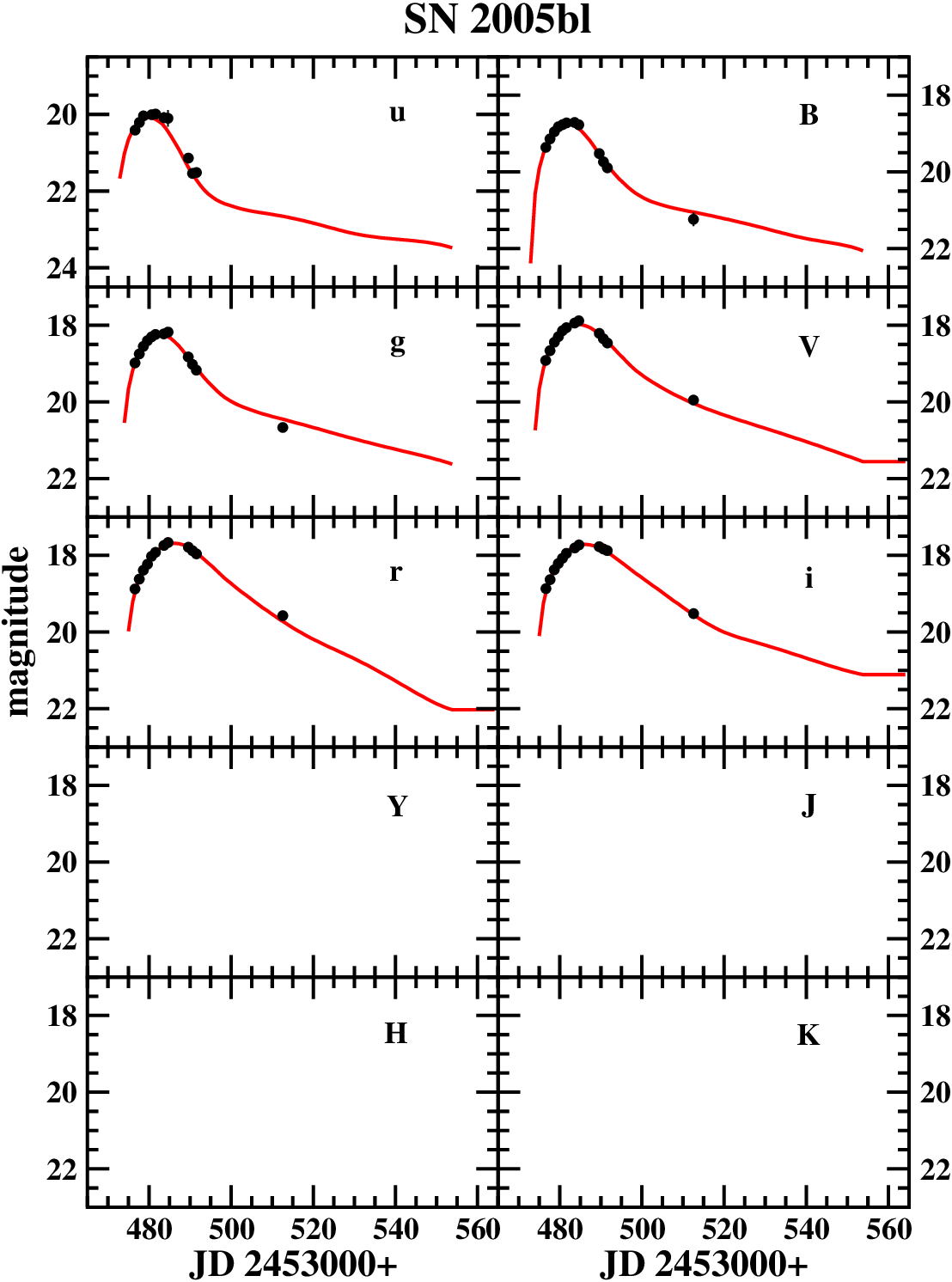}
  {\center Contreras {\it et al.} Fig. \ref{fig:flcurves}}

\end{figure}

\clearpage
\newpage

\begin{figure}[t]

  \plottwo{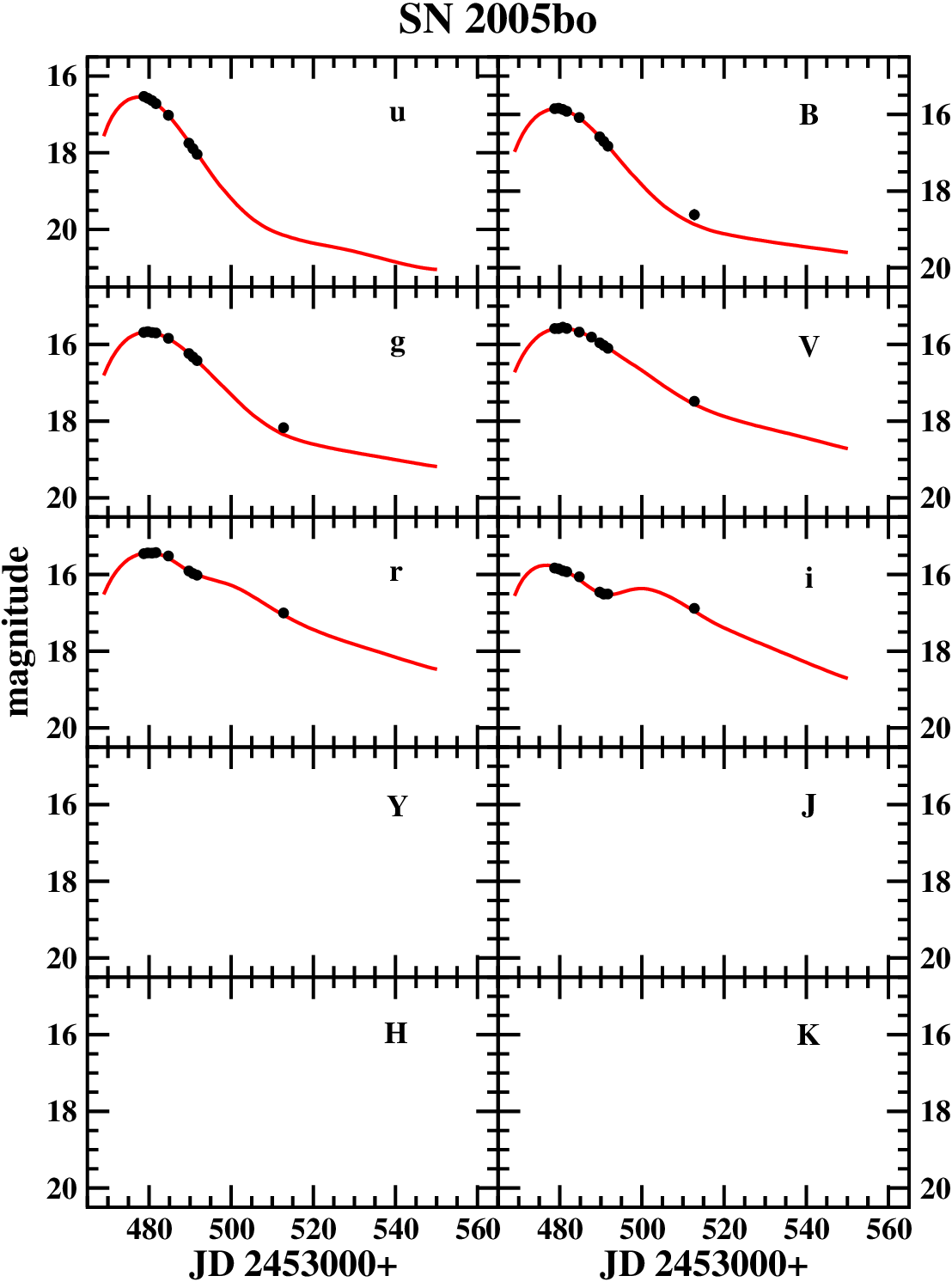}{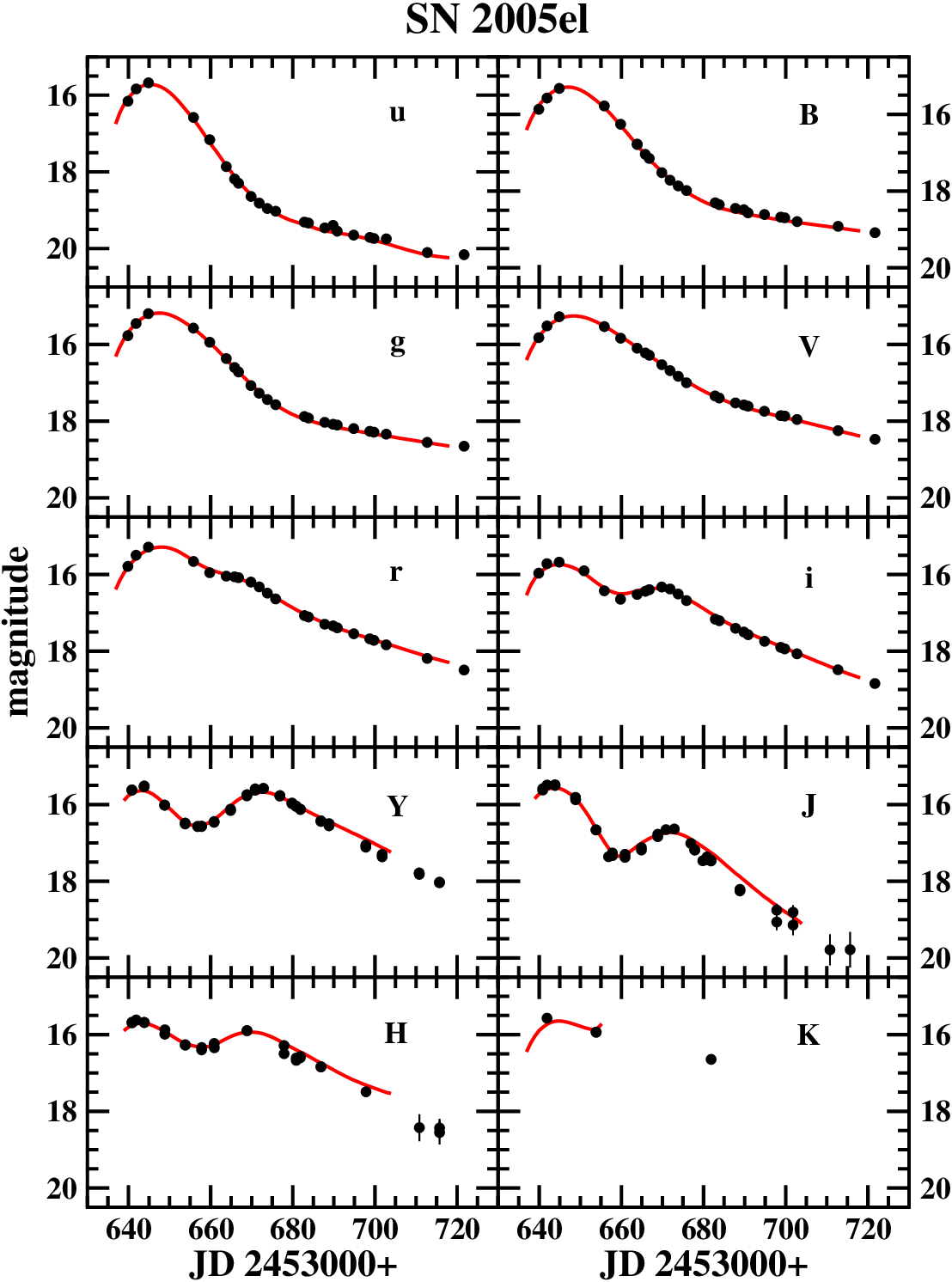}
  \newline
  \newline
  \plottwo{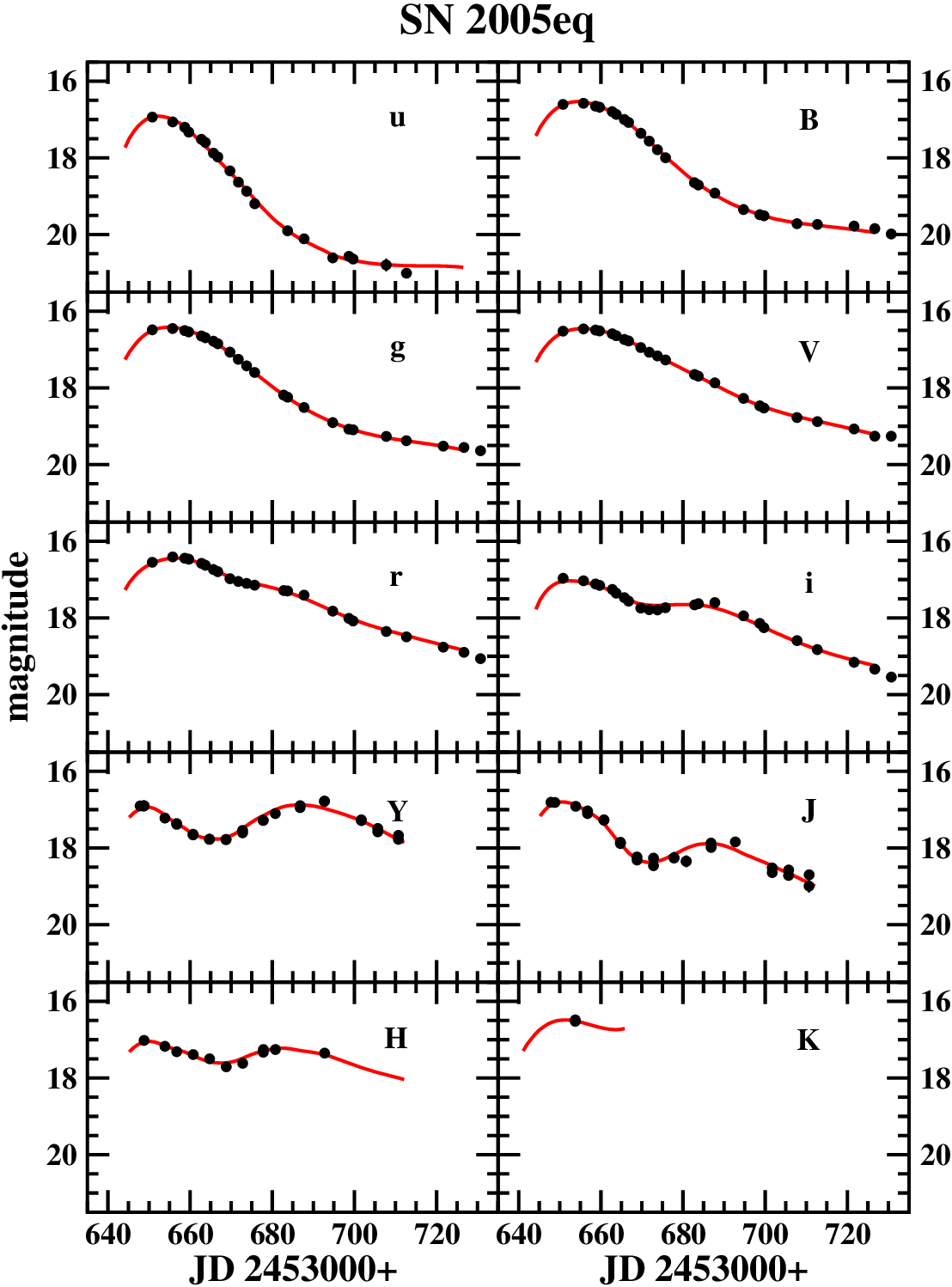}{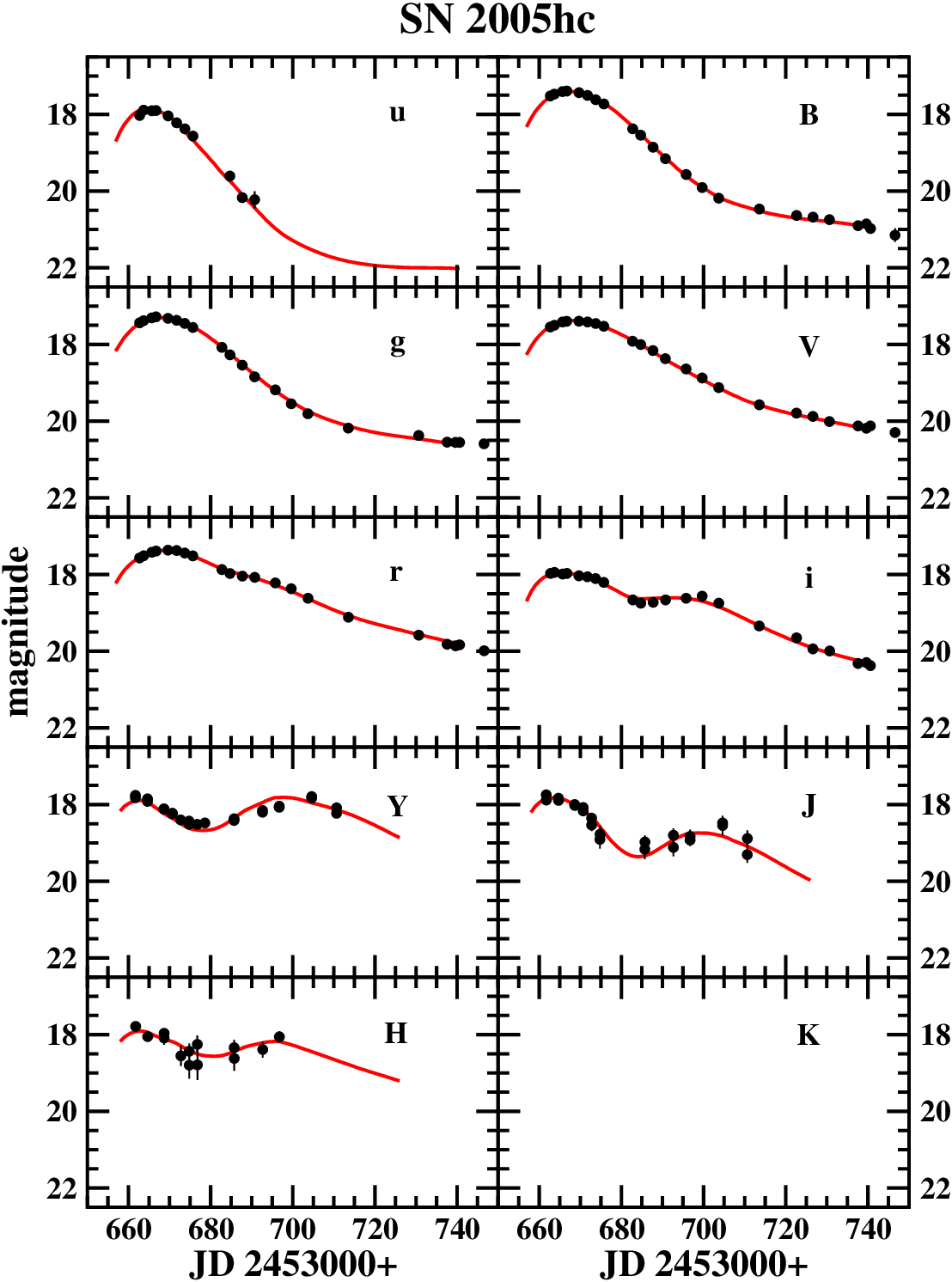}
  {\center Contreras {\it et al.} Fig. \ref{fig:flcurves}}

\end{figure}

\clearpage
\newpage

\begin{figure}[t]

  \plottwo{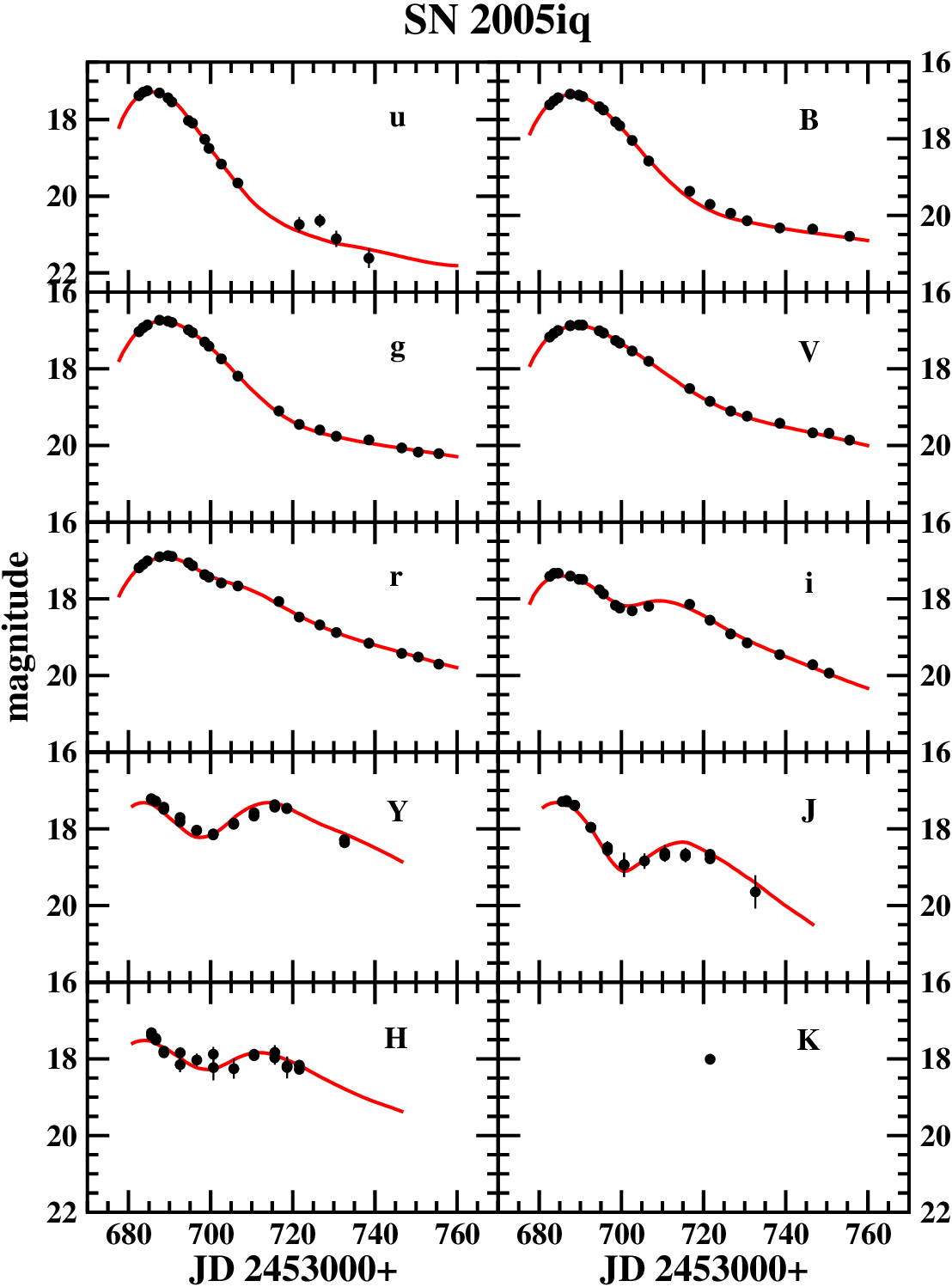}{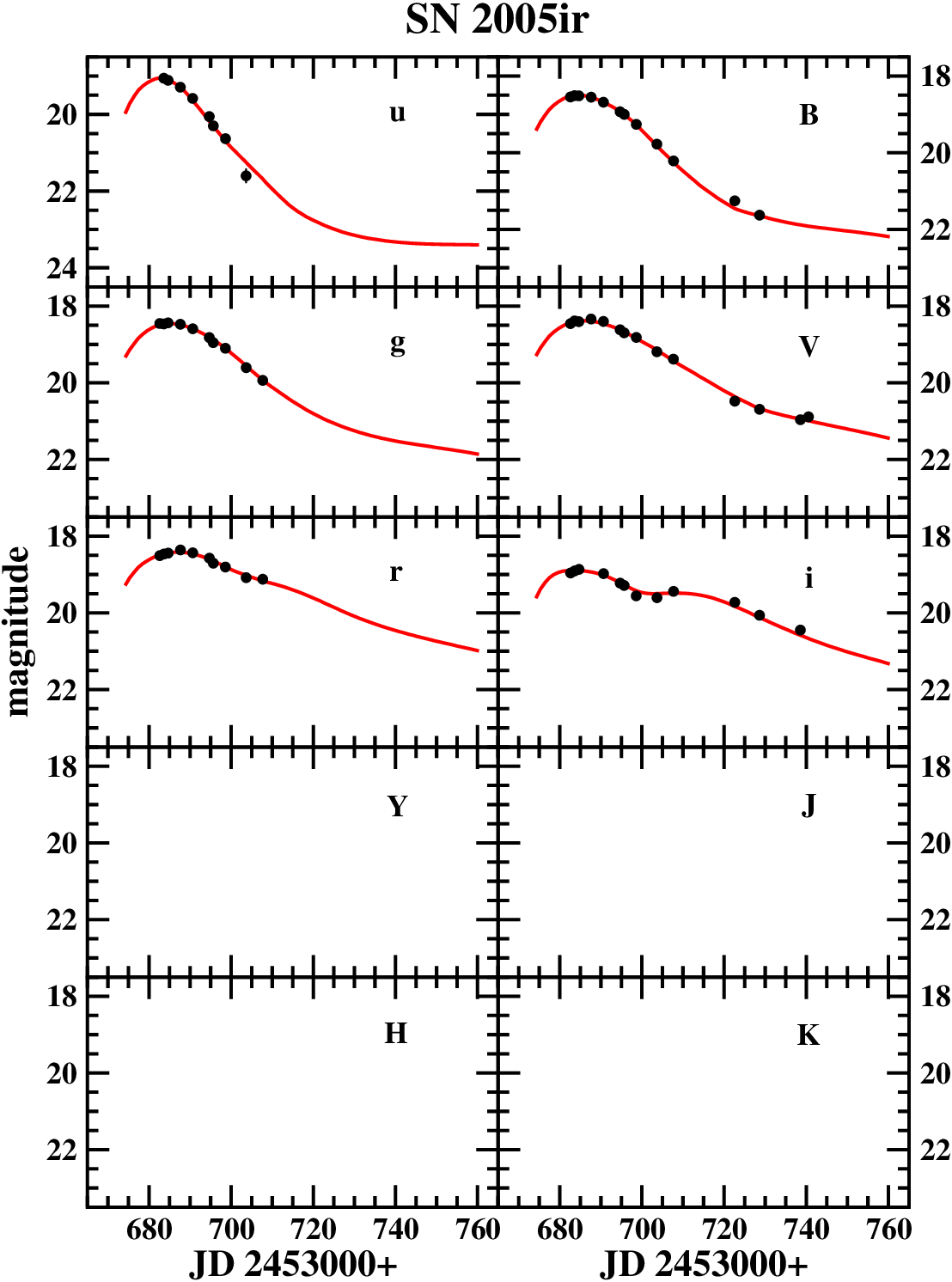}
  \newline
  \newline
  \plottwo{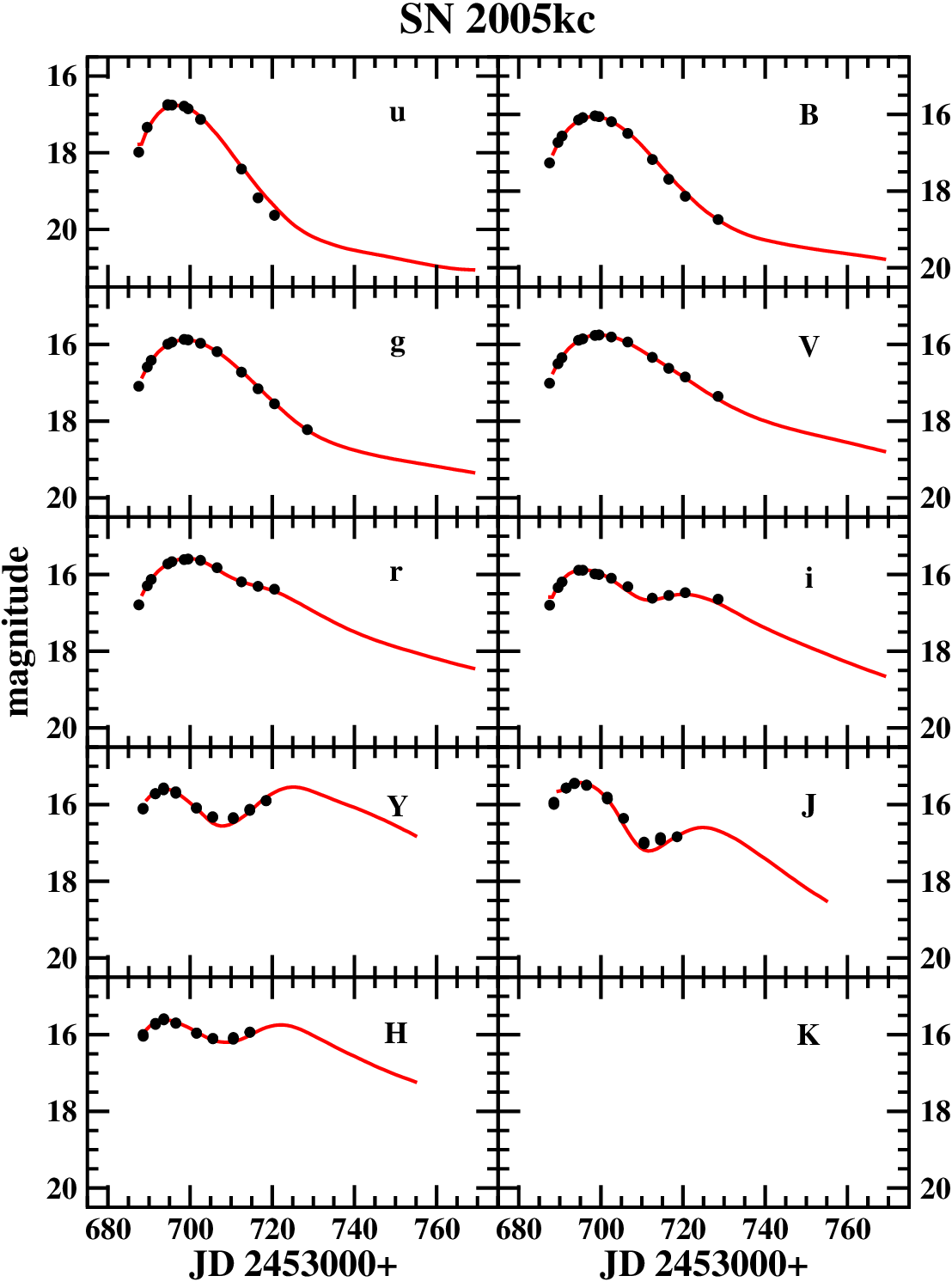}{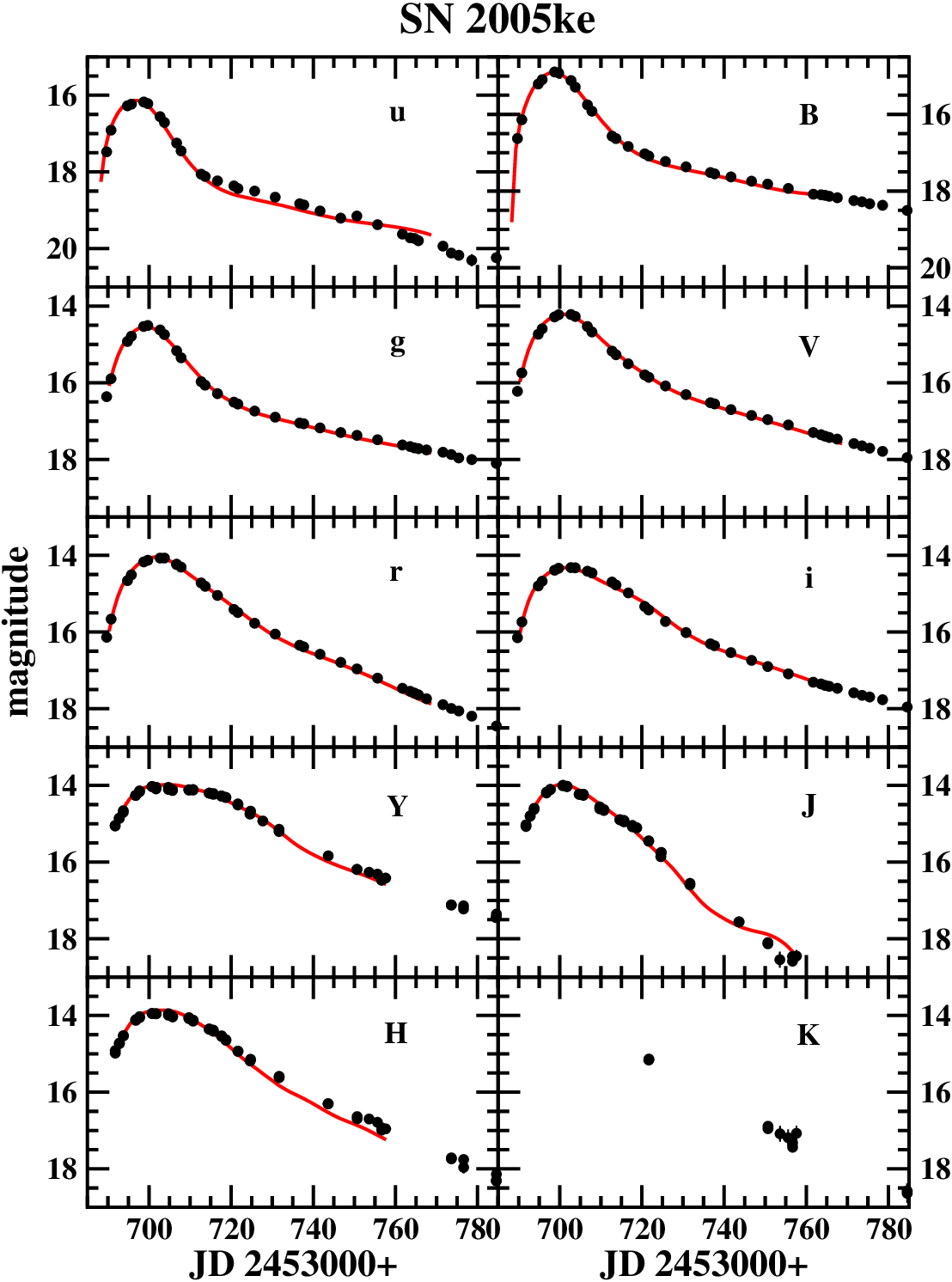}
  {\center Contreras {\it et al.} Fig. \ref{fig:flcurves}}

\end{figure}

\clearpage
\newpage

\begin{figure}[t]

  \plottwo{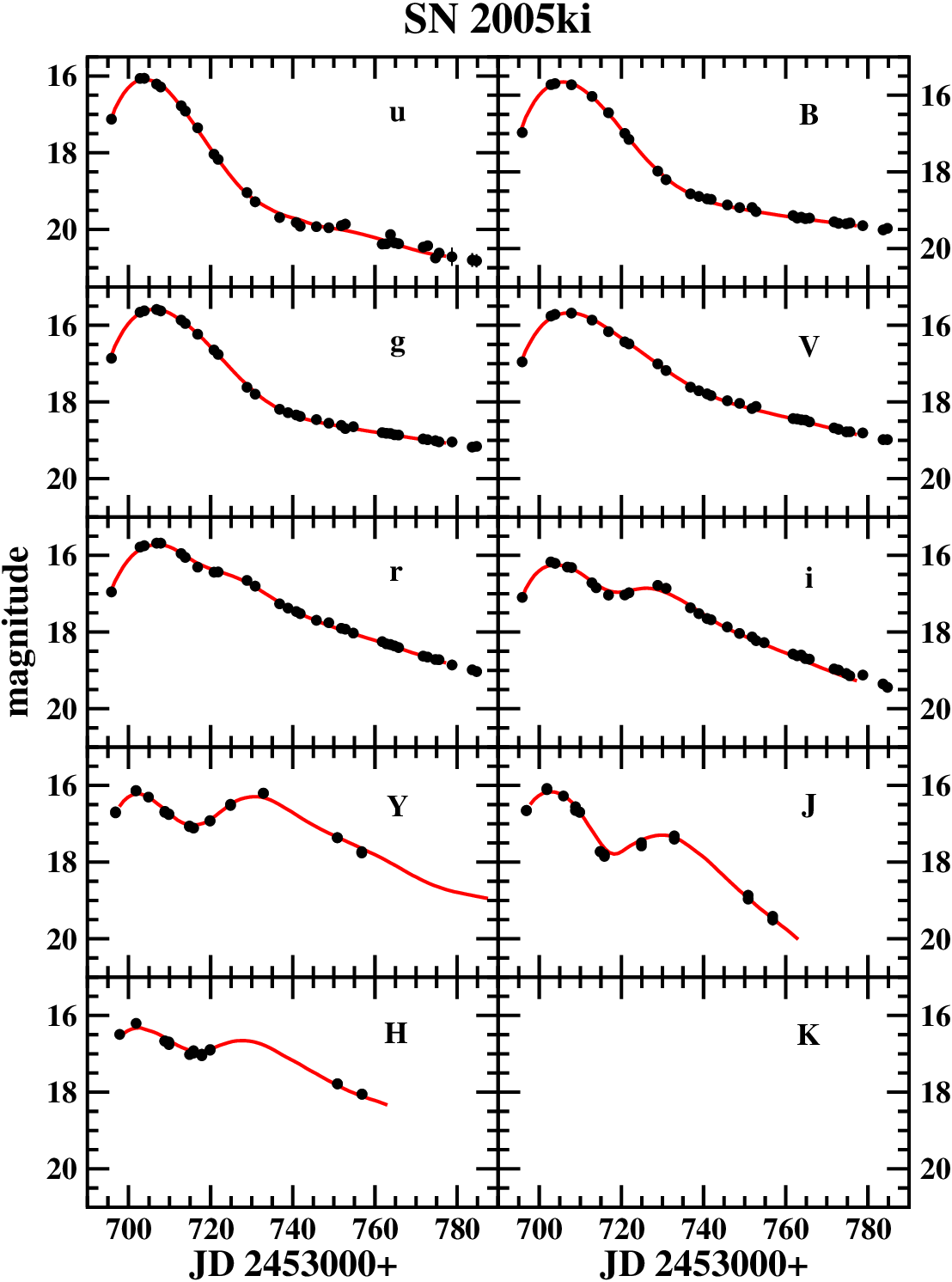}{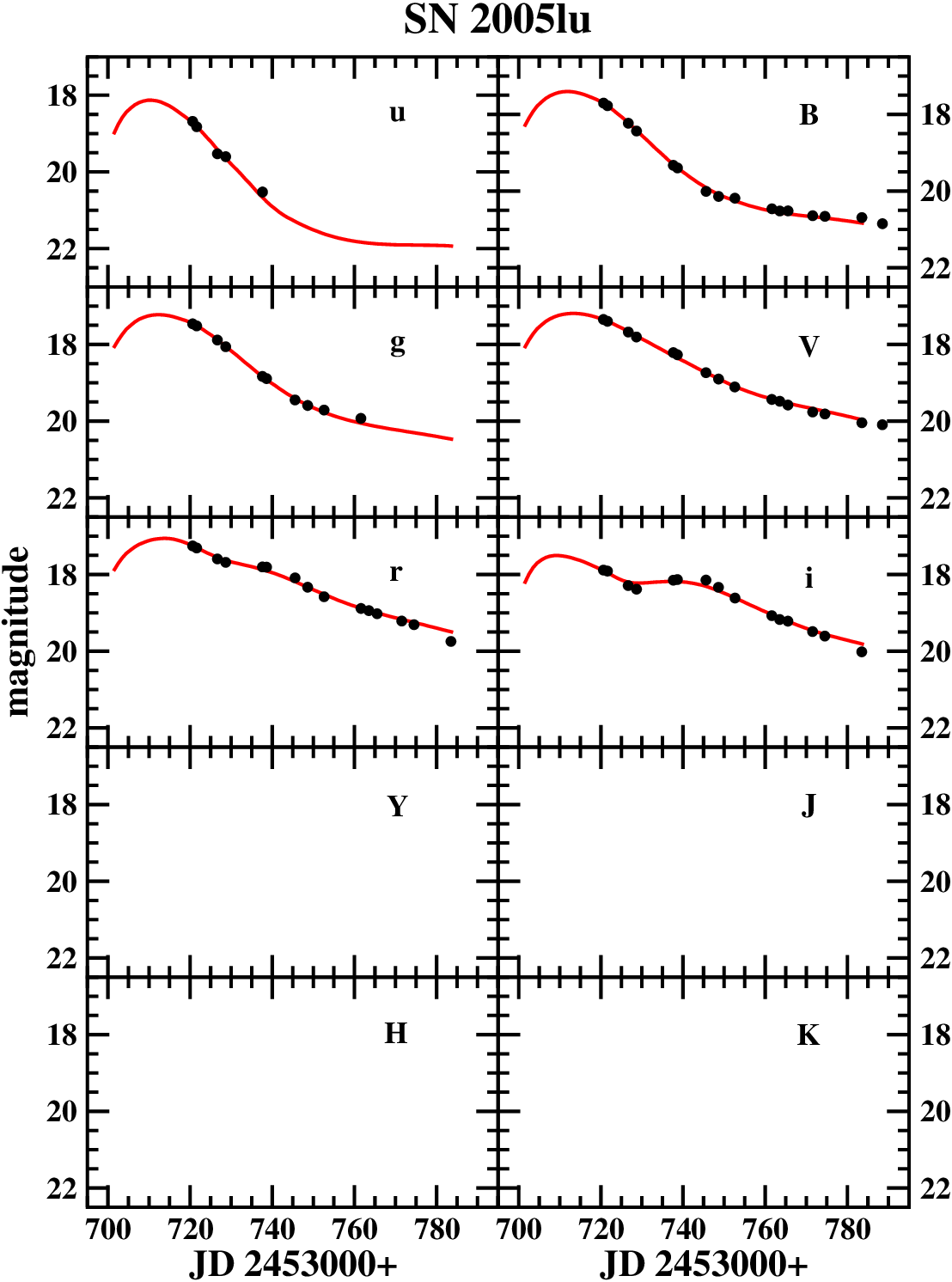}
  \newline
  \newline
  \plottwo{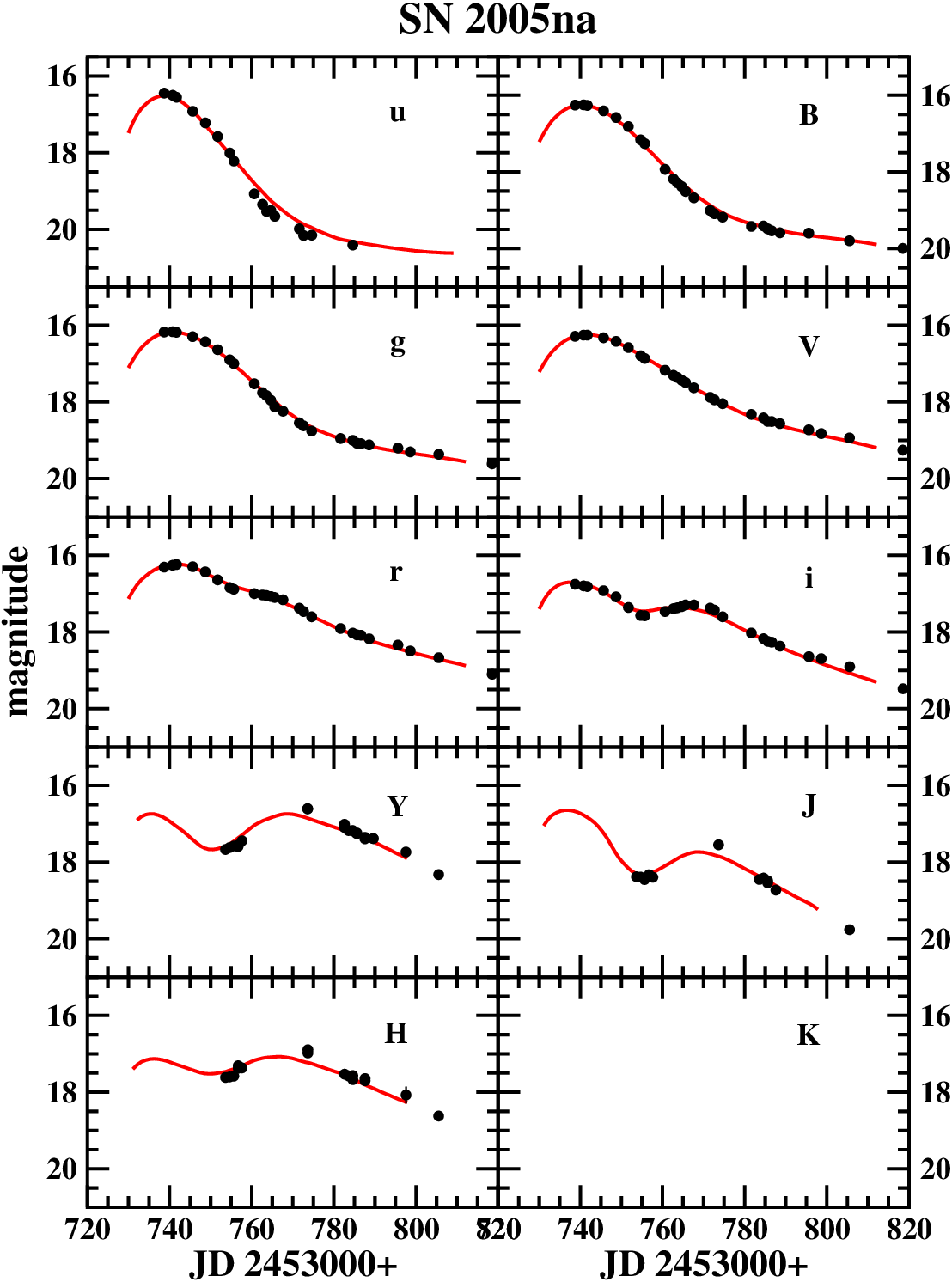}{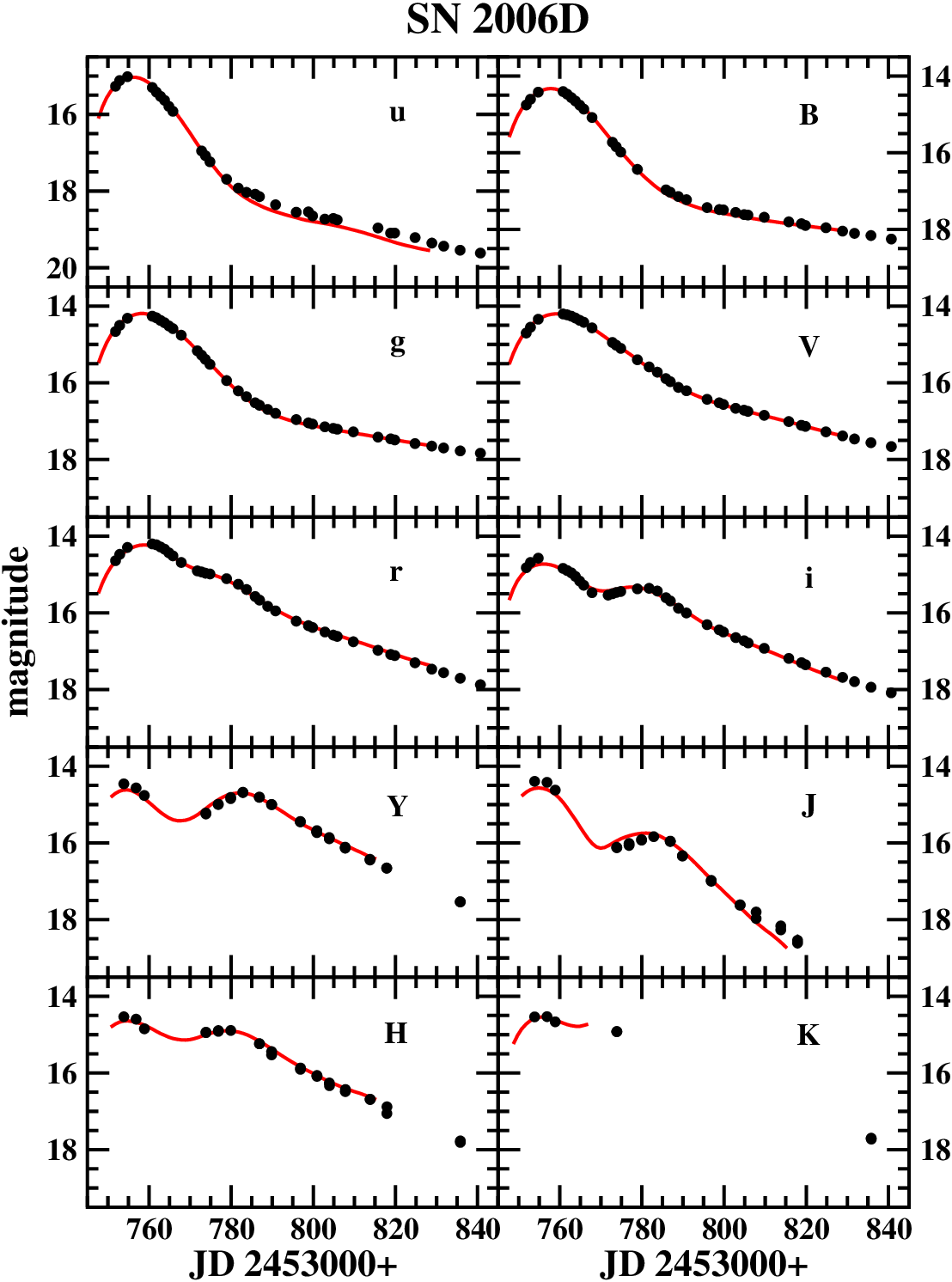}
  {\center Contreras {\it et al.} Fig. \ref{fig:flcurves}}

\end{figure}

\clearpage
\newpage

\begin{figure}[t]

  \plottwo{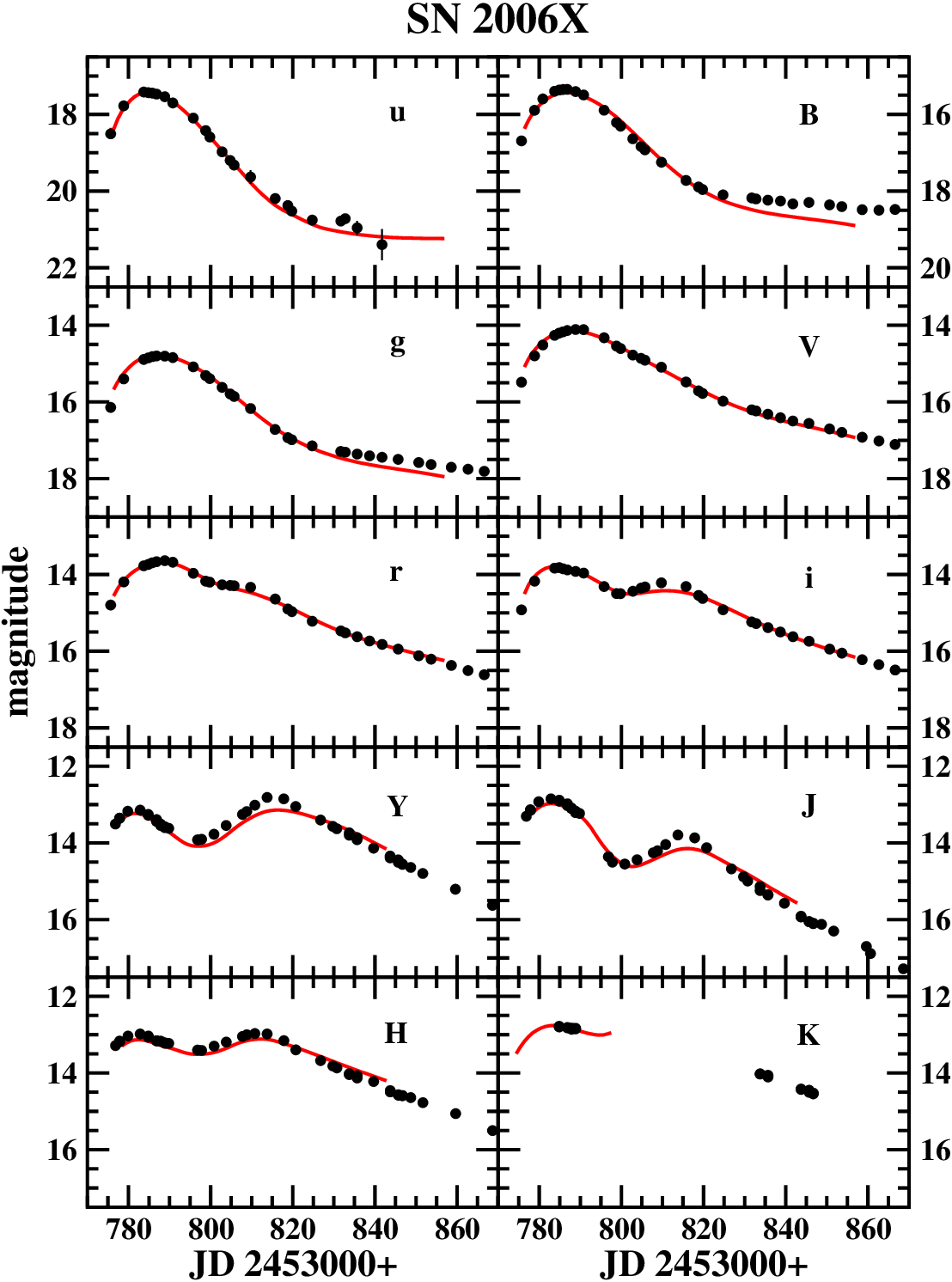}{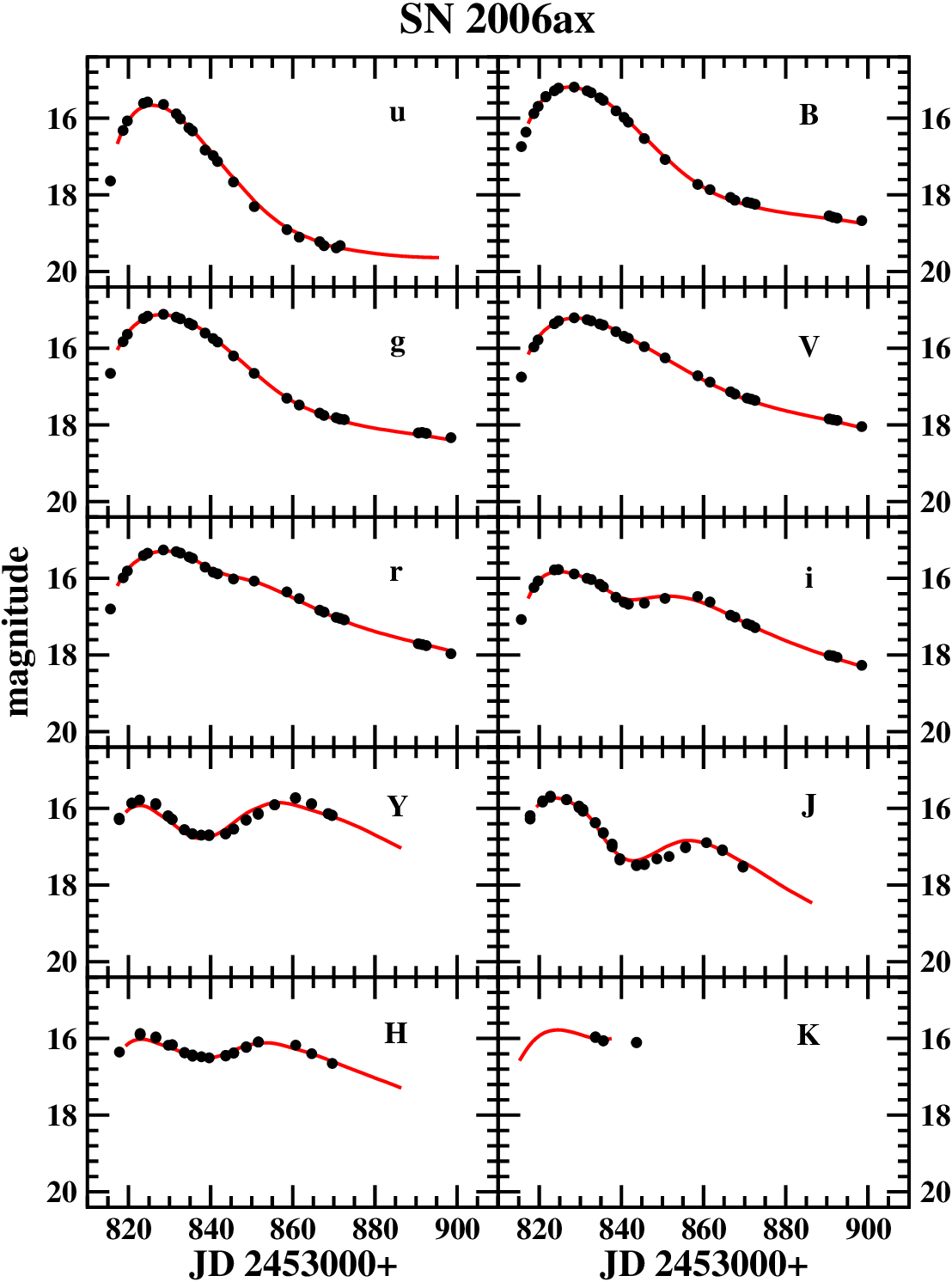}
  \newline
  \newline
  \plottwo{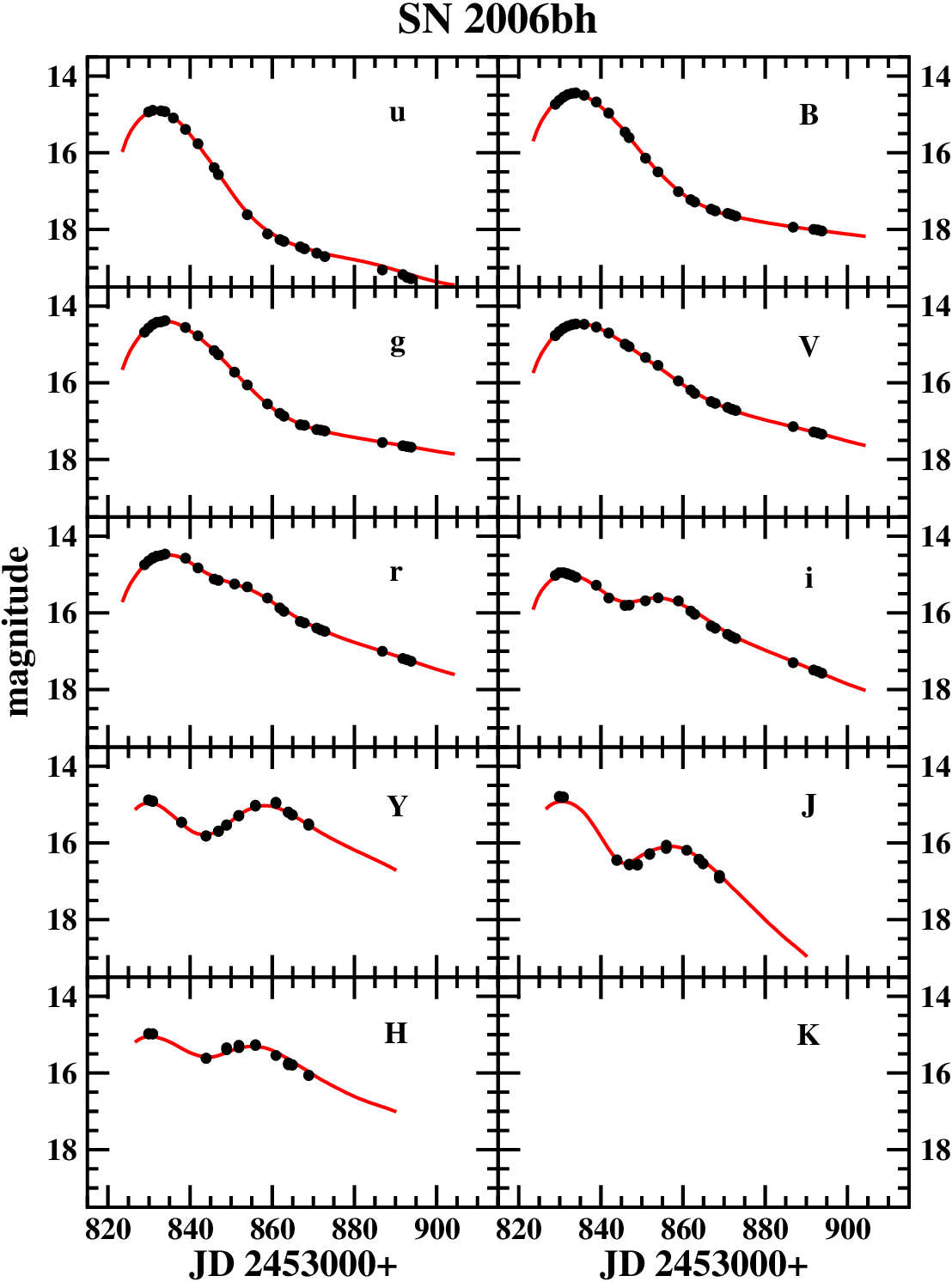}{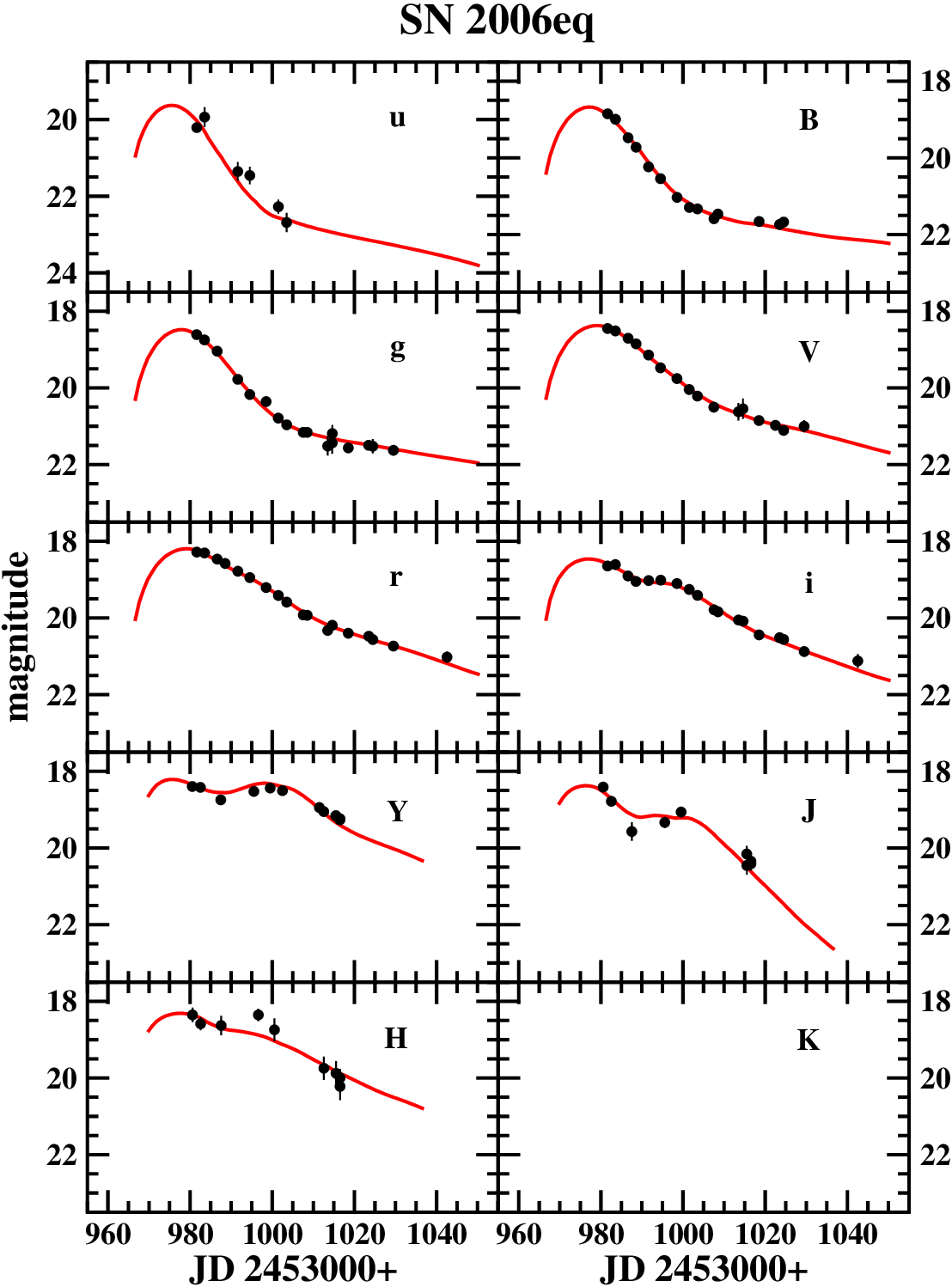}
  {\center Contreras {\it et al.} Fig. \ref{fig:flcurves}}

\end{figure}

\clearpage
\newpage

\begin{figure}[t]

  \plottwo{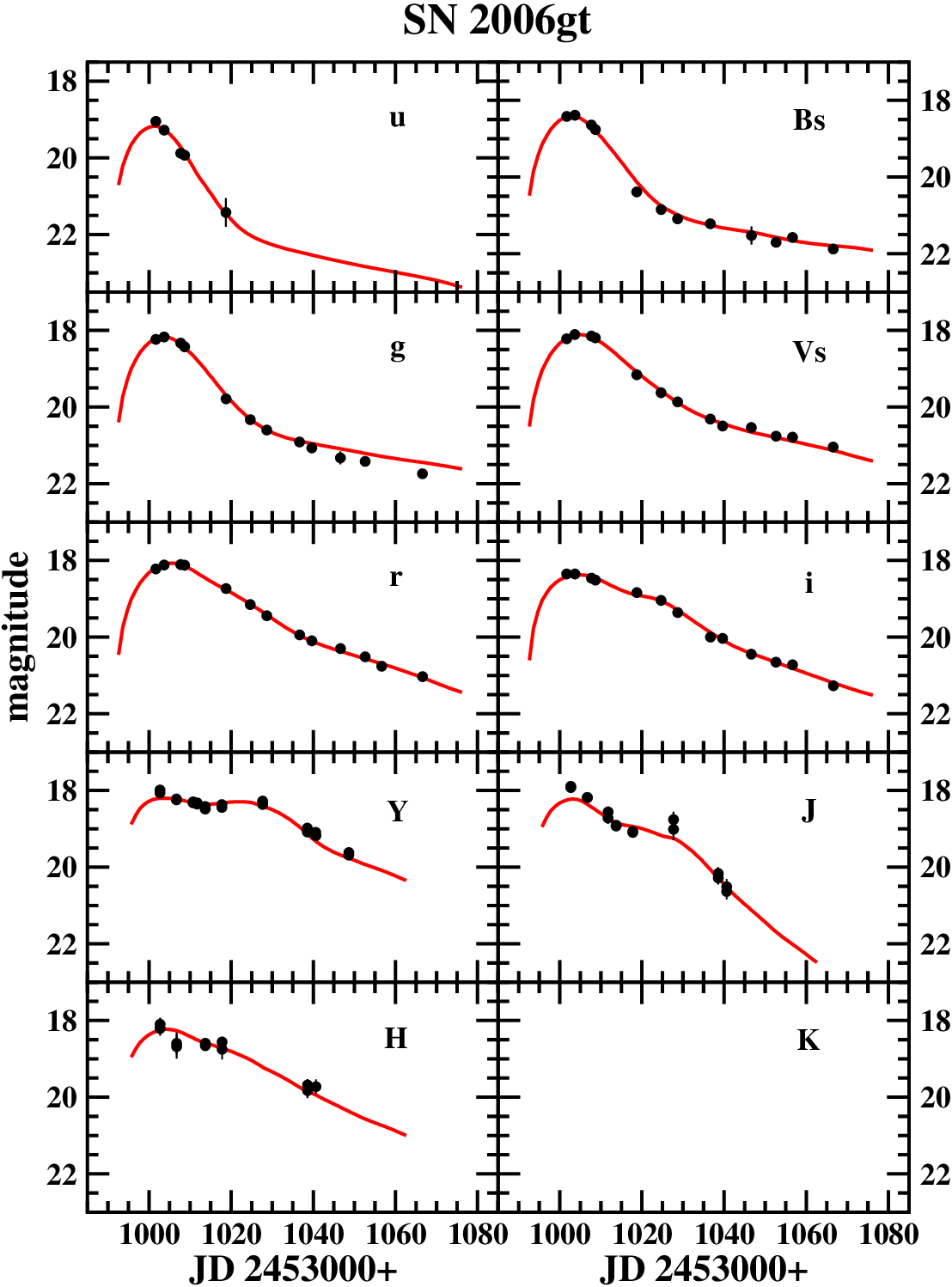}{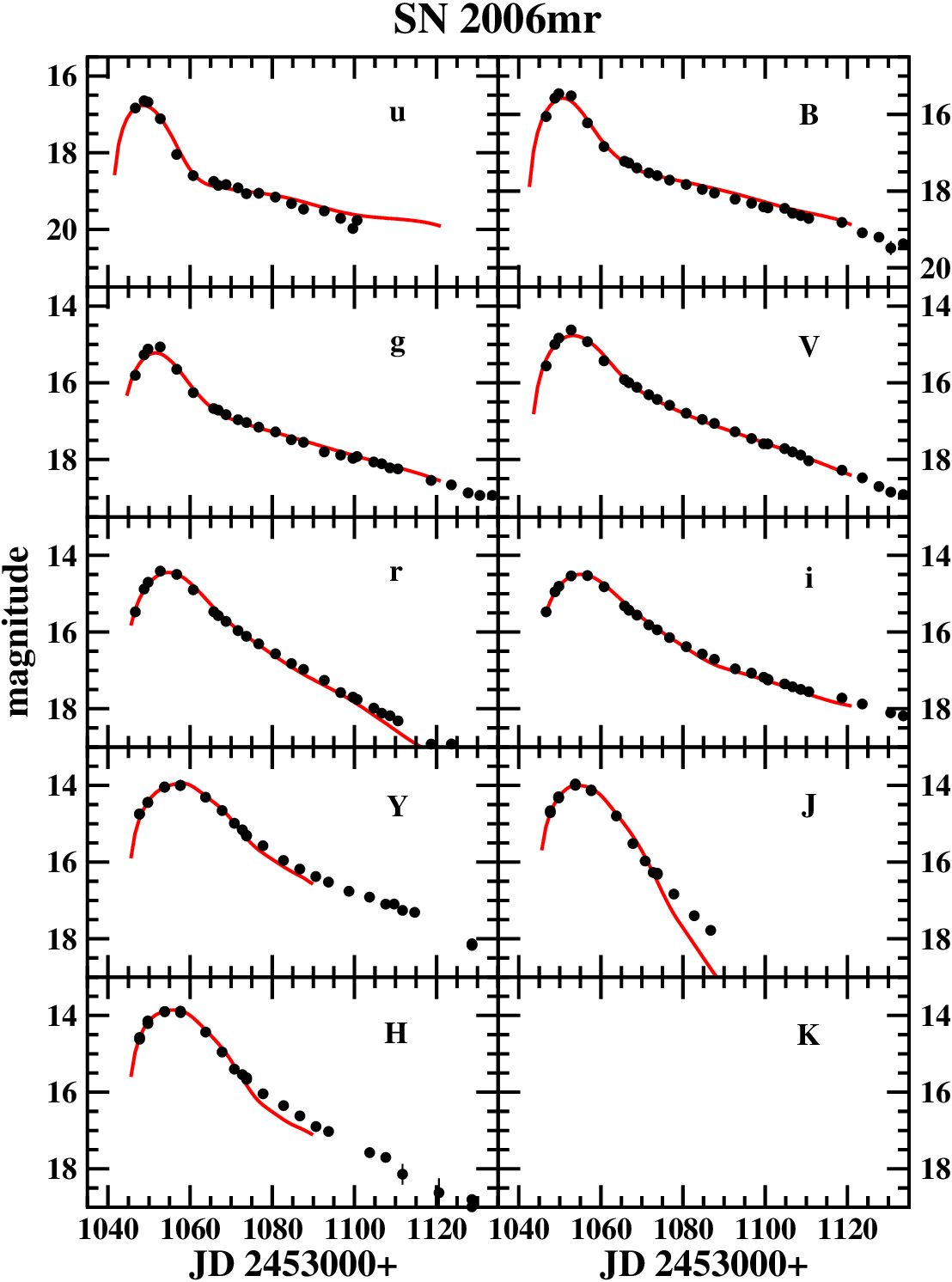}
  \newline
  \newline
  \epsscale{.46} \plotone{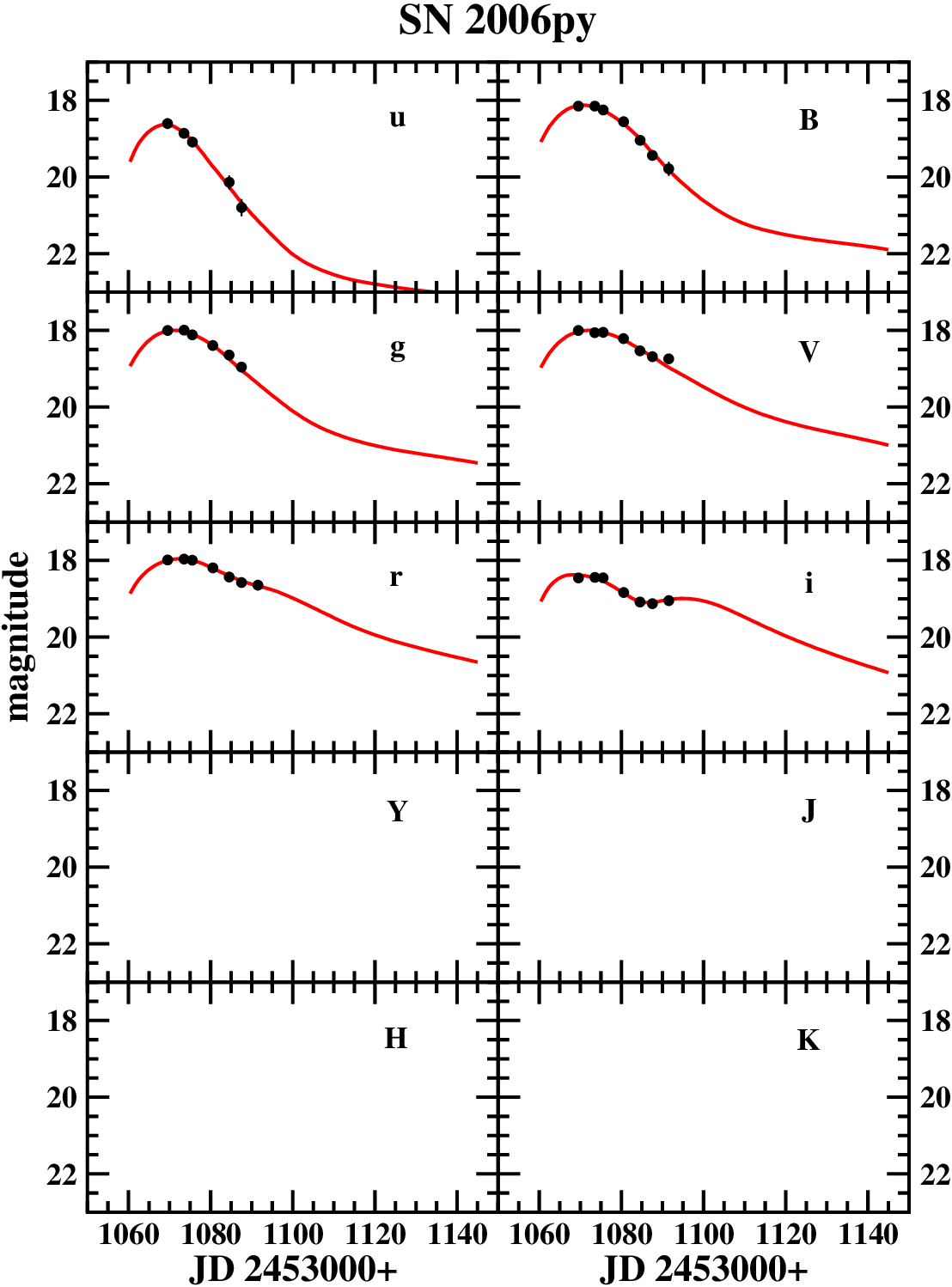}
  \epsscale{1.}
  {\center Contreras {\it et al.} Fig. \ref{fig:flcurves}}

\end{figure}

\clearpage
\newpage

% Synthetic filter color term plot
\begin{figure}
\plotone{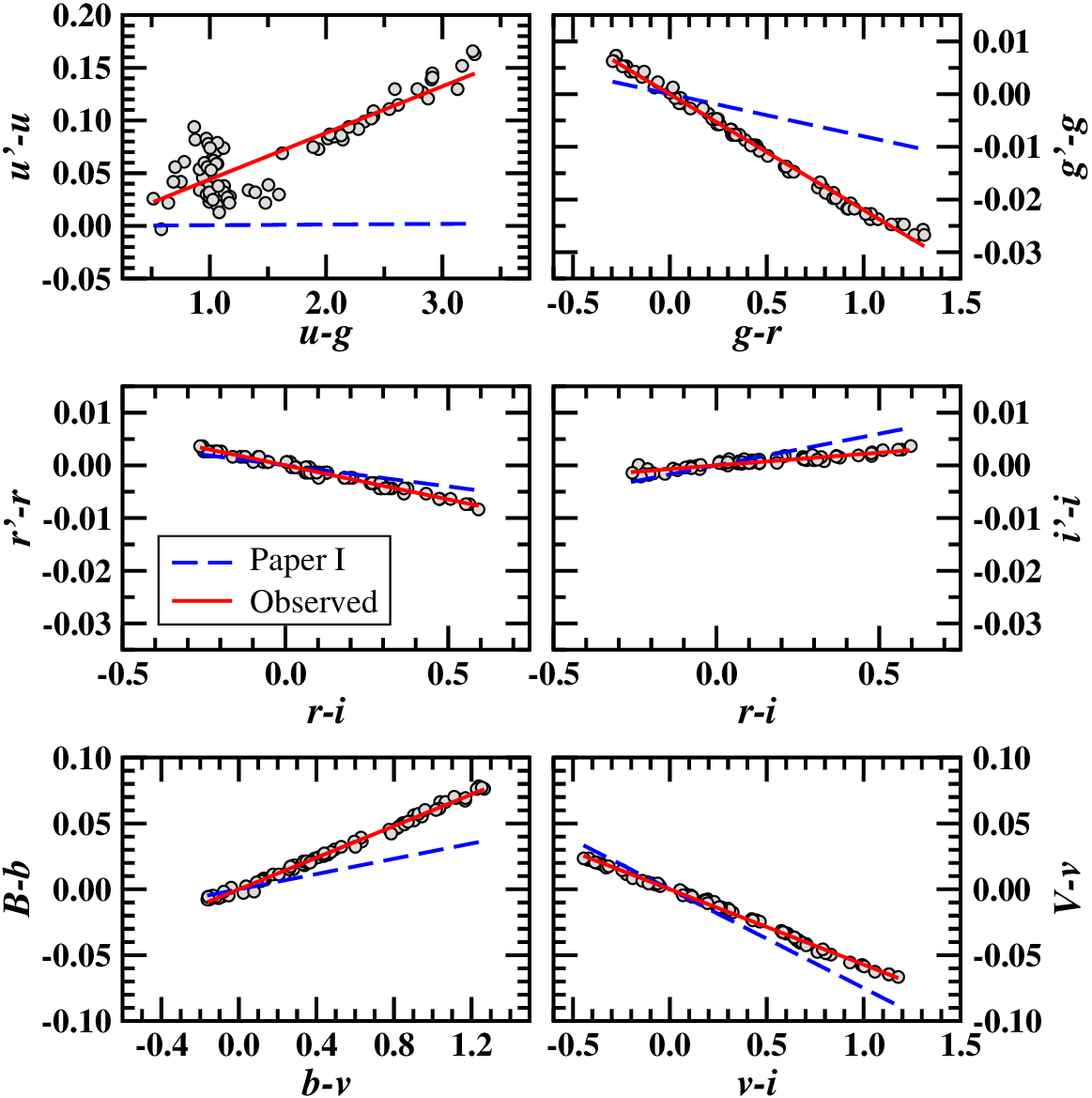}
{\center Contreras {\it et al.} Fig. \ref{fig:fcterms}}
\end{figure}
% Filter plot
\begin{figure}
\plotone{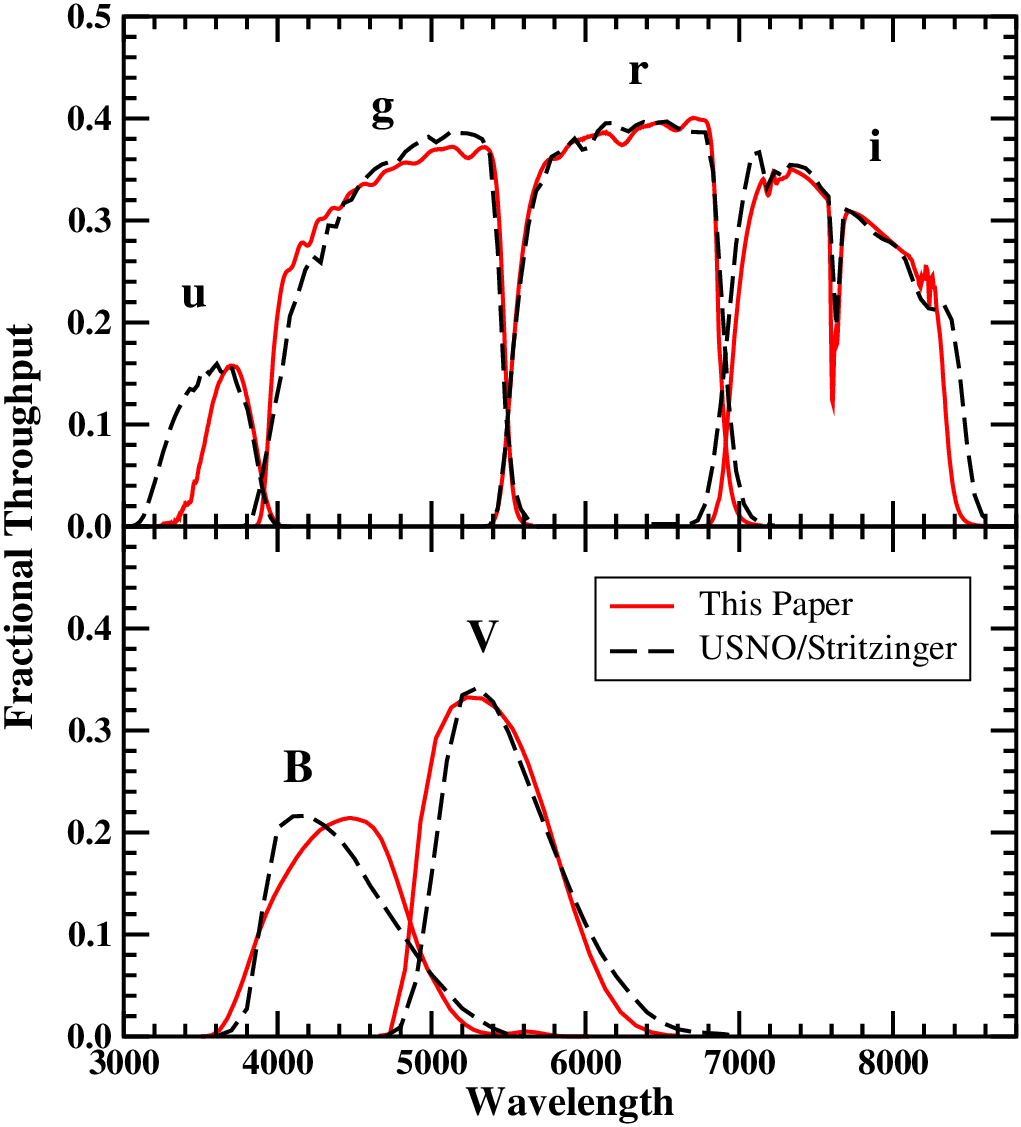}
{\center Contreras {\it et al.} Fig. \ref{fig:filters}}
\end{figure}
\clearpage

% Tables
\begin{deluxetable} {lccllcll}
%\rotate
%\setlength{\tabcolsep}{0.02in} 
\tabletypesize{\tiny}
\tablecolumns{15}
\tablewidth{0pt}
\tablecaption{General Properties of the 35 Type Ia Supernovae \label{tab:sne1}}
\tablehead{%
\colhead{} & 
\colhead{SN}   & 
\colhead{SN}   &
\colhead{Host} & 
\colhead{}   & 
\colhead{}   & 
\colhead{Discovery} &
\colhead{}   \\
\colhead{SN} &
\colhead{$\alpha (2000)$} & 
\colhead{$\delta (2000)$} & 
\colhead{Galaxy} &
\colhead{Morphology\tablenotemark{a}} & 
\colhead{$z_{\rm helio}$\tablenotemark{a}} & 
\colhead{Reference} &
\colhead{Discovery Group/Individual\tablenotemark{b}}
}
\startdata
 2004dt & 02:02:12.77 &-00:05:51.5&NGC 799           &  (R')SB(s)a         &   0.0197 & IAUC 8386 & LOSS              \\ 
 2004ef & 22:42:10.02 &+19:59:40.4&UGC 12158         &  Sb                 &   0.0310 & IAUC 8399 & Boles; Armstrong  \\
 2004eo & 20:32:54.19 &+09:55:42.7&NGC 6928          &  SB(s)ab            &   0.0157 & IAUC 8406 & Itagaki           \\
 2004ey & 21:49:07.81 &+00:26:39.2&UGC 11816         &  SB(rs)c            &   0.0158 & IAUC 8419 & Tenagra~II        \\
 2004gc & 05:21:49.95 &+06:40:33.7&ARP 327 NED04     &  \nodata            &   0.0321 & IAUC 8442 & del Olmo et al.; Tenagra~II \\
 2004gs & 08:38:23.18 &+17:37:39.8&MCG +03-22-020    & S0?                 &   0.0267 & IAUC 8453 & LOSS              \\
 2004gu & 12:46:24.72 &+11:56:56.1&FGC 175A          &  \nodata            &   0.0459 & IAUC 8454 & ROTSE-III         \\
 2005A  & 02:30:43.25 &-02:56:19.8&NGC 958           &  SB(rs)c            &   0.0191 & IAUC 8459 & LOSS              \\
\enddata
\tablenotetext{a}{Host galaxy morphologies and heliocentric redshifts taken 
  from the NASA/IPAC Extragalactic Database (NED).}
\tablenotetext{b}{References/URLs: LOSS \citep[see][and references therein]{filippenko05}; 
  Tenagra~II (http://www.tenagraobservatories.com/Discoveries.htm); ROTSE-III \citep{quimby06}); 
  Puckett (http://www.cometwatch.com); BRASS (http://brass.astrodatabase.net); Perth \citep{williams97}; 
  SDSS~II \citep{frieman08}; CROSS (http://wwww.cortinasetelle.it/snindex.htm)}

\end{deluxetable}

\clearpage
\begin{deluxetable} {lccccccl}
%\rotate
%\setlength{\tabcolsep}{0.02in} 
\tabletypesize{\tiny}
\tablecolumns{15}
\tablewidth{0pt}
\tablecaption{Photometric/Spectroscopic Properties of the 35 Type Ia Supernovae \label{tab:sne2}}
\tablehead{
\colhead{} & 
\colhead{$\Delta m_{15}(B)$\tablenotemark{a}} & 
\colhead{Opt.} &  
\colhead{NIR.} &
\colhead{Opt. Std.} & 
\colhead{NIR. Std.} &
\colhead{Spectral}&
\colhead{$t(B_{\rm max})$}\\
\colhead{SN} &
\colhead{(mag)} &
\colhead{Epochs} &
\colhead{Epochs} &
\colhead{Epochs} &
\colhead{Epochs} &
\colhead{Subtype\tablenotemark{b}}&
\colhead{Days\tablenotemark{c}}}
\startdata
 2004dt & $\cdots$  & 48 & $\cdots$  & 5  & 1       & normal     & $+$23.6 \\ 
 2004ef & 1.39(01)  & 49 & 4         & 23 & 3       & normal     & $-$1.8  \\
 2004eo & 1.36(03)  & 42 & 9         & 14 & 3       & normal     & $-$10.3\\
 2004ey & 0.97(01)  & 33 & 10        & 9  & 4       & normal     & $-$0.8\\
 2004gc & 1.09(06)  & 29 & $\cdots$  & 7  & $\cdots$& normal     & $+$11.1\\
 2004gs & 1.54(02)  & 56 & 12        & 6  & 5       & normal     & $-$3.4\\
 2004gu & 0.80(04)  & 31 & 8         & 4  & 3       & 06gz-like & $-$3.1\\
 2005A  & 1.16(06)  & 37 & 11        & 6  & 7       & normal     & $+$25.4\\
\enddata
\tablenotetext{a}{Uncertainties are given in parentheses in hundreds of a magnitude.}
\tablenotetext{b}{Most likely spectral subtype as determined by the
  Supernova Identification (SNID) code \citep{blondin07}.}
\tablenotetext{c}{Epoch of spectrum (with respect to time of $B$-band maximum) used to classify spectral type with SNID.}

\end{deluxetable}

\clearpage
\begin{deluxetable} {lllcccccccccc}
\rotate
\tabletypesize{\tiny}
\tablecolumns{13}
\tablewidth{0pt}
\tablecaption{Optical and near-infrared photometry of the local sequences in the standard system\label{tab:optir_stds}}
\tablehead{
\colhead{} &
\colhead{}  &
\colhead{}  &
\colhead{$u'$}   &
\colhead{$g'$}   &
\colhead{$r'$}   &
\colhead{$i'$}   &
\colhead{$B$}   &
\colhead{$V$}  &
\colhead{$Y$}   &
\colhead{$J$}   &
\colhead{$H$}  &
\colhead{$K_s$}\\
\colhead{STAR} &
\colhead{$\alpha(2000)$}  &
\colhead{$\delta(2000)$}  &
\colhead{(mag)}   &
\colhead{(mag)}   &
\colhead{(mag)}   &
\colhead{(mag)}   &
\colhead{(mag)}   &
\colhead{(mag)}  &
\colhead{(mag)}   &
\colhead{(mag)}   &
\colhead{(mag)}  &
\colhead{(mag)}}
\startdata
%#########################  SN04dt  #########################
\multicolumn{13}{c}{\bf 2004dt}\\
01 & 02:02:09.97 & -00:08:42.02 & 17.480(013)  & 15.486(009)  & 14.656(007)  & 14.259(007)  & 15.989(008)  & 15.015(013) &     \nodata      &     \nodata      &     \nodata      &     \nodata      \\
02 & 02:02:10.06 & -00:06:49.80 & 18.876(009)  & 16.778(003)  & 16.022(003)  & 15.771(003)  & 17.285(004)  & 16.343(005) & 14.998(020)  & 14.696(020)  &              & 14.101(020) \\
03 & 02:01:57.62 & -00:02:08.23 & 16.346(008)  & 15.071(007)  & 14.633(007)  & 14.474(008)  & 15.419(004)  & 14.786(004) &     \nodata      &     \nodata      &     \nodata      &     \nodata      \\
04 & 02:02:04.30 & -00:02:30.68 & 18.613(007)  & 16.125(004)  & 15.070(004)  & 14.617(004)  & 16.703(004)  & 15.549(007) &     \nodata      &     \nodata      &     \nodata      &     \nodata      \\
05 & 02:02:05.35 & -00:02:05.48 & 16.377(009)  & 15.384(004)  & 15.185(003)  & 15.136(003)  & 15.619(004)  & 15.243(005) &     \nodata      &     \nodata      &     \nodata      &     \nodata      \\
06 & 02:02:08.16 & -00:02:01.70 & 19.695(015)  & 16.822(003)  & 15.486(006)  & 14.274(010)  & 17.623(004)  & 16.082(015) &     \nodata      &     \nodata      &     \nodata      &     \nodata      \\
07 & 02:02:10.50 & -00:00:49.75 & 16.850(016)  & 15.205(011)  &              &              & 15.655(011)  &             &     \nodata      &     \nodata      &     \nodata      &     \nodata      \\
\enddata
\tablecomments{Uncertainties given in parentheses in thousandths of a
  magnitude correspond to an rms of the magnitudes obtained on at least 3
  photometric nights, with a minimum uncertainty of 0.015 mag for an
  individual measurement.}
%\tablenotetext{a}{}
\end{deluxetable}

\clearpage
\begin{deluxetable} {lcccccccccccc}
%\rotate
\setlength{\tabcolsep}{0.02in}
\tabletypesize{\scriptsize}
\tablecolumns{7}
\tablewidth{0pt}
\tablecaption{Optical Photometry of 35 SNe~Ia in the Swope Natural System\label{tab:opt_sne}}
\tablehead{
\colhead{JD} &
\colhead{$u$ }  &
\colhead{$g$ }   &
\colhead{$r$ }   &
\colhead{$i$ }   &
\colhead{$B$ }   &
\colhead{$V$ }  \\
\colhead{2453000+} &
\colhead{(mag)}  &
\colhead{(mag)}  &
\colhead{(mag)}  &
\colhead{(mag)}  &
\colhead{(mag)} &
\colhead{(mag)}   }
\startdata

\multicolumn{7}{c}{{\bf 2004dt}}\\ 
 249.77& 16.630(013)& 15.593(004)& 15.660(004)& 16.069(006)& 15.769(006)& 15.576(006) \\
 250.79& 16.782(012)& 15.661(008)& 15.701(007)& 16.110(009)& 15.848(008)& 15.629(009) \\
 251.81& 16.886(011)& 15.768(008)& 15.753(009)& 16.110(009)& 15.949(009)& 15.672(009) \\
 252.85& 17.030(011)& 15.835(005)& 15.786(004)& 16.049(007)& 16.047(006)& 15.719(005) \\
 253.77& 17.150(011)& 15.910(004)& 15.778(004)&  \nodata   & 16.151(013)& 15.774(005) \\
 256.72& 17.627(009)& 16.237(005)& 15.803(005)& 16.021(007)& 16.568(006)& 15.960(010) \\
 257.71& 17.767(013)& 16.355(008)& 15.837(010)& 16.032(011)& 16.705(008)& 16.016(009) \\
\enddata
\tablecomments{Uncertainties given in parentheses in thousandths of a magnitude.}
%\tablenotetext{a}{}
\end{deluxetable}

\clearpage
\begin{deluxetable} {lcccccccccccc}
%\rotate
%\setlength{\tabcolsep}{0.02in}
\tabletypesize{\scriptsize}
\tablecolumns{5}
\tablewidth{0pt}
\tablecaption{Near-infrared Photometry of 25 SNe~Ia \label{tab:ir_sne}}
\tablehead{
\colhead{JD}  &
\colhead{$Y$}   &
\colhead{$J$}   &
\colhead{$H$}   &
\colhead{$K_s$}   &
\colhead{}   \\
\colhead{2453000+}  &
\colhead{(mag)}   &
\colhead{(mag)}   &
\colhead{(mag)}   &
\colhead{(mag)}   &
\colhead{Telescope\tablenotemark{a}}   }
\startdata

\multicolumn{6}{c}{{\bf 2004ef}}\\ 
 269.68& 17.606(031)& 17.958(048)& 18.013(094)& 17.515(114)&DUP \\
 277.62& 17.769(014)& 18.710(049)& 17.793(040)& 17.391(088)&DUP \\
 281.58& 17.518(011)& 18.425(023)& 17.709(023)& 17.294(040)&DUP \\
 328.56& 19.095(038)&  \nodata   & 19.553(177)&  \nodata   &DUP \\
\multicolumn{6}{c}{{\bf 2004eo}}\\ 
 269.58& 16.459(018)& 16.058(007)& 16.128(044)& 16.117(052)&DUP \\
 270.55& 16.210(017)& 15.912(028)& 16.041(021)& 15.899(012)&DUP \\
\enddata
\tablecomments{Uncertainties given in parentheses in thousandths of a
  magnitude.}
\tablenotetext{a}{DUP, SWO and BAA correspond to the telescopes du Pont,
 Swope, and Baade, respectively.}
\end{deluxetable}

\clearpage
\begin{deluxetable} {lcccl}
\tabletypesize{\scriptsize}
\tablecolumns{5}
\tablewidth{0pt}
\tablecaption{$YJH$ Measurements of Feige 16\label{tab:feige16}}
\tablehead{
\colhead{} & 
\colhead{$Y$} & 
\colhead{$J$} &
\colhead{$H$} & 
\colhead{}\\
\colhead{System} & 
\colhead{(mag)} & 
\colhead{(mag)} &
\colhead{(mag)} & 
\colhead{Comment}}
\startdata
LCO & 12.347(016) & 12.356(016) & 12.334(019) & measured \\
MKO & \nodata & 12.382(018) & 12.359(011) & transformed to LCO \\
\enddata
\tablecomments{Uncertainties given in parentheses in thousandths of a
  magnitude.}
\end{deluxetable}

\clearpage
\begin{deluxetable} {cc}
\tabletypesize{\scriptsize}
\tablecolumns{2}
\tablewidth{0pt}
\tablecaption{Photometric Zero Points of CSP Bandpasses\label{tab:zpts}}
\tablehead{
\colhead{Filter} & 
\colhead{Zero Point}}
\startdata
$u$ & 12.400 \\
$g$ & 14.432 \\
$r$ & 14.136 \\
$i$ & 13.687 \\
$B$ & 13.669 \\
$V$ & 13.774 \\
$Y$ & 12.687 \\
$J$ & 12.853 \\
$H$ & 12.555 \\
$K_s$ & 11.967 \\
\enddata
\end{deluxetable}


\begin{thebibliography}{}
\bibitem[Aldering et al.(2002)]{aldering02}
Aldering, G., et al. 2002, SPIE, 4836, 61

\bibitem[Allen et al.(2007)]{allen07} Allen,~S. W., et al. 2007, \mnras,
  383, 879

\bibitem[Astier et al.(2006)]{astier06} Astier,~P., et al. 2006, \aap, 447, 31  

\bibitem[Bessell(1990)]{bessell90} Bessell,~M.~S. 1990, \pasp, 102, 1181

\bibitem[Bessell, Castelli, \& Plez(1998)]{bessell98} 
Bessell,~M.~S., Castelli, F., \& Plez, B. 1998, \pasp, 102, 1181

\bibitem[Blondin \& Tonry(2007)]{blondin07} 
Blondin,~S., \& Tonry,~J.~L. 2007, \apj, 666, 1024

\bibitem[Bohlin \& Gilliland(2004a)]{bohlin04} Bohlin,~R.~C., \& 
Gilliland,~R.~L. 2004, \aj, 127, 3508

\bibitem[Bohlin \& Gilliland(2004b)]{bohlin04b} Bohlin,~R.~C., \& 
Gilliland,~R.~L. 2004, \aj, 128, 3053

\bibitem[Burns et al.(2009)]{burns09} Burns,~C., et al. 2009, in preparation

\bibitem[Cleveland(1979)]{cleveland79} Cleveland, W.~S. 1979, Journal of the 
American Statistical Association, 74, 829

\bibitem[Cohen et al.(1999)]{cohen99} Cohen,~M., Walker,~R.~G., Carter,~B.,
Hammersley,~P., Kidger,~M., \& Noguchi,~K. 1999, \aj, 117, 1864

\bibitem[Elias et al.(1982)]{elias82} Elias,~J.~H., Frogel,~J.~A., 
Matthews,~K., \& Neugebauer,~G. 1982, \aj, 87, 1029

\bibitem[Feige(1958)]{feige58} Feige,~J. 1958, \apj, 128, 267

\bibitem[Filippenko(2005)]{filippenko05} Filippenko,~A. 2005, in The Fate of the
Most Massive Stars, ed. R. Humphreys \& K. Stanek (San Francisco: ASP), p. 33

\bibitem[Filippenko, Li, \& Treffers(2009)]{filippenko09}
Filippenko, A., Li, W. D., \& Treffers, R. R. 2009, in prep.

\bibitem[Filippenko et al.(2001)]{filippenko01}
Filippenko, A., Li, W. D., Treffers, R. R., Modjaz, M. 2001, ASPC, 246, 121

\bibitem[Folatelli et al.(2006)]{folatelli06} Folatelli,~G., et al. 2006, 
  \apj, 641, 1039

\bibitem[Folatelli et al.(2009)]{folatelli09} Folatelli,~G., et al. 2009,
  submitted

\bibitem[Freedman et al.(2009)]{freedman09} 
Freedman,~W.~L., et al. 2009, ApJ, in press

\bibitem[Frieman et al.(2008)]{frieman08}
Frieman, J. A., et al. 2008, \aj, 135, 338

\bibitem[Fukugita et al.(1996)]{fukugita96} Fukugita,~M., Ichikawa,~T., 
  Gunn,~J.~E., Doi,~M., Shimasaku,~K., Schneider,~D.~P. 1996, \aj, 111, 1748

\bibitem[Garavini et al.(2005)]{garavini05}
Garavini, G., et al. 2005, \aj, 130, 2278

\bibitem[Giannantonio et al.(2008)]{giannantonio08}
Giannantonio, T., et al. 2008, \prd, 77,  [12]12350

\bibitem[Hamuy et al.(1992)]{hamuy92} 
Hamuy,~M., et al. 1992, \pasp, 104, 533

\bibitem[Hamuy et al.(1996)]{hamuy96} 
Hamuy,~M., et al. 1996, \aj, 112, 2408

\bibitem[Hamuy et al.(2006)]{hamuy06} 
Hamuy,~M., et al. 2006, \pasp, 118, 2 (H06)

\bibitem[Hicken et al.(2007)]{hicken07}
Hicken, M., et al. 2007, \apjl, 669, L17 

\bibitem[Hicken et al.(2009)]{hicken09}
Hicken, M., et al. 2009, \apj, 700, 331

\bibitem[Hillenbrand et al.(2002)]{hillenbrand02} Hillenbrand,~L.~A., 
  Foster,~J.~B., Persson,~S.~E., \& Matthews,~K. 2002, \pasp, 114, 708

\bibitem[Jha et al.(2006)]{jha06} Jha,~S., et al. 2006, \aj, 131, 527

\bibitem[Johnson \& Morgan(1953)]{johnson53}
Johnson, H. L., Morgan, W. W. 1953, \apj, 117, 313

\bibitem[Krisciunas et al.(2003)]{krisciunas03} 
Krisciunas, K., et al. 2003, \aj, 125, 166

\bibitem[Krisciunas et al.(2004)]{krisciunas04} Krisciunas, K., Phillips,
  M.~M., \& Suntzeff, N.~B. 2004, \apj, 602, L81

\bibitem[Landolt(1983)]{landolt83} Landolt,~A.~U. 1983, \aj, 88, 439

\bibitem[Landolt(1992)]{landolt92} Landolt,~A.~U. 1992, \aj, 104, 340

\bibitem[Leggett et al.(2006)]{leggett06} Leggett,~S.~K., et al. 2006,
  \mnras, 373, 781 

\bibitem[Madore(2001)]{madore01}
Madore, B. 2001, private communication

\bibitem[Maeda et al.(2009)]{maeda09} Maeda,~K., Kawabata,~K., Li,~W.
   Tanaka,~M., Mazzali,~P.~A., Hattori,~T., Nomoto,~K., \&
   Filippenko,~A.~V. 2009, \apj, 690, 1745

\bibitem[Perlmutter et al.(1999)]{perlmutter99} Perlmutter,~S., et al. 1999, 
  \apj, 517, 565

\bibitem[Persson et al.(1998)]{persson98}Persson,~S.~E., Murphy,~D.~C., 
  Krzeminski,~W., Roth,~M., \& Rieke,~M.~J. 1998, \aj, 116, 2475

\bibitem[Persson et al.(2004)]{persson04}Persson,~S.~E., Madore,~B.~F., 
  Krzeminski,~W., Freedman,~W.~L., Roth,~M., \& Murphy,~D.~C. 2004, \aj, 
  128, 2239

\bibitem[Phillips(1993)]{phillips93} Phillips, M.~M. 1993, \apj, 413, L105

\bibitem[Phillips et al.(2006)]{phillips06} 
Phillips,~M.~M., et al. 2006, \aj, 131, 2615

\bibitem[Phillips et al.(2007)]{phillips07}
Phillips,~M.~M., et al. 2007, \pasp, 119, 360

\bibitem[Prieto et al.(2007)]{prieto07}
Prieto, J. L., et al. 2007, arXiv:0706.4088

\bibitem[Quimby(2006)]{quimby06} Quimby,~R.~M. 2006, Ph.~D. thesis, Univ. Texas

\bibitem[Riess et al.(1998)]{riess98}
Riess,~A.~G., et al. 1998, \aj, 116, 1009

\bibitem[Riess et al.(1999)]{riess99}
Riess,~A.~G., et al. 1999, \aj, 117, 707 

\bibitem[Schlegel, Finkbeiner, \& Davis(1998)]{schlegel98} 
Schlegel,~D.~J., Finkbeiner,~D.~P., \& Davis,~M. 1998, \apj, 500, 525 

\bibitem[Schweizer et al.(2008)]
{schweizer08} Schweizer, F., et al. 2008, \aj, 136, 1482

\bibitem[Smith et al.(2002)]{smith02} 
Smith,~J.~A., et al. 2002, \aj,  123, 2121

\bibitem[Stritzinger et al.(2002)]{stritzinger02} 
Stritzinger,~M., et al. 2002, \aj, 124, 2100

\bibitem[Stritzinger et al.(2005)]{stritzinger05} Stritzinger,~M.,
  et al. 2005, \pasp, 117, 810
  
  \bibitem[Stritzinger et al.(2009)]{stritzinger09} Stritzinger,~M.,
  et al. 2009, \apj, 696, 713

\bibitem[Suntzeff(2000)]{suntzeff00} Suntzeff,~N.~B. 2000, in Cosmic
Explosions, ed. S. S. Holt \& W. W. Zhang (New York: AIP), 65

\bibitem[Taubenberger et al.(2008)]{taubenberger08} Taubenberger, S., et al.
  2008, \mnras, 385, 75
  
\bibitem[Williams(1997)]{williams97} Williams, A.~J. 1997, Pub. Astron. Soc. Australia, 14, 208

\bibitem[Wood-Vasey et al.(2007)]{wood-vasey07} Wood-Vasey,~W.~M., et al. 2007,
  \apj, 666, 694

\bibitem[Wood-Vasey et al.(2008)]{wood08} 
Wood-Vasey,~W.~M., et al. 2008, \apj, 689, 377 

\end{thebibliography}
\end{document}